\documentclass[11pt,a4paper]{article}
\usepackage{amsmath,amssymb,bbm}
\usepackage{latexsym}
\usepackage{graphicx}
\usepackage{setspace}
\usepackage{cite}

\renewcommand{\theequation}{\thesection.\arabic{equation}}
\renewcommand{\thefootnote}{\fnsymbol{footnote}}
\newlength{\extraspace}
\newlength{\extraspaces}
\newcommand{\be}{\begin{equation}}
\newcommand{\ee}{\end{equation}}
 
\newcommand{\ba}{\begin{eqnarray}}
\newcommand{\ea}{\end{eqnarray}}

\newcommand{\bas}{\begin{eqnarray*}}
\newcommand{\eas}{\end{eqnarray*}}
 
\newcommand{\bea}{\begin{eqnarray}}
\newcommand{\eea}{\end{eqnarray}}

\newcounter{subequation}[equation]
\makeatletter

\expandafter\let\expandafter
\reset@font\csname reset@font\endcsname

\def\subeqnarray{\arraycolsep1pt
    \def\@eqnnum\stepcounter##1{\stepcounter{subequation}%
        {\reset@font\rm(\theequation\alph{subequation})}}
\jot5mm     \eqnarray}

\def\subarray{\arraycolsep1pt
    \def\@eqnnum\stepcounter##1{\stepcounter{subequation}%
        {\reset@font\rm(\alph{subequation})}}
\jot5mm     \eqnarray}

\makeatother
\newcommand{\Zom}{\mathbb{Z}}
\newcommand{\one}{\mathbbm{1}}

\newcommand{\ra}{\rightarrow}

\newcommand{\is}{ &\! =\! & }
\newcommand{\nonum}{\nonumber \\[1.5mm]}
\newcommand{\sspace}{\makebox[1cm]{ }}
\newcommand{\bspace}{\makebox[2cm]{ }}

\newcommand{\half}{{\textstyle{\frac{1}{2}}}}

\newcommand{\Tr}{{\rm Tr}}

\renewcommand{\th}{{\theta}}
\newcommand{\eps}{\epsilon}
\newcommand{\lb}{\lambda}
\newcommand{\om}{\omega}

\newcommand{\dd}{{\partial}}

\newcommand{\cB}{{\cal B}}

\newcommand{\cD}{{\cal D}}
\newcommand{\cE}{{\cal E}}

\newcommand{\cJ}{{\cal J}}
\newcommand{\cL}{{\cal L}}

\newcommand{\cM}{{\cal M}}
\newcommand{\cO}{{\cal O}}

\newcommand{\cR}{{\cal R}}
\newcommand{\cS}{{\cal S}}

\newcommand{\cZ}{{\cal Z}}

\newcommand{\p}{\partial}
\newcommand{\I}{\mathrm{i}}
\newcommand{\Db}{\bar{D}}

\newcommand{\bg}{\bar{g}}
\newcommand{\ub}{\bar{u}}

\newcommand{\mn}{{\mu\nu}}

\newcommand{\m}{\mu}
\newcommand{\e}{\mathrm{e}}

\begin{document}
%------------------------------------------------------------------------------

\thispagestyle{empty}
\begin{flushright} \small
MZ-TH/12-01
\end{flushright}
\bigskip

\begin{center}
 {\LARGE\bfseries   Quantum Einstein Gravity\footnote{To appear in the
 \textit{New Journal of Physics} special issue on Quantum Einstein Gravity.}\\[10mm] 
%\LARGE\bfseries   in two lines  %\\[1.5ex]
%\LARGE\bfseries  
}

Martin Reuter and Frank Saueressig \\[3mm]
{\small\slshape
Institute of Physics, University of Mainz\\
Staudingerweg 7, D-55099 Mainz, Germany \\[1.1ex]
{\upshape\ttfamily reuter@thep.physik.uni-mainz.de} \\
{\upshape\ttfamily saueressig@thep.physik.uni-mainz.de} }\\
\end{center}
\vspace{10mm}

\hrule\bigskip

\centerline{\bfseries Abstract} \medskip
\noindent
We give a pedagogical introduction to the basic ideas and concepts of the Asymptotic Safety program in Quantum Einstein Gravity. Using the continuum approach based upon the effective average action, we summarize the state of the art of the field with a particular focus on the evidence supporting the existence of the non-trivial renormalization group fixed point at the heart of the construction. As an application, the multifractal structure of the emerging space-times is discussed in detail. In particular, we compare the continuum prediction for their spectral dimension with Monte Carlo data from the Causal Dynamical Triangulation approach.
\bigskip
\hrule\bigskip
\newpage

\begin{spacing}{1.5}
% -------------------- Introduction -------------------
\section{Introduction}
\label{sect:1}
\renewcommand{\thefootnote}{\arabic{footnote}}
\setcounter{footnote}{0}
%------------------------------------------------------
Finding a consistent and fundamental quantum theory
for gravity is still one of the most challenging
open problems in theoretical high energy physics to date \cite{QGbooks}.
As is well known, the perturbative quantization of the classical description for gravity, 
General Relativity, results in a non-renormalizable quantum theory \cite{tHooft:1974bx,Goroff:1985sz,vandeVen:1991gw}. 
One possible lesson drawn from this result may assert that gravity constitutes an effective field theory valid at low energies, whose UV completion requires 
the introduction of new degrees of freedom and symmetries. This is the path followed, e.g., by string
theory. In a less radical approach, one retains the fields and symmetries known from
General Relativity and conjectures that gravity constitutes a fundamental theory
at the non-perturbative level. One proposal along this line is
the Asymptotic Safety scenario \cite{livrev,oliverbook} which, motivated by gravity in $2+\eps$ dimensions \cite{Christensen:1978sc,Gastmans:1977ad}, was
initially put forward by Weinberg \cite{wein,Weinproc1}. The key ingredient in this construction 
is a non-Gaussian fixed point (NGFP) of the gravitational renormalization group (RG) flow, 
which controls the behavior of the theory at high energies and renders physical quantities safe from unphysical divergences. 

The primary tool for investigating this scenario is the functional renormalization group equation (FRGE) for gravity \cite{mr}, which constitutes the spring-board for the detailed analysis of the gravitational RG flow at the non-perturbative level 
\cite{mr,percadou,oliver1,frank1,oliver2,oliver3,oliver4,souma,frank2,prop,perper1,codello,litimgrav,essential,r6,MS1,Codello:2008vh,creh1,creh2,creh3,JE1,HD1,HD2,JEUM,Rahmede:2011zz,Benedetti:2010nr}.\footnote{Independent support for the Asymptotic Safety conjecture comes from a 2-dimensional symmetry reduction of the gravitational path-integral \cite{max}.} The FRGE defines a Wilsonian RG flow on a theory space which consists of all diffeomorphism invariant functionals of the metric $g_{\mu \nu}$ and yielded substantial evidence for the existence and predictivity of the NGFP underlying the Asymptotic Safety conjecture. The theory emerging from this construction, Quantum Einstein Gravity (henceforth denoted ``QEG''), defines a consistent and predictive quantum theory for gravity within the framework of quantum field theory. We stress that QEG is not a quantization of classical General Relativity: its bare action corresponds to a non-trivial fixed point of the RG flow and is a \emph{prediction} therefore. 

The approach of \cite{mr} employs the effective average
action $\Gamma_k$ \cite{avact,ym,avactrev,ymrev} which has crucial 
advantages as compared to other
continuum implementations of the Wilsonian RG flow \cite{bagber}. 
In particular, the RG scale dependence of $\Gamma_k$ is
governed by the FRGE \cite{avact}
\be\label{E9}
k \p_k \Gamma_k[\Phi, \bar{\Phi}] = \frac{1}{2} {\rm Str} \left[ 
\left( \frac{\delta^2 \Gamma_k}{\delta \Phi^A \delta \Phi^B} + \cR_k \right)^{-1} k \p_k \cR_k 
\right] \, . 
\ee
Here $\Phi^A$ is the collection of all dynamical fields considered and $\bar{\Phi}^A$ denotes their background counterparts.
Moreover $\cR_k$ is a matrix-valued infrared cutoff, which provides a $k$-dependent mass-term for fluctuations with momenta $p^2 \ll k^2$, while vanishing for $p^2 \gg k^2$.
Solutions of the flow equation give rise to 
 families of effective field theories $\{ \Gamma_k[g_{\mu \nu}], 0 \le k < \infty \}$ 
labeled by the coarse graining scale $k$. 
The latter property opens the door to a rather direct extraction of physical 
information from the RG flow, at least in single-scale cases: If the physical
 process under consideration involves 
 a single typical momentum scale $p_0$ only, it can be described by a tree-level 
evaluation of $\Gamma_k[g_{\mu \nu}]$, with $k = p_0$.\footnote{ 
The precision which can be achieved by this effective field theory 
description depends on the size of the fluctuations relative 
to mean values. If they turn out large, or if more than one scale is involved, it might be necessary to go beyond the tree-level 
analysis.}
 
A striking consequence of the scale-dependence in $\Gamma_k$ is the observation that the effective QEG space-time should have certain features in common with a fractal \cite{oliver1,oliver2}. The 
property underlying this assertion is  that the effective field equations derived from the gravitational average action equip every given smooth space-time manifold with, in principle, infinitely many different (pseudo) Riemannian structures, one for each coarse graining scale \cite{jan1,jan2}. Thus, very much like in the famous example of the coast line of England \cite{mandel}, the proper length on a QEG space-time depends on the ``length of the yardstick'' used to measure it. Earlier on similar fractal properties had already been found in other quantum gravity theories, in particular near dimension 2 \cite{ninomiya}, in a non-asymptotically safe model \cite{Floreanini:1993na}, and by analyzing the conformal anomaly \cite{nino}.

In ref.\ \cite{oliver1} the consequences of this scale-dependence for the 4-dimensional graviton propagator has been studied in the regime of asymptotically large momenta and it has been found that near the Planck scale a kind of dynamical dimensional reduction occurs. As a consequence of the NGFP controlling the UV behavior of the theory, the 4-dimensional graviton propagator essentially behaves 2-dimensional on microscopic scales. Subsequently, the ``finger prints'' of the NGFP on the fabric of the effective QEG space-times have been discussed in \cite{oliver2}, where it was shown that Asymptotic Safety induces a characteristic self-similarity of space-time on length-scales below the Planck length $\ell_{\rm PL}$. In particular the spectral dimension of the effective QEG space-times in the asymptotic scaling regime implied by the NGFP becomes $d_s = d/2$ with $d$ the classical space-time dimension \cite{oliver1, oliver2,oliverfrac}. Based on this observation it was argued in a cosmological context that the geometry fluctuations originating from the scaling regime can give rise to a scale free spectrum of primordial density perturbations responsible for structure formation \cite{cosmo1,entropy}.

Along a different line of investigations, the Causal Dynamical Triangulation (CDT) approach has been developed and first Monte Carlo simulations were performed \cite{ajl1,ajl2,ajl34,ajl5,Benedetti:2009ge,Kommu:2011wd}, see \cite{Ambjorn:2009ts} for a recent review. In this framework one attempts to compute quantum gravity partition functions by numerically constructing the continuum limit of an appropriate statistical mechanics system. From the perspective of the latter, this limit amounts to a second order phase transition. If CDT and its counterpart QEG, formulated in the continuum by means of the average action, belong to the same universality class\footnote{For the time being this is merely a conjecture, of course, albeit a very natural one.} one may expect that the phase transition of the former is described by the non-trivial fixed point underlying the Asymptotic Safety of the latter.

Remarkably, ref. \cite{ajl34} reported results which indicated that the 4-dimensional CDT space-times, too, undergo a dimensional reduction from four to two dimensions as one ``zooms'' in on short distances. In particular it had been demonstrated that the spectral dimension $d_s$ measured in the CDT simulations has the very same limiting behaviors, $4 \rightarrow 2$, as in QEG. Therefore it was plausible to assume that both approaches indeed ``see'' the same continuum physics. 

This interpretation became problematic, however, when it turned out that the Monte Carlo data corresponds to a regime where the cutoff length inherent in the triangulations is still significantly larger than the Planck length. The situation became even more puzzling when ref.\ \cite{Benedetti:2009ge} carried out CDT simulations for $d=3$ macroscopic dimensions, which favor a value near $d_s=2$ on the shortest length-scale probed. Furthermore, the authors of ref.\ \cite{laiho-coumbe} reported simulations within the {\it euclidean} dynamical triangulation (EDT) approach in $d=4$, where the 
 spectral dimension dropped from 4 to about 1.5. Obviously, both of these observations conflict the QEG expectations if one interprets the latter dimension as the value in the continuum limit.

In order to resolve this puzzle, \cite{frankfrac} computed several types of scale dependent effective dimensions, specifically the spectral dimension $d_s$ and the walk dimension $d_w$ for the effective QEG space-times. Surprisingly, the analysis revealed a further regime which exhibits the phenomenon of dynamical dimensional reduction on length scales slightly {\it larger} than $\ell_{\rm PL}$. There the spectral dimension is even smaller than near the fixed point, namely $d_s=4/3$ in the case of 4 dimensions classically. Moreover it was found that the (3-dimensional) results reported in \cite{Benedetti:2009ge} are in good agreement with QEG. This analysis also confirmed the supposition \cite{Benedetti:2009ge} that the shortest possible length scale achieved in the simulations is not yet close to the Planck length. Rather the Monte Carlo data probes the transition between the classical and the newly discovered ``semi-classical'' regime.

It is intriguing that Loop Quantum Gravity and spin foam models also show indications for a similar dimensional reduction \cite{modesto2008,Modesto:2009kq,modesto-caravelli2009}, with some hints for an intermittent regime where the spectral dimension is smaller than in the deep ultraviolet. In ref.\ \cite{carlip} an argument based upon the strong coupling limit of the Wheeler-DeWitt equation was put forward as a possible explanation of this dimensional reduction. Within non-commutative geometry Connes et al.\ \cite{ncgeom} interpreted the dynamical dimensional reduction to $d_s = 2$, which was observed in QEG, in the context of the derivation of the Standard Model from a spectral triple. In fact, from the data encoded in a spectral triple, its Dirac operator in particular, one can compute a type of spectral dimension of the resulting non-commutative space which is closely related to the one considered here. Also for standard fractals such as Cantor sets, it has been possible to find spectral triples representing them and to compute the corresponding dimensions \cite{spec-trip-frac}. 

Furthermore, a number of model systems (quantum sphere, $\kappa$-Minkowski space, etc.) give rise to a similar reduction as fully fledged quantum gravity  \cite{Dario-kappa}. Among other developments, these findings also motivated the investigation of physics on {\it prescribed} fractal space-times. In refs. \cite{calcagni-applications,calcagni-etal} a fractional differential calculus \cite{calcagni-reviews} was employed in order to incorporate fractal features, and in  \cite{dunne-phot}  recent exact results on spectral zeta-functions on certain fractals \cite{dunne-complexdim} were used to study the thermodynamics of photons on fractals. In ref.\ \cite{hill} matter quantum field theories were constructed 
and renormalized on a fractal background. The almost universal appearance of fractional properties of space-time and its accessibility in various, a priori different, approaches to quantum gravity make the generalized notions of dimensionality
which we are going to review here a valuable tool in comparing the physics content of these different formulations.

The present article is intended to provide the necessary background for understanding the
frontier developments in Asymptotic Safety. In the next section we follow \cite{livrev} and
 introduce the general concepts related to the
Wilsonian picture of renormalization: theory space, renormalization group flows, non-perturbative renormalizability,  
and the non-perturbative approximation scheme of truncating theory space. 
The construction of the effective average action and its FRGE for gravity \cite{mr} is reviewed in section \ref{sect:3}.
In section 4 we illustrate the most commonly used 
approximation scheme of ``truncating theory space''
by means of a simple example, the so-called Einstein-Hilbert 
truncation. Section \ref{sect:6} contains a state of the art summary of the results obtained
using truncated flow equations, with an emphasis on the question
as to whether there exists a non-trivial fixed point for the average action's
RG flow. If so, QEG could be established as a fundamental theory of
quantum gravity which is non-perturbatively renormalizable and ``asymptotically
safe'' from unphysical divergences. The remainder of the review is dedicated
to the discussion of the multifractal space-times emerging from QEG: section \ref{sect:7} reviews 
the structures underlying these fractal properties, the scale-dependent space-time metrics.
Subsequently, the spectral, walk, and Hausdorff dimensions introduced in appendix \ref{sect:7c}
 are computed in the framework of QEG in sections \ref{sect:7d} and  \ref{sect:7e}.
 In particular subsection \ref{sect:7f} compares the QEG results for the spectral dimension with
the one found in the CDT program. 
We close with a short summary and some concluding remarks in section \ref{sect:7g}.

%%%%%%%%%%%%%%%%%%%%%%%%%%%%%%%%%%%%%%%%%%%%%%%%%%%%%%%%%%%%%%%%%%%%%%%
\section{RG flows, theory space, and Asymptotic Safety}
\setcounter{equation}{0}
\label{sect:2}
%%%%%%%%%%%%%%%%%%%%%%%%%%%%%%%%%%%%%%%%%%%%%%%%%%%%%%%%%%%%%%%%%%%%%%%

\begin{figure}[t]
\centering
\includegraphics[width=.9\columnwidth]{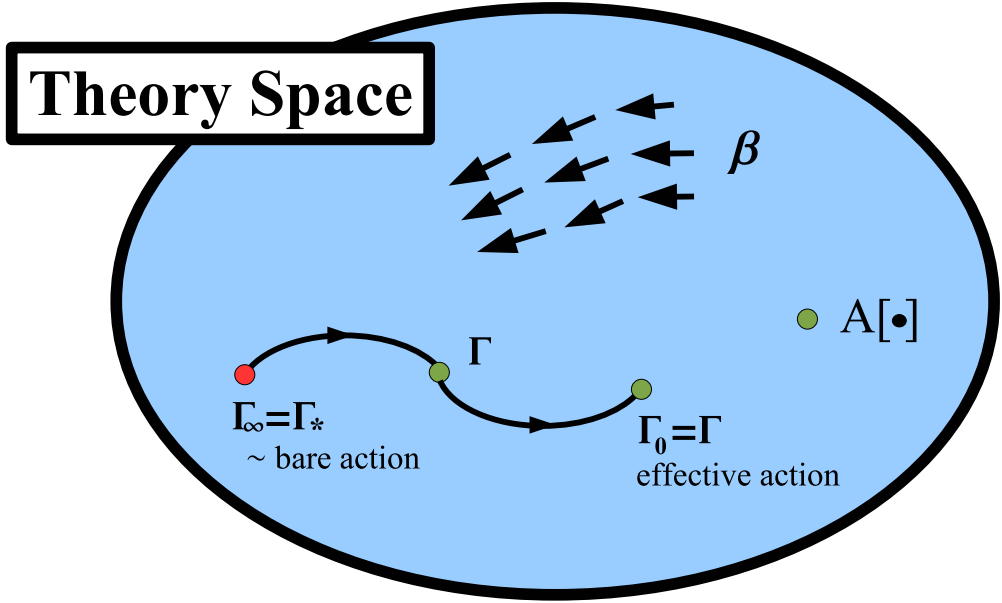}
\caption{\small The points of theory space are action functionals $A[\, \cdot \,]$. The RG equation defines a vector field $\vec \beta$ on this space; its integral curves are the RG trajectories $k \mapsto \Gamma_k$. They emanate from the fixed point action $\Gamma_*$ and end at the standard effective action $\Gamma$.}
\label{theoryspace}
\end{figure}

The key idea of the Wilsonian renormalization group is the description of a physical system in terms of a one-parameter family of effective actions, each valid at a certain energy scale $k$. The RG flow of the theory thereby connects the effective descriptions at different scales. The arena in which this dynamics takes place is the ``theory space'' sketched in fig.\ \ref{theoryspace}. In order to describe it, we first specify our ``theory'' by fixing its field content $\Phi(x)$ and, possibly, imposing certain symmetry requirements (a $\Zom_2$-symmetry for a single scalar, or diffeomorphism invariance if $\Phi$ denotes the space-time metric, for instance). The theory space corresponding to this theory consists of all (action) functionals $A: \Phi \mapsto A[\Phi]$ depending on this set of fields and compatible with the symmetry requirements. Thus the theory space $\{A[\, \cdot \,]\}$ is fixed once the field content and the symmetries are given.

Let us assume we can find a set of ``basis functionals''
$\{ P_\alpha[ \, \cdot \, ] \}$ so that every point of theory space has an expansion of the form
\be\label{Aexpansion}
A[\Phi, \bar{\Phi}] = \sum_{\alpha = 1}^\infty \, \ub_\alpha \, P_\alpha [\Phi, \bar{\Phi}] \, .
\ee
The basis $\{ P_\alpha[ \, \cdot \, ] \}$ will include both local field monomials and non-local
invariants and we may use the ``generalized couplings'' $\{ \bar u_\alpha , \alpha = 1,2, \cdots \}$
as local coordinates. More precisely, the theory space is coordinatized by the subset of 
``essential couplings'', i.e., those coordinates which cannot be absorbed by a field reparameterization.

Geometrically speaking the FRGE for the effective average action, eq.\ \eqref{E9},
 defines a vector field $\vec \beta$ on theory space. The integral curves along this vector field are the ``RG trajectories'' $k \mapsto \Gamma_k$ parameterized by the scale $k$. They start, for $k \ra \infty$, at the microscopic action $S$ and terminate at the ordinary effective action at $k=0$. The natural orientation of the trajectories is from higher to lower scales $k$, the direction of increasing ``coarse graining''. Expanding $\Gamma_k$ as in \eqref{Aexpansion},
\be\label{Gexpansion}
\Gamma_k[\Phi, \bar{\Phi}] = \sum_{\alpha = 1}^\infty \, \ub_\alpha(k) \, P_\alpha [\Phi, \bar{\Phi}] \, ,
\ee
the trajectory is described by infinitely many ``running couplings'' $\ub_\alpha(k)$. Inserting \eqref{Gexpansion} into the FRGE we obtain a system of infinitely many coupled differential equations for the $\ub_\alpha$'s:
\be\label{rgeqn1}
k \partial_k \, \ub_\alpha(k) = \overline{\beta}_\alpha(\ub_1 , \ub_2 , \cdots ; k) \; , \quad \alpha = 1,2,\cdots \, .
\ee
Here the ``beta functions'' $\overline{\beta}_\alpha$ arise by 
expanding the trace on the RHS of the FRGE in terms of $\{ P_\alpha[\, \cdot \, ] \}$, i.e.,
$\tfrac{1}{2} \Tr \left[ \cdots \right] = \sum_{\alpha = 1}^\infty \overline{\beta}_\alpha(\ub_1 , \ub_2 , \cdots ; k) P_\alpha[\Phi, \bar{\Phi}]$. The expansion coefficients $\overline{\beta}_\alpha$ have the 
interpretation of beta functions similar to those of perturbation 
theory, but not restricted to relevant couplings. In standard field theory
jargon one would refer to $\ub_\alpha(k =\infty)$ as the ``bare'' parameters and to 
$\ub_\alpha(k =0)$ as the ``renormalized'' or ``dressed'' parameters. 
   
The notation with the bar on $\ub_\alpha$ and $\overline{\beta}_\alpha$ 
is to indicate that we are still dealing with dimensionful 
couplings. Usually the flow equation is reexpressed in terms of the 
dimensionless couplings 
\be\label{dimlessu}
u_\alpha \equiv k^{-d_\alpha} \, \ub_\alpha \, ,
\ee
where $d_\alpha$ is the canonical mass dimension of $\ub_\alpha$. Correspondingly the essential $u_\alpha$'s are used as 
coordinates of theory space. The resulting 
RG equations 
\be 
k \dd_k u_\alpha(k) = \beta_\alpha(u_1, u_2, \cdots ) 
\label{E17}
\ee
are a coupled system of autonomous differential equations. 
The $\beta_\alpha$'s have no explicit $k$-dependence and define 
a ``time independent'' vector field on theory space. The RG trajectories
arise as solutions or, equivalently, integral curves of \eqref{E17}.

Based on these structures, the concept of renormalization can be understood as follows.
The boundary of theory space depicted in fig.\ \ref{theoryspace} is meant to separate points with coordinates $\{u_\alpha, \alpha = 1,2,\cdots\}$ with all the essential couplings $u_\alpha$ well defined, from points with undefined, divergent couplings. The basic task of renormalization theory consists in constructing an ``infinitely long'' RG trajectory which lies entirely within this theory space, i.e., a trajectory which neither leaves theory space (that is, develops divergences) in the UV limit $k \rightarrow \infty$ nor in the IR limit $k \rightarrow 0$. Every such trajectory defines one possible quantum theory.

The key idea of Asymptotic Safety is to perform the UV limit $k \rightarrow \infty$ at a fixed point $\{u_\alpha^*, \alpha = 1,2,\cdots\} \equiv u^*$ of the RG flow. The fixed point is a zero of the vector field $\vec \beta \equiv (\beta_\alpha)$, i.e., $\beta_\alpha(u^*) = 0$ for all $\alpha = 1,2,\cdots$. The RG trajectories have a low ``velocity'' near a fixed point because the $\beta_\alpha$'s are small there and directly at the fixed point the running stops completely. As a result, one can ``use up'' an infinite amount of  RG time near/at the fixed point if one bases the quantum theory on a trajectory which runs into a fixed point for $k \rightarrow \infty$. The fact, that, in the UV limit  the trajectory ends at a fixed point, an ``inner point'' of theory space giving rise to a well behaved action functional,  ensures that the trajectory does not escape from theory space, i.e., does not develop pathological properties such as divergent couplings. For $k \rightarrow \infty$ the resulting quantum theory is ``asymptotically safe''  from unphysical divergences.

In a sense, standard perturbation theory takes the $k \rightarrow \infty$-limit at the Gaussian fixed point (GFP), i.e., a fixed point where $u_\alpha^* = 0, \forall \alpha = 1,2,\ldots$. This construction is the one   underlying asymptotic freedom. More general, however, one can also use a non-Gaussian fixed point (NGFP) for letting $k \rightarrow \infty$, where, by definition, not all of the coordinates $u^*_\alpha$ vanish. In the context of gravity, Weinberg \cite{wein} proposed that the UV completion of the theory is precisely given by such a NGFP, which therefore constitutes the essential ingredient in the Asymptotic Safety program.

\begin{figure}[t]
\centering
\includegraphics[width=.67\columnwidth]{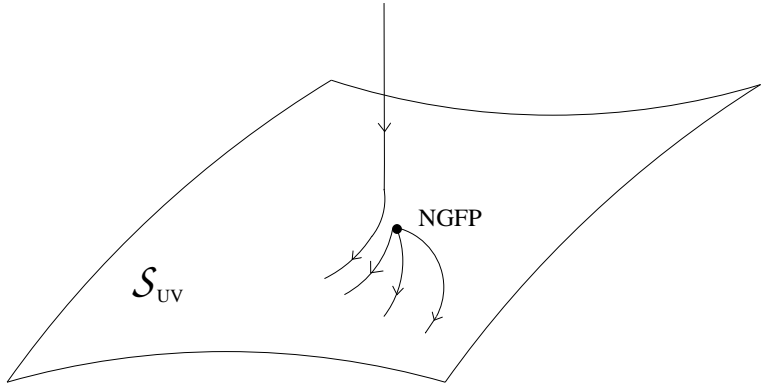}
\caption{\small Schematic picture of the UV critical hypersurface $\cS_{\rm UV}$ of the NGFP. It is spanned by RG trajectories emanating from the NGFP as the RG scale $k$ is lowered. Trajectories not in the surface are attracted towards $\cS_{\rm UV}$ as $k$ decreases. (The arrows point in the direction of decreasing $k$, from the ``UV'' to the ``IR''.)}
\label{UVsurface}
\end{figure}

Given a NGFP, an important concept is its {\it UV critical hypersurface} $\cS_{\rm UV}$, or synonymously, its {\it unstable manifold}. By definition, it consists of all points of theory space which are pulled into the NGFP by the inverse RG flow, i.e., for {\it in}creasing $k$.
Its dimensionality ${\rm dim}\left({\cal S}_{\rm UV}\right)\equiv \Delta_{\rm UV}$
is given by the number of attractive (for {\it in}creasing cutoff $k$) 
directions in the space of couplings.

For the RG equations \eqref{E17}, 
the linearized flow near the fixed point is governed by the Jacobi matrix
${\bf B}=(B_{\alpha \gamma})$, $B_{\alpha \gamma}\equiv\partial_\gamma  \beta_\alpha(u^*)$:
\begin{eqnarray}
\label{H2}
k\,\partial_k\,{u}_\alpha(k)=\sum\limits_\gamma B_{\alpha \gamma}\,\left(u_\gamma(k)
-u_{\gamma}^*\right)\;.
\end{eqnarray}
The general solution to this equation reads
\begin{eqnarray}
\label{H3}
u_\alpha(k)=u_{\alpha}^*+\sum\limits_I C_I\,V^I_\alpha\,
\left(\frac{k_0}{k}\right)^{\theta_I}
\end{eqnarray}
where the $V^I$'s are the right-eigenvectors of ${\bf B}$ with eigenvalues 
$-\theta_I$, i.e., $\sum_{\gamma} B_{\alpha \gamma}\,V^I_\gamma =-\theta_I\,V^I_\alpha$. Since ${\bf B}$ is not symmetric in general the $\theta_I$'s are not guaranteed to be real. We
assume that the eigenvectors form a complete system though. Furthermore, $k_0$ 
is a fixed reference scale, and the $C_I$'s are constants of integration. The quantities $\theta_I$ are referred to as \emph{critical exponents}
since when the renormalization group is applied to critical 
phenomena (second order phase transitions) the traditionally 
defined critical exponents are related to the $\theta_I$'s in a 
simple way \cite{avactrev}.

If $u_\alpha(k)$ is to describe a trajectory in $\cS_{\rm UV}$,
 $u_\alpha(k)$ must 
approach $u_{\alpha}^*$ in the limit
$k\rightarrow\infty$ and therefore we must set $C_I=0$ for all $I$ with 
${\rm Re}\,\theta_I<0$. Hence the dimensionality $\Delta_{\rm UV}$ equals the 
number of ${\bf B}$-eigenvalues with a negative real part, i.e., the number of
$\theta_I$'s with ${\rm Re}\,\theta_I >0$. The corresponding eigenvectors 
span the tangent space to
$\cS_{\rm UV}$ at the NGFP.
If we {\it lower} the cutoff for a generic trajectory with all
$C_I$ nonzero, only $\Delta_{\rm UV}$ 
``relevant'' parameters
corresponding to the eigendirections tangent to $\cS_{\rm UV}$ grow 
(${\rm Re}\, \theta_I > 0$), while the remaining ``irrelevant'' couplings 
pertaining to the eigendirections normal to $\cS_{\rm UV}$ decrease 
(${\rm Re}\, \theta_I < 0$). Thus near the NGFP a generic trajectory 
is attracted towards $\cS_{\rm UV}$, see fig.\ \ref{UVsurface}.

Coming back to the Asymptotic Safety construction, let us now 
use this fixed point in order to take the limit $k \ra \infty$. 
The trajectories which define an infinite cutoff limit are 
special in the sense that all irrelevant couplings are set to zero: $C_I = 0$ 
if ${\rm Re} \, \theta_I < 0$. These conditions place the trajectory 
exactly on $\cS_{\rm UV}$. There is a $\Delta_{\rm UV}$-parameter family 
of such trajectories, and the experiment must decide which one is 
realized in Nature. 
Therefore the predictive power of the theory increases with decreasing
 dimensionality of ${\cal S}_{\rm UV}$, i.e., number of UV attractive eigendirections of the
NGFP. If $\Delta_{\rm UV} < \infty$, the quantum field 
theory thus constructed is comparable to and as predictive as a perturbatively
renormalizable model with $\Delta_{\rm UV}$ ``renormalizable couplings'', i.e.,
couplings relevant at the GFP, see \cite{livrev} for a more detailed discussion.

Up to this point our discussion did not involve any approximation.
In practice, however, it is usually impossible to find exact solutions 
to the flow equation. As a way out, one could evaluate
the trace on the RHS of the FRGE by expanding it with respect to 
some small coupling constant, for instance, thus recovering the familiar 
perturbative beta functions. A more interesting option which gives rise 
to non-perturbative approximate solutions is to truncate the theory 
space $\{A[\,\cdot\,]\}$. The basic idea is to project the RG flow onto a
finite dimensional subspace of theory space. The subspace should be chosen 
in such a way  that the projected flow encapsulates the essential physical 
features of the exact flow on the full space.

Concretely the projection onto a truncation subspace is performed 
as follows. One makes an ansatz of the form 
$ 
\Gamma_k[\Phi, \bar{\Phi}] = \sum_{i=1}^N {\ub}_i(k) P_i[\Phi, \bar{\Phi}]\,,
%\label{E14}
$
where the $k$-independent functionals
$\{P_i[\, \cdot \,], i=1,\cdots,N \}$ form a `basis' on the subspace selected. 
For a scalar field $\phi$, say, examples include pure potential terms 
$\int d^dx \phi^m(x)$, 
$\int d^dx \phi^n(x) \ln \phi^2(x)$, $\cdots$, a standard kinetic 
term $\int \! d^dx (\dd \phi)^2$, higher order derivative terms 
$\int \! d^dx \, \phi \left({\dd^2} \right)^n \phi$, 
 $\cdots$, and non-local terms like 
$\int \!d^dx \, \phi \ln(-\dd^2) \phi$, $\cdots$.   
Even if $S = \Gamma_{\infty}$ is simple, a standard $\phi^4$ action,
say, the evolution from $k =\infty$ downwards will generate such     
terms, a priori only constrained by symmetry requirements.  
The difficult task in practical RG applications consists in 
selecting a set of $P_i$'s which, on the one hand, is generic enough 
to allow for a sufficiently precise description of the physics one 
is interested in, and which, on the other hand, is small enough to be 
computationally manageable. 

The projected RG flow is described by a set of ordinary (if $N < \infty$) 
differential equations for the couplings $\ub_i(k)$. They arise as follows.
 Let us assume we 
expand the $\Phi$-dependence of $\frac{1}{2}{\rm Tr}[\cdots]$ 
(with the ansatz for $\Gamma_k[\Phi, \bar{\Phi}]$ inserted) in a basis
$\{P_{\alpha}[\, \cdot \,]\}$ of the {\it full} theory space which contains  
the $P_i$'s spanning the truncated space as a subset: 
\be 
\frac{1}{2} {\rm Tr}[\cdots] =  
\sum_{\alpha =1}^{\infty} \overline{\beta}_{\alpha}(\ub_1, \cdots, \ub_N;k) 
\, P_{\alpha}[\Phi, \bar{\Phi}]     
=      
\sum_{i =1}^N \overline{\beta}_i(\ub_1, \cdots, \ub_N;k) 
\, P_i[\Phi, \bar{\Phi}] + {\rm rest}\,. 
\label{E15}
\ee
Here the ``rest'' contains all terms outside the truncated theory 
space; the approximation consists in neglecting precisely
those terms. Thus, equating (\ref{E15}) to the LHS of the flow equation,      
$\dd_t \Gamma_k = \sum_{i=1}^N \dd_t \ub_i(k) P_i$, the linear independence 
of the $P_i$'s implies the coupled system of ordinary differential 
equations 
\be 
\dd_t \ub_i(k) = \overline{\beta}_i(\ub_1,\cdots , \ub_N;k)\,,
\quad i = 1, \cdots, N\,.
\label{E16}
\ee
Solving (\ref{E16}) one obtains an {\it approximation} to the 
exact RG trajectory projected onto the chosen subspace. Note that 
this approximate trajectory does, in general, not coincide with 
the projection of the exact trajectory, but if the subspace 
is well chosen, it will not be very different from it. In fact, the most 
non-trivial problem in using truncated flow equations is to
 find and justify a truncation subspace which should be as low dimensional 
as possible to make the calculations feasible, 
but at the same time large enough to describe at least qualitatively 
the essential physics. We shall return to the issue of 
testing the quality of a given truncation later on. 

%%%%%%%%%%%%%%%%%%%%%%%%%%%%%%%%%%%%%%%%%%%%%%%%%%%%%%%%%%%%%%%%%
\section{The effective average action for gravity}
\label{sect:3}
\setcounter{equation}{0}
%%%%%%%%%%%%%%%%%%%%%%%%%%%%%%%%%%%%%%%%%%%%%%%%%%%%%%%%%%%%%%%%%

In the case of QEG, ideally, we would like theory space to consist of functionals
$A[g_{\mu\nu}]$ depending on a symmetric tensor field, the metric, in a diffeomorphism
invariant way. Given a theory space, the form of the
FRGE and, as a result, the vector field $\vec \beta$ are completely fixed.
However, in the case of gravity it is much harder to make this idea work in
a concrete way as compared to a simple matter field theory on a non-dynamical
space-time, for instance. The reasons are of both conceptual and technical nature:

\noindent
\textbf{(1)} The theory of quantum gravity we are aiming at should be formulated in a
background independent way. It should \emph{explain} rather than presuppose the existence and
the properties of space-time. Hence no special space-time manifold, in particular no special
causal, let alone Riemannian, structure should play a distinguished role at the fundamental
level of the theory.
In particular it should also cover ``exotic'' phases in which the metric is degenerate or
has no expectation value at all. Since almost our entire repertoire of quantum field theory
methods applies only to the case of an externally given space-time manifold, usually Minkowski
space, this is a severe conceptual problem. In fact, it is \emph{the} central challenge for
basically all approaches to quantum gravity, in one guise or another \cite{QGbooks}.

\noindent
\textbf{(2)} The second difficulty, while less deep, is important for the practical applicability
of the RG methods to be developed. It occurs already in the standard functional integral
quantization of gauge or gravity theories, and is familiar 
from Yang-Mills theories. If one gauge-fixes the functional 
integral with an ordinary (covariant) gauge fixing condition
like $\dd^{\mu} A_{\mu}^a =0$, couples the (non-abelian) gauge field $A_{\mu}^a$ to a source, 
and constructs the ordinary effective action, the resulting 
functional $\Gamma[A_{\mu}^a]$ is {\it not} invariant under 
the gauge transformations of $A_{\mu}^a$, $A_{\mu}^a \mapsto 
A_{\mu}^a + D_{\mu}^{ab}(A) \, \om^b$. Only at the level of physical   
quantities constructed from $\Gamma[A_{\mu}^a]$, 
S-matrix elements for instance, gauge invariance is recovered. 

\noindent
\textbf{(3)} A more profound problem is related to the fact that in a gauge
theory a ``coarse graining'' based on a naive Fourier decomposition 
of $A_{\mu}^a(x)$ is not gauge covariant and hence not physical.
In fact, if one were to gauge transform a slowly varying 
$A_{\mu}^a(x)$ using a parameter function $\om^a(x)$ with a fast 
$x$-variation, a gauge field with a fast $x$-variation would arise
which, however, still describes the same physics. 
In a non-gauge theory the coarse graining is performed by 
expanding the field in terms of eigenfunctions of the (positive) 
operator $-\dd^2$ and declaring its eigenmodes `long' or `short' 
wavelength depending on whether the corresponding eigenvalue $p^2$ is smaller 
or larger than a given $k^2$. In a gauge theory the best one can do 
in installing this procedure is to expand with respect to 
the {\it covariant} Laplacian or a similar operator, and then 
organize the modes according to the size of their eigenvalues.
While gauge covariant, this approach sacrifices to some extent 
the intuition of a Fourier coarse graining in terms of slow and fast modes. 
Analogous remarks apply to theories of gravity covariant under 
general coordinate transformations. 

The key idea which led to a solution of all three problems
was the use of the background field method \cite{dewitt-books}. At first sight it seems to be
a contradiction in terms to use background fields in order to achieve ``background independence''.
However, actually it is not, since the background metric introduced, $\bar{g}_{\mu\nu}$, is kept
\emph{completely arbitrary}, and no physics ever may depend on it. In fact, it becomes a second argument
of $\Gamma_k[g_{\mu\nu},\bar{g}_{\mu\nu}]$ and may be freely chosen from the same function space the
dynamical metric ``lives'' in. The use of a background field also opens the door for complying with the
requirement of a gauge invariant effective action. It is well known \cite{back,joos} that there exist
special gauge choices, the so-called background gauge fixing conditions, that make the (ordinary) effective
action a gauge or diffeomorphism invariant functional of its arguments (including the background fields!).
As it turned out \cite{ym,mr} this technique also lends itself for implementing a covariant IR cutoff, and it is
at the core of the effective average action for Yang-Mills theories \cite{ym,ymrev} and for gravity \cite{mr}.
In the following we briefly  review the effective average action for gravity which has been introduced in ref.\ \cite{mr}.

The ultimate goal is to give meaning to an integral over `all' 
metrics $\gamma_{\mu\nu}$ of the form $\int \! \cD \gamma_{\mu\nu} \,
\exp\{ - S[\gamma_{\mu\nu}] + {\rm source \; terms}\}$ whose 
bare action $S[\gamma_{\mu\nu}]$ is invariant under general 
coordinate transformations,
\be 
\delta \gamma_{\mu\nu} = \cL_v \gamma_{\mu \nu} \equiv
v^{\rho} \dd_{\rho} \gamma_{\mu\nu} 
+ \dd_{\mu} v^{\rho} \gamma_{\rho \nu} + 
\dd_{\nu} v^{\rho} \gamma_{\rho \mu} \,,
\label{F1}
\ee
where $\cL_v$ is the Lie derivative with respect to the vector 
field $v^{\mu}\dd_{\mu}$. To start with we consider $\gamma_{\mu\nu}$
to be a Riemannian metric and assume that $S[\gamma_{\mu\nu}]$ is positive 
definite. Heading towards the background field formalism, the 
first step consists in decomposing the variable of integration 
according to $\gamma_{\mu\nu} = \bg_{\mu\nu} + h_{\mu\nu}$, 
where $\bg_{\mu\nu}$ is a fixed background metric. Note that 
we are not implying a perturbative expansion here, $h_{\mu\nu}$ 
is not supposed to be small in any sense. After the background split 
the measure $\cD \gamma_{\mu\nu}$ becomes $\cD h_{\mu\nu}$ 
and the gauge transformations which we have to gauge-fix read 
\be 
\delta h_{\mu\nu} = \cL_v \gamma_{\mu\nu} = 
\cL_v( \bg_{\mu\nu} + h_{\mu\nu})\,,\quad 
\delta \bg_{\mu\nu}= 0\,.
\label{F2}
\ee
Picking an a priori arbitrary gauge fixing condition $F_{\mu}(h;\bg) =0$ 
the Faddeev-Popov trick can be applied straightforwardly \cite{back}. 
Upon including an IR cutoff $\Delta_k S[h,C,\bar{C};\bar{g}]$ 
(cfg.\ eq.\ \eqref{F12} below) we are led to 
the following $k$-dependent generating functional $W_k$ for the 
connected Green functions:
\ba 
&& \exp\left\{ W_k[t^{\mu\nu},\sigma^{\mu}, \bar{\sigma}_{\mu}; 
\bg_{\mu\nu}] \right\} = \int \! \cD h_{\mu\nu} \cD C^\mu \cD \bar{C}_\mu
\,\exp\Big\{ -S[\bar{g}+h]-S_{\rm gf}[h;\bar{g}]
\nonum
&& \bspace \bspace 
-S_{\rm gh}[h,C,\bar{C};\bar{g}]-\Delta_k S[h,C,\bar{C};\bar{g}]
     -S_{\rm source} \Big\}\,.
\label{F3}
\ea 
Here $S_{\rm gf}$ denotes the gauge fixing term    
\be 
S_{\rm gf}[h;\bg]=\frac{1}{2\alpha}\int \! d^dx 
\sqrt{\bg}\,\bg^{\mn} F_\mu F_\nu\,,
\label{F4}
\ee
and $S_{\rm gh}$ is the action for the corresponding Faddeev--Popov
ghosts $C^\mu$ and $\bar{C}_\mu$:
\be
S_{\rm gh}[h,C,\bar{C};\bg]=
-\kappa^{-1}\int \! d^dx \sqrt{\bg} \,\bar{C}_\mu\, \bg^{\mu\nu}
\,
\frac{\partial F_\nu}{\partial h_{\alpha\beta}}
\,\cL_C\left(\bg_{\alpha\beta}+h_{\alpha\beta}\right)\,.
\label{F5}
\ee
The Faddeev--Popov action $S_{\rm gh}$ is obtained along the
same lines as in Yang--Mills theory: one applies a gauge
transformation (\ref{F2}) to $F_{\mu}$ and replaces the 
parameters $v^{\mu}$ by the ghost field $C^{\mu}$. The 
integral over $C^{\mu}$ and $\bar{C}_{\mu}$ exponentiates the 
Faddeev-Popov determinant $\det[\delta F_{\mu}/\delta v^{\nu}]$.
In (\ref{F3}) we coupled $h_{\mu\nu}, \,C^{\mu}$ and $\bar{C}_{\mu}$ 
to sources $t^{\mu\nu},\,\bar{\sigma}_{\mu}$ and $\sigma^{\mu}$,
respectively: 
$
%\label{F6}
S_{\rm source} =  -\int \!d^dx \, \sqrt{\bg}
\Big\{ t^{\mu\nu} h_{\mu\nu} +\bar{\sigma}_\mu C^\mu +\sigma^\mu 
\bar{C}_\mu \Big\} \,.
%\ee 
$
The $k$- and source-dependent expectation values of 
$h_{\mu\nu},\, C^{\mu}$ and $\bar{C}_{\mu}$ are then given by 
\be
\label{F7}
\bar{h}_{\mu\nu} = \frac{1}{\sqrt{\bg}}\frac{\delta W_k}{\delta t^{\mu\nu}}
\qquad , \qquad
\xi^\mu=\frac{1}{\sqrt{\bg}}\frac{\delta W_k}{\delta \bar{\sigma}_\mu}
\qquad , \qquad
\bar{\xi}_\mu=\frac{1}{\sqrt{\bg}}\frac{\delta W_k}{\delta\sigma^\mu}\,.
\ee
As usual we assume that one can invert the relations (\ref{F7}) 
and solve for the sources $(t^{\mu\nu}\,,\, \sigma^\mu \, , \, 
\bar{\sigma}_\mu )$ as functionals of 
$(\bar{h}_{\mu\nu} \, , \, \xi^\mu \, , \, \bar{\xi}_\mu )$ and,
parametrically, of $\bg_{\mu\nu}$. The Legendre transform 
$\widetilde{\Gamma}_k$ of $W_k$ reads 
\be
\label{F8}
\widetilde{\Gamma}_k[\bar{h},\xi,\bar{\xi}; \bg]
= \int \! d^dx \, \sqrt{\bg}
\left\{ t^{\mu\nu} \bar{h}_{\mu\nu} +
\bar{\sigma}_\mu \xi^\mu + \sigma^\mu\bar{\xi}_\mu
\right\} -W_k[t,\sigma,\bar{\sigma}; \bg]\,.
\ee
This functional inherits a parametric $\bg_{\mu\nu}$-dependence from 
$W_k$. 

As mentioned earlier for a generic gauge fixing condition the 
Legendre transform (\ref{F8}) is not a diffeomorphism invariant  
functional of its arguments since the gauge breaking under the 
functional integral is communicated to $\widetilde{\Gamma}_k$ via
the sources. While  $\widetilde{\Gamma}_k$ does indeed describe the correct
`on-shell' physics satisfying all constraints coming from BRST invariance, it is not invariant off-shell 
\cite{back,joos}. The situation is different for the class of 
gauge fixing conditions of the background type. While --  
as any gauge fixing condition must -- they break the invariance under 
(\ref{F2}) they are  chosen to be invariant under the so-called
background gauge transformations 
\be 
\delta h_{\mu\nu} = \cL_v h_{\mu\nu} \,,\sspace 
\delta \bg_{\mu\nu} = \cL_v \bg_{\mu\nu} \,. 
\label{F9}
\ee
The complete metric $\gamma_{\mu\nu} = g_{\mu\nu} + h_{\mu\nu}$ 
transforms as $\delta \gamma_{\mu\nu} = \cL_v \gamma_{\mu\nu}$ both 
under (\ref{F9}) and under (\ref{F2}). The crucial difference 
is that the (`quantum') gauge transformations (\ref{F2}) keep $\bg_{\mu\nu}$
 unchanged  so that the entire 
change of $\gamma_{\mu\nu}$ is ascribed to $h_{\mu\nu}$. This is 
the point of view one adopts in a standard perturbative calculation 
around flat space where one fixes $\bg_{\mu\nu} = \eta_{\mu\nu}$ and 
allows for no variation of the background. In the present 
construction, instead, we leave $\bg_{\mu\nu}$ unspecified but insist on
covariance under (\ref{F9}). This will lead to a completely 
background covariant formulation.        

Clearly there exist many possible gauge fixing terms 
$S_{\rm gf}[h;\bg]$ of the form (\ref{F4}) which break 
(\ref{F2}) and are invariant under (\ref{F9}). A convenient 
choice which has been employed in practical calculations 
is the one-parameter family of gauge conditions
\be\label{F10}
F_\mu[h;\bg] = \sqrt{2} \, \kappa \, \big( \Db^\nu \, h_{\mu \nu} 
- \varpi \, \Db_\mu \, h^\nu{}_{\nu} \big) \, , 
\ee
parameterized by $\varpi$. The covariant derivative $\bar{D}_{\mu}$ involves the Christoffel 
symbols $\bar{\Gamma}^{\rho}_{\mu\nu}$ of the background metric. 
Note that (\ref{F10}) is linear in the quantum field 
$h_{\alpha \beta}$. For $\varpi = \half$, \eqref{F10} reduces to the 
background version of the harmonic coordinate condition \cite{back}:
on a flat background with $\bg_{\mu\nu} = \eta_{\mu\nu}$ 
the condition $F_{\mu} =0$ becomes the familiar harmonic coordinate 
condition, $\dd^{\mu} h_{\mu\nu} = 
\frac{1}{2} \dd_{\nu} h_{\mu}^{\; \mu}$. In eqs.\ (\ref{F10}) and (\ref{F5}) 
$\kappa$ is an arbitrary constant with the dimension of a mass. We shall 
set $\kappa \equiv (32 \pi \bar{G})^{-1/2}$ with $\bar{G}$ a constant reference value of 
Newton's constant. The ghost action for the gauge 
condition (\ref{F10}) reads 
\be
\label{F11a}
S_{\rm gh}[h,C,\bar{C};\bg]=-\sqrt{2}\int \!d^dx \,  \sqrt{\bg}
\,\bar{C}_\mu \cM [g,\bg]^\mu{}_\nu C^\nu
\ee
with the Faddeev--Popov operator
\be\label{F11b}
\cM[g,\bg]^{\mu}{}_{\nu} = \Db^\rho \, g^\mu{}_\nu \, D_\rho + \Db^\rho \, g_{\rho \nu} D^\mu - 2 \,  \varpi \, \Db^\mu \, \bg^{\rho \sigma} \, g_{\rho \nu} D_\sigma \, .
\ee
It will prove crucial that for every background-type 
choice of $F_{\mu}$, $S_{\rm gh}$ is invariant under (\ref{F9}) together with 
\be 
\delta C^{\mu} = \cL_v C^{\mu} \,,\sspace 
\delta \bar{C}_{\mu} = \cL_v \bar{C}_{\mu} \,. 
\label{F11c}
\ee

The essential piece in eq.~(\ref{F3}) is the IR cutoff for the gravitational
field $h_{\mu\nu}$ and for the ghosts. It is taken to be of the form
\be
\label{F12}
\Delta_k S
= \frac{\kappa^2}{2}\int\! d^dx \, \sqrt{\bg}\, h_{\mu\nu}
\cR^{\rm grav}_k[\bg]^{\mu\nu \rho\sigma}h_{\rho\sigma}
  +\sqrt{2}\int d^dx\, \sqrt{\bg}\, \bar{C}_\mu \cR^{\rm gh}_k[\bg]C^\mu\,.
\ee
The cutoff operators $\cR^{\rm grav}_k$ and $\cR^{\rm gh}_k$ serve the purpose
of discriminating between high--momentum and low--momentum modes.
Eigenmodes of $-\bar{D}^2$ with eigenvalues $p^2\gg k^2$ are integrated out
 without any suppression whereas modes with small eigenvalues
$p^2\ll k^2$ are suppressed.
The operators $\cR^{\rm grav}_k$ and $\cR^{\rm gh}_k$ have the structure
$%\be
%\label{F13a}
\cR_k[\bg]=\cZ_k k^2 R^{(0)}(-\bar{D}^2/k^2)\,,
%\ee
$
where the dimensionless function $R^{(0)}$ interpolates between $R^{(0)}(0)=1$ and $R^{(0)}(\infty)=0$.
A convenient choice is, e.g., the exponential cutoff 
$R^{(0)}(w)=w[\exp(w)-1]^{-1}$ or the optimized cutoff $R^{(0)}(w)=(1-w) \theta(1-w)$, where $w = p^2/k^2$.
The factors $\cZ_k$ are different for the graviton and the ghost cutoff. 
They are determined by the condition that all fluctuation spectra are cut off 
at precisely the same $k^2$, such that $\cR_k$ combines with $\Gamma_k^{(2)}$ 
to the inverse propagator $\Gamma_k^{(2)} + \cR_k = 
\cZ_k( p^2 + k^2) + \cdots$, as it is necessary if the IR cutoff 
is to give rise to a $({\rm mass})^2$ of size $k^2$ rather than 
$(\cZ_k)^{-1} k^2$.  
Following this condition the ghost $\cZ_k \equiv Z^{\rm gh}_k$ is a pure number, whereas for
the metric fluctuation $\cZ_k \equiv \cZ^{\rm grav}_k$ is a
tensor, constructed only from the background metric $\bg_{\mu\nu}$.

A feature of $\Delta_k S$ which is essential from a practical point of view 
is that the modes of $h_{\mu\nu}$ and the ghosts are organized according 
to their eigenvalues with respect to the {\it background} Laplace 
operator $\bar{D}^2 = \bg^{\mu\nu} \bar{D}_{\mu} \bar{D}_{\nu}$ rather 
than $D^2 = g^{\mu\nu} D_{\mu} D_{\nu}$, which would pertain to the full 
quantum metric $\bg_{\mu\nu} + h_{\mu\nu}$. Using $\bar{D}^2$, the 
functional $\Delta_k S$ is quadratic in the quantum field $h_{\mu\nu}$,
while it becomes extremely complicated if $D^2$ is used instead. 
The virtue of a quadratic $\Delta_k S$ is that it gives 
rise to a flow equation which contains only {\it second} functional 
derivatives of $\Gamma_k$ but no higher ones. The flow equations resulting 
from the cutoff operator $D^2$ are prohibitively complicated and 
can hardly be used for practical computations. A second property 
of $\Delta_k S$ which is crucial for our purposes is that it is invariant 
under the background gauge transformations (\ref{F9}) with (\ref{F12}). 

Having specified all the ingredients which enter the functional  
integral (\ref{F3}) for the generating functional $W_k$ we can 
write down the final definition of the effective average action 
$\Gamma_k$. It is obtained from the Legendre transform 
$\widetilde{\Gamma}_k$ by subtracting the cutoff action 
$\Delta_k S$ with the classical fields inserted:
\be
\label{F14}
\Gamma_k[\bar{h},\xi,\bar{\xi}; \bg]
= \widetilde{\Gamma}_k[\bar{h},\xi,\bar{\xi}; \bg]
- \Delta_k S[\bar{h}, \xi , \bar{\xi} ; \bg]\,.
\ee
It is convenient to define the expectation value of the quantum metric $\gamma_{\mu \nu}$,
\be
\label{F15}
g_{\mu\nu}(x) \equiv \bg_{\mu\nu}(x) +\bar{h}_{\mu\nu}(x)\,,
\ee
and consider $\Gamma_k$ as a functional of $g_{\mu\nu}$ rather than 
$\bar{h}_{\mu\nu}$:
\be
\label{F16}
\Gamma_k[g_{\mu\nu} ,\bg_{\mu\nu} , \xi^\mu , \bar{\xi}_\mu]
\equiv \Gamma_k[g_{\mu\nu} -\bg_{\mu\nu} , \xi^\mu , \bar{\xi}_\mu; 
\bg_{\mu\nu}]\,.
\ee

So, what did we gain going through this seemingly complicated 
background field construction, eventually ending up 
with an action functional which depends on {\it two} metrics even?
The main advantage of this setting is that the corresponding 
functionals $\widetilde{\Gamma}_k$, and as a result $\Gamma_k$, 
are  invariant under general coordinate transformations
where all its arguments transform as tensors of the corresponding rank:
\be
\label{F17}
\Gamma_k[\Phi +\cL_v \Phi] = \Gamma_k[\Phi]\,,
\qquad  \qquad
\Phi \equiv \left\{
  g_\mn , \bg_\mn , \xi^\mu , \bar{\xi}_\mu \right\}\,.
\ee
Note that in (\ref{F17}), contrary to the ``quantum gauge transformation''
(\ref{F2}), also the background metric transforms as an ordinary
tensor field: $\delta\bg_\mn = \cL_v \bg_\mn$.
Eq.~(\ref{F17}) is a consequence of
\be
\label{F18}
W_k\left[\cJ + \cL_v \cJ\right] = W_k\left[\cJ\right]\,,
\quad \quad \cJ \equiv \left\{
  t^\mn , \sigma^\mu , \bar{\sigma}_\mu ;\, \bg_\mn \right\}\,.
\ee
This invariance property follows from (\ref{F3}) if one performs
a  compensating transformation (\ref{F9}), (\ref{F12}) on the integration 
variables $h_\mn$, $C^\mu$ and $\bar{C}_\mu$ and uses the invariance 
of $S[\bg + h],\,S_{\rm gf},\,S_{\rm gh}$ and $\Delta_k S$. 
At this point we assume that the functional measure in (\ref{F3}) 
is diffeomorphism invariant.  
     
Since the $\cR_k$'s vanish for $k=0$, the limit $k \ra 0$ of 
$\Gamma_k[g_\mn, \bg_\mn, \xi^{\mu}, \bar{\xi}_{\mu}]$ brings us 
back to the standard effective action functional which still 
depends on two metrics, though. The ``ordinary'' effective action 
$\Gamma[g_\mn]$ with one metric argument is obtained from this 
functional by setting $\bg_\mn = g_\mn$, or equivalently 
$\bar{h}_\mn =0$ \cite{back,joos}:
\be 
\Gamma[g] \equiv 
\lim_{k \ra 0} \Gamma_k[g,\bg = g, \xi =0, \bar{\xi} =0] = 
\lim_{k \ra 0} \Gamma_k[\bar{h} =0,\xi =0, \bar{\xi} =0; g = \bg]\,. 
\label{F19}
\ee   
This equation brings about the ``magic property'' of the background
field formalism: a priori the 1PI $n$-point 
functions of the metric are obtained by an $n$-fold functional 
differentiation of $\Gamma_0[\bar{h},0,0;\bg_\mn]$ with respect 
to $\bar{h}_\mn$. Hereby $\bg_\mn$ is kept fixed; it acts simply as 
an externally prescribed function which specifies the form of the 
gauge fixing condition. Hence the functional $\Gamma_0$ 
and the resulting {\it off-shell} Green functions do depend on 
$\bg_\mn$, but the {\it on-shell} Green functions, related 
to observable scattering amplitudes, do not depend on 
$\bg_\mn$. In this respect $\bg_\mn$ plays a role similar to the gauge 
parameter $\alpha$ in the standard approach. Remarkably, the same on-shell 
Green functions can be obtained by differentiating the functional 
$\Gamma[g_\mn]$ of (\ref{F19}) with respect to $g_\mn$, or 
equivalently $\Gamma_0[\bar{h} =0, \xi=0, \bar{\xi} =0;
\bg = g]$, with respect to its $\bg$ argument. In this context,
`on-shell' means that the metric satisfies the effective 
field equation $\delta \Gamma_0[g]/\delta g_\mn =0$.

With (\ref{F19}) and its $k$-dependent counterpart 
\be
\bar{\Gamma}_k[g_\mn] \equiv \Gamma_k[g_\mn, g_\mn, 0,0]\,
\label{F20}
\ee
we succeeded in constructing a diffeomorphism invariant 
generating functional for gravity: thanks to (\ref{F17}) 
$\Gamma[g_\mn]$ and $\bar{\Gamma}_k[g_\mn]$ are invariant
 under general coordinate transformations $\delta g_\mn
= \cL_v g_\mn$. However, there is  a price to be paid for their
invariance: the simplified functional $\bar{\Gamma}_k[g_\mn]$ does not 
satisfy an exact RG equation, basically because it contains insufficient
information. The actual RG evolution has to be performed 
at the level of the functional $\Gamma_k[g,\bg,\xi,\bar{\xi}\,]$.
Only {\it after} the evolution one may set $\bg = g,\, \xi =0, 
\bar{\xi} =0$. As a result, the actual theory space 
of QEG, $\{A[g,\bg,\xi,\bar{\xi} \,] \}$, consists of 
functionals of all four variables, $g_\mn, \bg_\mn, \xi^{\mu}, 
\bar{\xi}_\mu$, subject to the invariance condition (\ref{F17}).

Taking a scale derivative of the regularized functional integral (\ref{F3}) and
reexpressing the result in terms of $\Gamma_k$ one finds the following FRGE \cite{mr}: 
\be
\label{F21}
\begin{split}
\partial_t \Gamma_k[\bar h, \xi, \bar\xi; \bg] =&
\frac{1}{2}\Tr
\left[\left(\Gamma^{(2)}_k + \widehat \cR_k\right)^{-1}_{\bar h \bar h}
\left(\partial_t \widehat \cR_k\right)_{\bar h \bar h} \right]
\\[2mm]
&-  \frac{1}{2}\Tr
\left[\left\{
 \left(\Gamma^{(2)}_k + \widehat \cR_k\right)^{-1}_{\bar \xi \xi}
-\left(\Gamma^{(2)}_k + \widehat \cR_k\right)^{-1}_{\xi \bar\xi}
\right\}
\left(\partial_t \widehat \cR_k\right)_{\bar \xi \xi} \right]\,. 
%\nonumber
\end{split}
\ee 
Here $\Gamma^{(2)}_k$ denotes the Hessian of $\Gamma_k$ with 
respect to the dynamical fields $\bar{h},\, \xi,\,\bar{\xi}$ at fixed 
$\bg$. It is a block matrix labeled by the fields $\varphi_i \equiv
\{\bar{h}_\mn,\,\xi^\mu, \bar{\xi}_\mu\}$:
\be
\label{F22}
\Gamma^{(2)\, ij}_k(x,y) \equiv 
\frac{1}{\sqrt{\bg(x)\bg(y)}} \,
\frac{\delta^2 \Gamma_k}{\delta \varphi_i(x)\delta \varphi_j(y)}\,.
\ee
(In the ghost sector the derivatives are understood as left derivatives.)
Likewise, $\widehat{\cR}_k$ is a block diagonal matrix  with entries 
$(\widehat{\cR}_k)_{\bar{h}\bar{h}}^{\mu\nu\rho\sigma} \equiv 
\kappa^2 (\cR_k^{\rm grav}[\bg])^{\mu\nu\rho\sigma}$ and 
$\widehat{\cR}_{\bar{\xi} \xi} = \sqrt{2} \cR_k^{\rm gh}[\bg]$.
Performing the trace in the position representation it includes 
an integration $\int\! d^dx \sqrt{\bg(x)}$ involving the
background volume element. For any cutoff which is qualitatively 
similar to the exponential cutoff the traces on the RHS of eq.~(\ref{F21}) 
are well convergent, both in the IR and the UV. The interplay
between the $\widehat \cR_k$ in the denominator and the  
factor $\partial_t \widehat \cR_k$ in the numerator thereby ensures that
 the dominant contributions come 
from a narrow band of generalized momenta centered around $k$. Large 
momenta are exponentially suppressed.

Besides the FRGE the effective average action also satisfies  an 
exact integro-differential equation,
which can be used to find the $k
\ra \infty$ limit of the average action:
\be
\Gamma_{k \ra \infty}[\bar{h},\xi,\bar{\xi};\bg] = 
S[\bg+\bar{h}] + S_{\rm gf}[\bar{h};\bg] 
+S_{\rm gh}[\bar{h},\xi,\bar{\xi};\bg]\,.
\label{F24}
\ee
Intuitively, this limit can be understood from the observation that for $k \rightarrow \infty$
all quantum fluctuation in the path integral are suppressed by an infinity
mass-term. Thus, in this limit no fluctuations are integrated out and
 $\Gamma_{k \ra \infty}$ agrees with the microscopic action $S$ supplemented by
 the gauge fixing and ghost actions. At the level of the functional 
$\bar{\Gamma}_k[g]$, eq.~(\ref{F24}) boils down to 
$\bar{\Gamma}_{k \ra \infty}[g] = S[g]$. However, as $\Gamma_k^{(2)}$ 
involves derivatives with respect to $\bar{h}_\mn$ (or equivalently 
$g_\mn$) at fixed $\bg_\mn$ it is clear that the evolution cannot 
be formulated entirely in terms of $\bar{\Gamma}_k$ alone. 

The background gauge invariance of $\Gamma_k$, expressed in eq.~(\ref{F17}),
is of enormous practical importance. It implies that if the 
initial functional does not contain non-invariant terms, 
the flow will not generate such terms. Very often this reduces the 
number of terms to be retained in a reliable truncation ansatz quite 
considerably. Nevertheless, even if the initial action is simple, the RG 
flow will generate all sorts of local and non-local terms in 
$\Gamma_k$ which are consistent with the symmetries. 

%%%%%%%%%%%%%%%%%%%%%%%%%%%%%%%%%%%%%%%%%%%%%%%%%%%%%%%%%%%%%%%%%
\section{Truncated flow equations}
\label{sect:4}
\setcounter{equation}{0}
%%%%%%%%%%%%%%%%%%%%%%%%%%%%%%%%%%%%%%%%%%%%%%%%%%%%%%%%%%%%%%%%%
Solving the FRGE (\ref{F21}) subject to the initial condition (\ref{F24}) 
is equivalent to (and in practice as difficult as) calculating the 
original functional integral over $\gamma_\mn$. It is therefore important 
to devise efficient approximation methods. The truncation of theory 
space is the one which makes maximum use of the FRGE reformulation of the 
quantum field theory problem at hand. 

As for the flow on the theory space $\{A[g,\bg,\xi,\bar{\xi}]\}$, 
a still very general truncation consists of neglecting the evolution of the 
ghost action by making the ansatz
\be
\label{G1}
\Gamma_k[g,\bg ,\xi ,\bar\xi]
= \bar\Gamma_k[g]+\widehat\Gamma_k[g,\bg]
+S_{\rm gf}[g-\bg ;\bg] +S_{\rm gh}[g-\bg ,\xi,\bar\xi ;\bg]\,,
\ee
where we extracted the classical $S_{\rm gf}$ and $S_{\rm gh}$
from $\Gamma_k$. The remaining functional depends on both 
$g_\mn$ and $\bg_\mn$. It is further decomposed as $\bar\Gamma_k+
\widehat\Gamma_k$ where $\bar\Gamma_k$ is defined as in (\ref{F20}) 
and $\widehat\Gamma_k$ contains the deviations for $\bg\neq g$. 
Hence, by definition, $\widehat\Gamma_k[g,g]=0$, and 
$\widehat\Gamma_k$ contains, in particular, quantum corrections to 
the gauge fixing term which vanishes for $\bg = g$, too. 
This ansatz satisfies the initial condition (\ref{F24}) 
if\footnote{See \cite{elisa1} for a detailed discussion of the relation between $S$ and $S_{\rm bare}$.}
\be
\bar\Gamma_{k\ra \infty} = S \qquad \mbox{and}\qquad
\widehat \Gamma_{k\ra \infty} =0\,.
\ee
Inserting (\ref{G1}) into the exact FRGE \eqref{F21}
one obtains an evolution equation on the truncated  space 
$\{ A[g,\bg]\}$:
\ba
\label{G2}
\partial_t\Gamma_k[g,\bg]
&=& \frac{1}{2}\Tr
\left[\left(
  \kappa^{-2}\Gamma^{(2)}_k[g,\bg] +\cR_k^{\rm grav}[\bg]
      \right)^{-1}
  \partial_t \cR^{\rm grav}_k[\bg]
\right]
\nonum 
&& -\Tr\left[\left(
  -\cM[g,\bg]+ \cR^{\rm gh}_k[\bg] \right)^{-1} \dd_t 
\cR^{\rm gh}_k[\bg] \right]\,.
\ea
This equation evolves the functional 
\be\label{eq:4.4}
\Gamma_k[g,\bg] \equiv  \bar\Gamma_k[g]+S_{\rm gf}[g-\bg;\bg]
    +\widehat\Gamma_k[g,\bg]\,.
\ee
Here $\Gamma^{(2)}_k$ denotes the Hessian of $\Gamma_k[g, \bg]$ with respect
to $g_\mn$ at fixed $\bg_\mn$ and $\cM$ is given in eq.\ \eqref{F11b}. 

The truncation ansatz (\ref{G1}) is still too general for practical 
calculations to be easily possible. The first truncation for 
which the RG flow has been found \cite{mr} is the ``Einstein-Hilbert
truncation'' which retains in $\bar{\Gamma}_k[g]$ only the terms 
$\int\! d^dx \, \sqrt{g}$ and $\int\! d^dx \, \sqrt{g} R$, already present in the  
in the classical action, with $k$-dependent coupling constants, 
and includes only the wave function renormalization in $\widehat{\Gamma}_k$:
\be
\label{G3}
\Gamma_k[g,\bg]  =  
2\kappa^2 Z_{Nk} \int\! d^dx \,\sqrt{g} \left\{ 
-R + 2\bar\lambda_k \right\}
 + \frac{ Z_{Nk}}{2 \alpha}  \int \!d^dx \,\sqrt{\bg} \, 
\bar g^\mn F_\mu F_\nu\,.
\ee
In this case the truncation subspace is 2-dimensional. The ansatz 
(\ref{G3}) contains two free functions of the scale, the running 
cosmological constant $\bar{\lb}_k$ and $Z_{Nk}$ or, equivalently,
the running Newton constant $G_k \equiv \bar{G}/Z_{Nk}$. Here $\bar{G}$ 
is a fixed constant, and $\kappa \equiv (32 \pi \bar{G})^{-1/2}$. 
As for the gauge fixing term, $F_{\mu}$ is given by eq.~(\ref{F10})     
with $\bar{h}_\mn \equiv g_\mn - \bg_\mn$ replacing $h_\mn$; it vanishes 
for $g = \bg$. The ansatz (\ref{G3}) has the general structure of 
(\ref{G1}) with 
\be\label{EHstruc}
\widehat{\Gamma}_k = (Z_{Nk} -1) S_{\rm gf} \, .
\ee
 Within the 
Einstein-Hilbert approximation the gauge fixing parameter $\alpha$ is 
kept constant. Here we shall set $\alpha =1$ and comment on generalizations 
later on.

Upon inserting the ansatz (\ref{G3}) into the flow 
equation (\ref{G2}) it boils down to a system of two ordinary differential 
equations for $Z_{Nk} $ and $\bar{\lb}_k$. Their derivation is rather 
technical, so we shall focus on the conceptual aspects here. 
In order to find $\dd_t Z_{Nk}$ and $\dd_t \bar{\lb}_k$ it is 
sufficient to consider (\ref{G2}) for $g_\mn = \bg_\mn$. In this case 
the LHS of the flow equation becomes 
$2 \kappa^2 \int \! d^dx \sqrt{g} [- R \dd_t Z_{Nk} + 
2 \dd_t( Z_{Nk} \bar{\lb}_k)]$. The RHS is assumed to admit an expansion 
in terms of invariants $P_{i}[g_\mn]$. In the Einstein-Hilbert truncation only two of them, 
$\int\! d^dx \, \sqrt{g}$ and $\int\! d^dx \, \sqrt{g} R$, need to be retained.
 They can be extracted from the traces in (\ref{G2}) 
by standard derivative expansion techniques. Equating the result to 
the LHS and comparing the coefficients of $\int\! d^dx \sqrt{g}$ and $\int\! d^dx \sqrt{g}
R$, a pair of coupled differential equations for $Z_{Nk}$ and 
$\bar{\lb}_k$ arises. It is important to note that, on the RHS, we may set 
$g_\mn = \bg_\mn$ only {\it after} the functional derivatives of 
$\Gamma^{(2)}_k$ have been obtained since they must be taken at 
fixed $\bg_\mn$.

As demonstrated explicitly in \cite{Codello:2008vh,Benedetti:2010nr}, this calculation can be performed without ever considering 
any specific metric $g_\mn = \bg_\mn$. This reflects the fact that 
the approach is background covariant. The RG flow is universal 
in the sense that it does not depend on any specific metric. 
In this respect gravity is not different from the more 
traditional applications of the renormalization group: the RG 
flow in the Ising universality class, say, has nothing to do 
with any specific spin configuration, it rather reflects the 
statistical properties of very many such configurations. 

While there is no conceptual necessity to fix the background 
metric, it nevertheless is sometimes advantageous from a computational 
point of view to pick a specific class of backgrounds. Leaving        
$\bg_\mn$ completely general, the calculation of the functional traces is 
very hard work usually. In principle there exist well known 
derivative expansion and heat kernel techniques which could 
be used for this purpose, but their application is an 
extremely lengthy and tedious task usually. Moreover, typically 
the operators $\Gamma_k^{(2)}$ and $\cR_k$ are of a complicated non-standard 
type so that no efficient use of the tabulated Seeley coefficients 
can be made. However, often calculations of this type simplify if 
one can assume that $g_\mn= \bg_\mn$ has specific properties. 
Since the beta functions are background independent we may therefore 
restrict $\bg_\mn$ to lie in a conveniently chosen class of 
geometries which is still general enough to disentangle the invariants
retained and at the same time simplifies the calculation. 

For the Einstein-Hilbert truncation the most efficient choice is a 
family of $d$-spheres $S^d(r)$, labeled by their radius $r$. 
These maximally symmetric backgrounds satisfy
\be\label{dsphere}
R_{\mu\nu} = \tfrac{1}{d} g_{\mu\nu} R \, , \qquad R_{\mu\nu\rho\sigma} = \tfrac{1}{d(d-1)}(g_{\mu\rho}g_{\nu\sigma} - g_{\mu\sigma}g_{\nu\rho})R \, ,
\ee
and, in particular, $D_{\alpha} R_{\mn \rho \sigma} =0$, so 
they give a vanishing value to all invariants constructed from 
$g = \bg$ containing covariant derivatives acting on curvature 
tensors. What remains (among the local invariants) are terms of the 
form $\int \! \sqrt{g} P(R)$, where $P$ is a polynomial in the 
Ricci scalar. Up to linear 
order in $R$ the two invariants relevant for the Einstein-Hilbert truncation 
are discriminated by the $S^d$ metrics as the latter scale    
differently with the radius of the sphere: $\int \! \sqrt{g} 
\sim r^d$, $\int \! \sqrt{g} R \sim r^{d-2}$. Thus, in order 
to compute the beta functions of $\bar{\lb}_k$ and $Z_{Nk}$ 
it is sufficient to insert an $S^d$ metric with arbitrary $r$ 
and to compare  the coefficients of $r^d$ and $r^{d-2}$.      
If one wants to do better and include the three quadratic 
invariants $\int \!R_{\mn \rho\sigma} R^{\mn \rho \sigma}$, 
$\int \! R_\mn R^\mn$, and $\int \! R^2$, the family $S^d(r)$ 
is not general enough to separate them; all scale like $r^{d-4}$   
with the radius. 

Under the trace we need the operator $\Gamma_k^{(2)}[\bar{h};\bg]$.
It is most easily calculated by Taylor expanding the truncation ansatz,
$\Gamma_k[\bg + \bar{h}, \bg] = \Gamma_k[\bg,\bg] 
+ O(\bar{h}) + \Gamma_k^{\rm quad}[\bar{h};\bg] + O(\bar{h}^3)$,
and stripping off the two $\bar{h}$'s from the quadratic term, 
$\Gamma_k^{\rm quad} = \frac{1}{2} \int \! \bar{h} \Gamma_k^{(2)} \bar{h}$. 
For $\bg_\mn$ the metric on $S^d(r)$ one obtains 
\ba
\label{G4}
\Gamma_k^{\rm quad}[\bar{h};\bg]
&=& 
\frac{1}{2}Z_{Nk} \kappa^2 \int \!d^dx\, \Bigg\{
  \widehat{h}_\mn
  \left[-\bar D^2-2\bar\lambda_k+C_T\bar{R} \right]\widehat{h}^\mn
\nonum 
&&
\bspace \sspace -\left(\frac{d-2}{2d}\right)\phi
  \left[-\bar D^2-2\bar\lambda_k+C_S\bar{R} \right]\phi \Bigg\}\,,
\ea
with
$C_T \equiv (d(d-3)+4)/(d(d-1))$, $C_S \equiv (d-4)/d$.
In order to partially diagonalize this quadratic form $\bar{h}_\mn$
has been decomposed into a traceless part $\widehat{h}_\mn$ and 
the trace part proportional to $\phi$: $\bar{h}_\mn = 
\widehat{h}_\mn + d^{-1} \bg_\mn \phi$, $\bg^\mn \widehat{h}_\mn =0$. 
Further, $\bar{D}^2 = \bg^\mn \bar{D}_\mu \bar{D}_\nu$ is the 
covariant Laplace operator corresponding to the background geometry,
and $\bar{R} = d(d-1)/r^2$ is the numerical value of the curvature 
scalar on $S^d(r)$. 

At this point we can fix the constants $\cZ_k$ which appear in the 
cutoff operators $\cR_k^{\rm grav}$ and $\cR_k^{\rm gh}$ of 
(\ref{F12}). They  should be adjusted in such a way that for every 
low--momentum mode the cutoff combines with the kinetic term of 
this mode to $-\bar D^2+k^2$ times a constant. Looking at (\ref{G4}) 
we see that the respective kinetic terms for $\widehat{h}_\mn$ 
and $\phi$ differ by a factor of $-(d-2)/2d$. This suggests the following 
choice:
\be
\label{G6}
\left(\cZ_k^{\rm grav}\right)^{\mn\rho\sigma}
= \left[   \left(\one -P_\phi\right)^{\mn\rho\sigma}
 -\frac{d-2}{2} P_\phi^{\mn\rho\sigma} \right] Z_{Nk}\,. 
\ee
Here $(P_\phi)_\mn{}^{\rho\sigma} =d^{-1} \bg_\mn\bg^{\rho\sigma}$
is the projector on the trace part of the metric.
For the traceless tensor (\ref{G6}) gives $\cZ_k^{\rm grav}=Z_{Nk} \one$, 
and for $\phi$ the different relative normalization is taken into account. (See ref.\ \cite{mr} for a detailed discussion of the subtleties related to this choice.)
Thus we obtain in the $\widehat{h}$ and the $\phi$-sector, respectively:  
\ba
\label{G7}
\left(   \kappa^{-2}\Gamma_k^{(2)}[g,g]+\cR_k^{\rm grav} 
\right)_{\widehat{h}\widehat{h}} 
\!\! &=& \!\!
Z_{Nk} \left[-D^2+k^2 R^{(0)}(-D^2/k^2)-2\bar\lambda_k+C_T R\right],
\\[2mm] 
\left( \kappa^{-2}\Gamma_k^{(2)}[g,g]+\cR_k^{\rm grav}
\right)_{\phi\phi}
\!\! &=& \!\! -\frac{d-2}{2d}
Z_{Nk} \left[-D^2+k^2 R^{(0)}(-D^2/k^2)-2\bar\lambda_k+C_S R\right]
\nonumber
\ea
From now on we may set $\bg=g$ and for simplicity we have omitted 
the bars from the metric and the curvature. Since we did not take
into account any renormalization effects in the ghost action
we set $Z_k^{\rm gh}\equiv1$ in $\cR_k^{\rm gh}$ and obtain
\be
\label{G8}
-\cM + \cR_k^{\rm gh} = - D^2 +k^2 R^{(0)}(-D^2/k^2)+C_V R\,,
\ee
with $C_V \equiv -1/d$. At this point the operator under the first trace on the RHS of 
(\ref{G2}) has become block diagonal, with the $\widehat{h} 
\widehat{h}$ and $\phi \phi$ blocks given by (\ref{G7}).
 Both block operators are expressible in terms of the 
Laplacian $D^2$, in the former case acting on traceless symmetric   
tensor fields, in the latter on scalars. The second trace in (\ref{G2}) 
stems from the ghosts; it contains (\ref{G8}) with $D^2$ acting 
on vector fields. 

It is now a matter of straightforward algebra to compute the first 
two terms in the derivative expansion of those traces, proportional 
to $\int\! d^d x\sqrt{g} \sim r^d$ and $\int\! d^d x\sqrt{g}R 
\sim r^{d-2}$. Considering the trace of an arbitrary function of the 
Laplacian, $W(-D^2)$, the expansion up to second order derivatives 
of the metric is given by 
\ba
\label{G9}
\Tr[W(-D^2)] &=& (4\pi)^{-d/2} {\rm tr}(I)
\Bigg\{ Q_{d/2}[W] \int \!d^dx \, \sqrt{g}
\nonum 
&&
\qquad\qquad\qquad
+\frac{1}{6} Q_{d/2-1}[W] \int \!d^dx\, \sqrt{g}R +O(R^2) \Bigg\}\,.
\ea
The $Q_n$'s are defined as 
\be
\label{G10}
Q_n[W] = \frac{1}{\Gamma(n)} \int_0^{\infty} dz\,z^{n-1} W(z)\,,
\ee
for $n >0$, and $Q_0[W] = W(0)$ for $n =0$. The trace ${\rm tr}(I)$ 
counts the number of independent field components. It equals 
$1,\,d,$ and $(d-1)(d+2)/2$, for scalars, vectors, and symmetric traceless 
tensors, respectively. The expansion (\ref{G9}) is easily derived
using standard heat kernel and Mellin transform techniques \cite{mr}. 

Using (\ref{G9}) it is easy to calculate the traces in (\ref{G2}) 
and to obtain the RG equations in the form $\dd_t Z_{Nk} = \cdots$
and $\dd_t (Z_{Nk} \bar{\lb}_k) = \cdots$. We shall not display them 
here since it is more convenient to rewrite them in terms of the 
dimensionless running cosmological constant and Newton constant, respectively:
\be \label{G11}
\lb_k \equiv k^{-2} \bar{\lb}_k\,, \qquad
 g_k \equiv k^{d-2} G_k \equiv k^{d-2} Z_{Nk}^{-1} \bar{G}\,.
\ee     
In terms of the dimensionless couplings $g$ and $\lb$ the RG equations
become a system of autonomous differential equations
\be\label{betaeq}
\dd_t g_k = \beta_g(g_k, \lb_k) \, , \qquad \dd_t \lambda_k = \beta_\lambda(g_k, \lb_k) \, ,
\ee
where 
\be\label{betafcts}
\begin{split}
 \beta_\lambda(g, \lambda) = & \, ( \eta_N - 2) \lambda
+ \half \left(4\pi\right)^{1-d/2} g  \\
& \times \left[ 2d(d+1) \Phi^1_{d/2}(- 2 \lambda) - 8d \Phi^1_{d/2}(0) - d (d+1) \eta_N \tilde{\Phi}^1_{d/2}(-2 \lambda) \right] \, , \\
\beta_g(g, \lambda) = & \, (d-2+\eta_N) g \, . 
\end{split}
\ee
Here the anomalous dimension of Newton's constant $\eta_N$ is given by
\be\label{G14}
\eta_N(g, \lambda) = \frac{g B_1(\lambda)}{1 - g B_2(\lambda)}
\ee
with the following functions of the dimensionless cosmological constant:
\be\label{Bns}
\begin{split}
B_1(\lambda) \equiv & \,  \tfrac{1}{3} \left( 4 \pi \right)^{1-d/2}  \Big[
d(d+1)\Phi^1_{d/2-1}(-2\lambda) - 6 d (d-1) \Phi^2_{d/2}(-2\lambda)    \\
& \qquad \qquad \qquad - 4 d \Phi^1_{d/2-1}(0) - 24 \Phi^2_{d/2}(0) \Big] \, ,\\
B_2(\lambda) \equiv & \, - \tfrac{1}{6} (4 \pi)^{1-d/2} \left[d(d+1) \tilde{\Phi}^1_{d/2-1}(-2\lambda) - 6 d (d-1) \tilde{\Phi}^2_{d/2}(-2 \lambda) \right] \, . 
\end{split}
\ee
The system (\ref{betaeq}) constitutes an
approximation to a 2-dimensional projection of the RG flow. 
Its properties, and in particular the domain of applicability 
and reliability  will be discussed 
in section \ref{sect:6}.

The ``threshold functions'' $\Phi$ and $\widetilde{\Phi}$ appearing  in \eqref{betafcts} 
and \eqref{Bns} are certain integrals involving the normalized 
cutoff function $R^{(0)}$: 
\ba
\label{G17}
\Phi^p_n(w) &\equiv& \frac{1}{\Gamma(n)}\int_0^\infty dz\,z^{n-1}
\frac{R^{(0)}(z)-z R^{(0)\,\prime}(z)}{[z+R^{(0)}(z)+w]^p}\,,
\nonum
\widetilde\Phi^p_n(w) &\equiv& \frac{1}{\Gamma(n)}\int_0^\infty dz\,
z^{n-1} \frac{R^{(0)}(z)}{[z+R^{(0)}(z)+w]^p}\,.
\ea 
They are defined for positive integers $p$, and $n >0$. 
While there are (few) aspects of the truncated RG flow which are 
independent of the cutoff scheme, i.e., independent of the 
function $R^{(0)}$, the explicit solution of the flow equation 
requires a specific choice of this function. 
In the literature various forms of $R^{(0)}$'s have been 
employed. E.g., the non-differentiable ``optimized cutoff'' \cite{opt} with
$R^{(0)}(w) = (1-w) \theta(1-w)$ allows for an analytic 
evaluation of the integrals \cite{litimgrav}
\be\label{phiopt}
\Phi^{{\rm opt;p}}_n(w) = \frac{1}{\Gamma(n+1)} \frac{1}{1+w} \, , \qquad 
\widetilde{\Phi}^{{\rm opt;p}}_n(w) = \frac{1}{\Gamma(n+2)} \frac{1}{1+w} \, . 
\ee
Easy to handle, but disadvantageous for high 
precision calculations is the sharp cutoff \cite{frank1} 
defined by $\cR_k(p^2) = \lim_{\hat{R} \ra \infty} \hat{R} 
\,\theta(1 - p^2/k^2)$, where the limit is to be taken after 
the $p^2$ integration. This cutoff also allows for an evaluation 
of the $\Phi$ and $\widetilde{\Phi}$ integrals in closed form. 
Taking $d=4$ as an example, eqs.~(\ref{betaeq}) boil down to the 
following simple system of equations:%
\footnote{To be precise, (\ref{G18}) corresponds to the sharp 
cutoff with $s=1$, see \cite{frank1}.}
\begin{subeqnarray} 
\dd_t \lb_k \is -(2 - \eta_N) \lb_k - \frac{g_k}{\pi} 
\Big[ 5\ln(1 - 2 \lb_k) - 2 \zeta(3) + \frac{5}{2} \eta_N\Big]\,,
\\
\dd_t  g_k \is (2 + \eta_N) \,  g_k\,,
\\
\eta_N \is - \frac{2 \, g_k}{ 6\pi + 5 \, g_k} 
\Big[ \frac{18}{1 - 2 \lb_k} +  5\ln(1 - 2 \lb_k) - 
\zeta(2) + 6 \Big]\,.
\label{G18}
\end{subeqnarray}

 In order to 
check the scheme (in)dependence of the results it is 
desirable to perform the calculation for a 
whole class of $R^{(0)}$'s. For this purpose the following one 
parameter family of exponential cutoffs has been used 
\cite{souma,oliver1,oliver2}: 
\be 
R^{(0)}(w;s) = \frac{sw}{e^{sw} -1}\,.
\label{G19}
\ee 
The precise form of the cutoff is controlled by the ``shape 
parameter'' $s$. For $s=1$, (\ref{G19}) coincides with the standard 
exponential cutoff. The exponential cutoffs are 
suitable for precision calculations, but the price to be  
paid is that their $\Phi$ and $\widetilde{\Phi}$ integrals 
can be evaluated only numerically. The same is true for 
a one-parameter family of shape functions with compact support 
which was used in \cite{oliver1,oliver2}. 

Above we illustrated the general ideas and constructions underlying 
gravitational RG flows by means of the simplest example, the 
Einstein-Hilbert truncation. In the literature various extensions 
have been investigated. The derivation and analysis of these more 
general flow equations, corresponding to higher dimensional 
truncation subspaces, is an extremely complex and computationally 
demanding problem in general. For this reason we cannot go into 
the technical details here and just mention some further developments. 

\noindent
{\bf (1)} The natural next step beyond the Einstein-Hilbert 
truncation consists in generalizing the functional 
$\bar{\Gamma}_k[g]$, while keeping the gauge fixing and 
ghost sector classical, as in (\ref{G1}). During the 
RG evolution the flow generates all possible diffeomorphism invariant terms 
in $\bar{\Gamma}_k[g]$ which one can construct from 
$g_\mn$. Both local and non-local 
terms are induced. The local invariants contain strings of 
curvature tensors and covariant derivatives acting upon them, with 
any number of tensors and derivatives, and of all possible index 
structures. The first truncation of this class which has been 
worked out completely \cite{oliver2,oliver3} is the 
$\boldsymbol{R^2}$\textbf{-truncation} defined by (\ref{G1}) with the same 
$\widehat{\Gamma}_k$ as before, and the $({\rm curvature})^2$ 
action 
\be 
\bar{\Gamma}_k[g] = \int\! d^dx \sqrt{g} 
\Big\{ (16 \pi G_k)^{-1} [-R + 2 \bar{\lb}_k] 
+ \bar{\beta}_k R^2 \Big\} \, .  
\label{G20}
\ee  
In this case the truncated theory space is 3-dimensional.
Its natural (dimensionless) coordinates are $(g, \lb ,\beta)$, 
where $\beta_k \equiv k^{4-d} \bar{\beta}_k$, and $g$ and 
$\lb$ defined in \eqref{G11}. Even though (\ref{G20}) 
contains only one additional invariant, the derivation 
of the corresponding RG equations is far more complicated 
than in the Einstein-Hilbert case. We shall summarize the 
results obtained with (\ref{G20}) \cite{oliver2,oliver3} 
in section \ref{sect:6.4}. 

\noindent
{\bf (2)} 
The natural \textbf{extension of the} $\boldsymbol{R^2}$\textbf{-truncation} consists of
including all gravitational four-derivative terms in the truncation subspace
\be\label{R2C2}
\bar{\Gamma}_k[g] =
\int d^4x \sqrt{g} \left[ 
(16 \pi G_k)^{-1} \left[ -R + 2 \bar{\lambda}_k \right]
- \frac{\omega_k}{3 \sigma_k} R^2 + \frac{1}{2 \sigma_k} C_{\mu\nu\rho\sigma} C^{\mu\nu\rho\sigma} 
+ \frac{\theta_k}{\sigma_k} E 
\right] \, , 
\ee
and adapting the classical gauge-fixing and ghost sectors to higher-derivative
gravity. Here $C_{\mu\nu\rho\sigma} C^{\mu\nu\rho\sigma} $ denotes 
the square of the Weyl tensor, and 
 $E = C_{\mu\nu\rho\sigma} C^{\mu\nu\rho\sigma} - 2 R_{\mu\nu} R^{\mu\nu} + \tfrac{2}{3} R^2$
is the integrand of the (topological) Gauss-Bonnet term in four dimensions. Using the FRGE
\eqref{F21}, the one-loop beta functions for higher-derivative gravity have recently been 
recovered in \cite{codello,maxpert,Saueressig:2011vn}, while first non-perturbative results have been
obtained in  \cite{HD1,HD2}. The key ingredient in the non-perturbative works was the generalization
of the background metric $\bg$ from the spherical symmetric background employed in
the $R^2$-truncation to a generic Einstein background. The former case results
in a flow equation for $\beta_k = - \frac{\omega_k}{3 \sigma_k} + \frac{\theta_k}{6 \sigma_k}$,
while working with a generic Einstein background metric allows to 
find the non-perturbative beta functions for two independent combinations of the coupling constants
\be\label{lincomb}
\beta_k = - \frac{\omega_k}{3 \sigma_k} + \frac{\theta_k}{6 \sigma_k} \, , \qquad
\gamma_k = \frac{1}{2 \sigma_k} + \frac{\theta_k}{\sigma_k} \, .
\ee
The results obtained from this so-called $R^2 + C^2$-truncation
are detailed in section \ref{sect:6.5}.

\noindent
{\bf (3)} The first steps towards analyzing the RG flow of QEG in the \textbf{ghost sector} have recently be undertaken in \cite{ Eichhorn:2009ah,Groh:2010ta,Eichhorn:2010tb}. In the first step, the classical ghost sector of the Einstein-Hilbert truncation \eqref{F11a} has been supplemented by a scale-dependent curvature ghost coupling $\bar{\zeta}_k$ in the form $\Gamma_{k}^{\rm R \, gh} = \bar{\zeta}_k \int d^dx \sqrt{g} \, \bar{\xi}^\mu \, R \,  \xi_\mu$ where it was found that $\bar \zeta_k = 0$ constitutes a fixed point of the RG flow, with $\bar \zeta_k$ being associated with an UV-attractive eigendirection. The backreaction of a non-trivial ghost-wavefunction renormalization on the flow of $g_k$ and $\lambda_k$ was subsequently studied in \cite{Groh:2010ta,Eichhorn:2010tb}, where it was established that the ghost-propagator is UV-suppressed by a negative anomalous dimension. Moreover, the phase-diagram of the ghost-improved Einstein-Hilbert truncation is almost identical to the one obtained without ghost-improvements shown in fig.\ \ref{fig0}.

\noindent
{\bf (4)} There are also partial results concerning the \textbf{gauge fixing term}. 
Even if one makes the ansatz (\ref{G3}) for $\Gamma_k[g,\bg]$ 
in which the gauge fixing term has the classical (or more 
appropriately, bare) structure one should treat its prefactor 
as a running coupling: $\alpha = \alpha_k$. The beta function of $\alpha$ 
has not been determined yet from the FRGE, but there is 
a simple argument which allows us to bypass this calculation. 

In non-perturbative Yang-Mills theory and in perturbative 
quantum gravity $\alpha = \alpha_k =0$ is known to be a fixed point 
for the $\alpha$ evolution. The following reasoning suggests 
that the same is true within the non-perturbative FRGE approach 
to gravity. In the standard functional integral the limit 
$\alpha \ra 0$ corresponds to a sharp implementation of the 
gauge fixing condition, i.e., $\exp(-S_{\rm gf})$ becomes 
proportional to $\delta[F_{\mu}]$. The domain of the $\int \! 
\cD h_\mn $ integration consists of those $h_\mn$'s 
which satisfy the gauge fixing condition exactly, $F_\mu =0$.
Adding the IR cutoff at $k$ amounts to suppressing some of the 
$h_\mn$ modes while retaining the others. But since all of them 
satisfy $F_\mu =0$, a variation of $k$ cannot change the domain 
of the $h_\mn$ integration. The delta functional $\delta[F_\mu]$ 
continues to be present for any value of $k$ if it was there
originally. As a consequence, $\alpha$ vanishes for all $k$, 
i.e., $\alpha =0$ is a fixed point of the $\alpha$ evolution \cite{alphafp}. 

Thus we can mimic the dynamical treatment of a running $\alpha$ 
by setting the gauge fixing parameter to the constant value
$\alpha =0$. The calculation for $\alpha =0$ is more complicated 
than at $\alpha =1$, but for the Einstein-Hilbert truncation 
the $\alpha$-dependence of $\beta_g$ and $\beta_\lb$, for 
arbitrary constant $\alpha$ has been found in \cite{falkenberg,
oliver1}. The $R^2$-truncations could be analyzed only in the 
simple $\alpha=1$ gauge, but the results from the Einstein-Hilbert 
truncation suggest the UV quantities of interest do not change 
much between $\alpha =0$ and $\alpha =1$ \cite{oliver1,oliver2}. 

\noindent
{\bf (5)} In refs.\ \cite{creh1,creh2,creh3} gravitational
RG flows have been explored in an approximation that goes beyond a truncation of
theory space. Here only the subsector of the basic path integral over the conformal degrees of freedom 
has been considered while all others were omitted. This leads to a scalar-like theory to which the same apparatus underlying the analysis of the full theory has been applied (average action, background decomposition). Remarkably, this
\textbf{conformally reduced gravity} (contrary to a standard 4-dimensional scalar theory) possesses a NGFP and a RG flow that
is qualitatively similar to that of full QEG. It was possible to establish the existence of the NGFP on an {\it infinite dimensional} theory space consisting of arbitrary potentials for the conformal factor. These somewhat unexpected results find their explanation \cite{creh1,creh2} by noting that the quantization scheme based upon the (conformal reduction of the) gravitational average action is {\it ``background independent''} in the sense that no special metric (flat space, etc.) plays a distinguished role.

\noindent
{\bf (6)} Up to now we considered pure gravity. As for as the general 
formalism, the \textbf{inclusion of matter fields} is straightforward.
The structure of the flow equation remains unaltered, except that 
now $\Gamma_k^{(2)}$ and $\cR_k$ are operators on the larger Hilbert
space of both gravity and matter fluctuations. In practice the 
derivation of the projected RG equations can be quite a 
formidable task, however, the difficult part being the 
decoupling of the various modes (diagonalization of 
$\Gamma_k^{(2)}$) which in most calculational schemes is 
necessary for the computation of the functional traces. 
Various matter systems, both interacting and non-interacting 
(apart from their interaction with gravity) have been studied 
in the literature \cite{percadou,grandamat,pires}. A rather 
detailed analysis has been performed 
by Percacci et al. In \cite{percadou,perper1} arbitrary multiplets of free 
(massless) fields with spin $0,1/2,1$ and $3/2$ were included. 
In  \cite{perper1} an interacting scalar theory coupled 
to gravity in the Einstein-Hilbert approximation was analyzed,
and a possible solution to the triviality and the hierarchy problem \cite{hier}
was a first application in this context.

\noindent
{\bf (7)} At the perturbative one-loop level, the flow of possibly {\bf non-local form factors} appearing in the curvature expansion of the effective average action has been studied in \cite{Codello:2010mj,Satz:2010uu}. For a a minimally coupled scalar field on a 2-dimensional curved space-time, the flow equation for the form factor in $\int d^dx \sqrt{g}R c_{k}(\Delta)R$ correctly reproduces the Polyakov effective action, while in $d=4$ this ansatz allows to recover the low energy effective action as derived in the effective field theory framework \cite{Donoghue:1994dn}.

\noindent
{\bf (8)} As yet, almost all truncations studied are of the ``single metric'' type
where $\Gamma_k$ depends on $\bar{g}_{\mu\nu}$ via the gauge fixing term only. The investigation
of genuine \textbf{bimetric truncations} with a nontrivial dependence on both $g_{\mu\nu}$ and
$\bar{g}_{\mu\nu}$ started only recently in \cite{elisa2,MRS}. Future work on the verification
of the NGFP has to go in this direction clearly.

\noindent
{\bf (9)} 
In ref.\ \cite{frank+friends} a first investigation of \textbf{Lorentzian gravity} was performed in a 3+1 split setting. In this case the flow equation \eqref{E9} has been formulated in terms of the ADM-decomposed metric degrees of freedom, which imprints a foliation structure on space-time and provides a preferred ``time-direction''. The resulting FRGE depends on an additional parameter $\epsilon$, which encodes the signature of the space-time metric. For a fixed truncation the resulting Lorentzian renormalization group flow turned out almost identical to the one obtained in the Euclidean case.

\noindent
{\bf (10)} A first step towards \textbf{new gravitational field variables}, different from
the metric, was taken in \cite{e-omega}. Employing the vielbein and spin connection as independent variables, the question of
Asymptotic Safety was reconsidered. In principle it is conceivable that an Einstein-Cartan type quantum
field theory is inequivalent to QEG. However, the actual calculations support the conjecture that
``Quantum Einstein-Cartan Gravity'' is asymptotically safe, too.\footnote{Also see \cite{Benedetti:2011nd} for a perturbative analysis of the running Immirzi parameter.}

%%%%%%%%%%%%%%%%%%%%%%%%%%%%%%%%%%%%%%%%%%%%%%%%%%%%%%%%%%%%%%%%%
\section{Average action approach to Asymptotic Safety}
\label{sect:6}
\begin{figure}[t]
\begin{center}
  \includegraphics[width=0.72\textwidth]{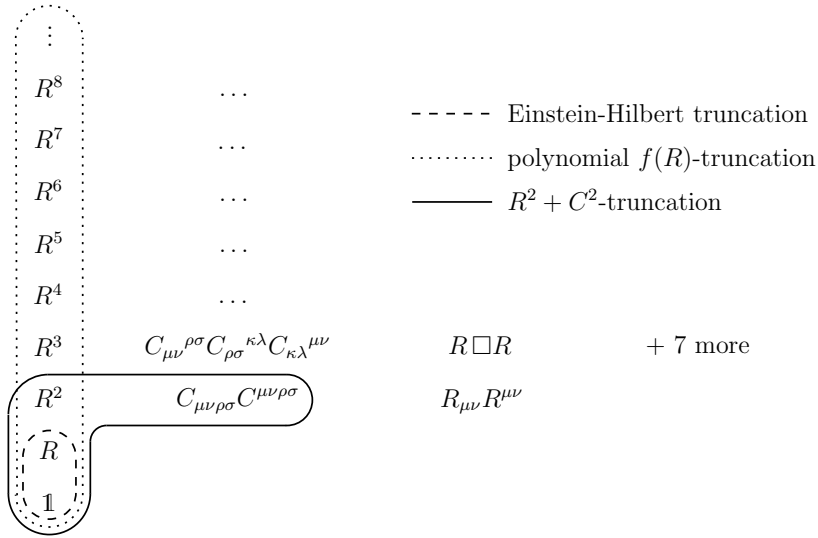}
\end{center}
  \caption{\small Overview of the various truncations employed in the systematic exploration of the theory space of QEG. The lines indicate the interaction monomials contained in the various truncation ans{\"a}tze for $\bar{\Gamma}_k[g]$, eq.\ \eqref{eq:4.4}. All truncations have confirmed the existence of a non-trivial UV fixed point of the gravitational RG flow.}
\label{Fig.ts}
\end{figure}

\setcounter{equation}{0}
%%%%%%%%%%%%%%%%%%%%%%%%%%%%%%%%%%%%%%%%%%%%%%%%%%%%%%%%%%%%%%%%%
Based on the exact flow equation \eqref{E9},
we now implement the ideas of the Asymptotic Safety construction
at the level of explicitly computable approximate RG flows on truncated theory spaces.
A summary of the truncations explored to date is provided in fig.\ \ref{Fig.ts}.
For a detailed derivation of the beta functions we refer
to \cite{mr} (Einstein-Hilbert truncation), \cite{oliver2} ($R^2$-truncation),
\cite{MS1} ($f(R)$-truncation), and \cite{HD2} for the $R^2 + C^2$-truncation,
respectively. 

%----------------------------------------------------------------------------------
%-------------------------------------------------------------------------------

%%%%%%%%%%%%%%%%%%%%%%%%%%%%%%%%%%%%%%%%%%%%%%%%%%%%%%%%%%%%%%%%%
\subsection{The Einstein-Hilbert truncation}
\label{sect:6.1}
\begin{figure}[t]
\centering
\includegraphics[width=.9\columnwidth]{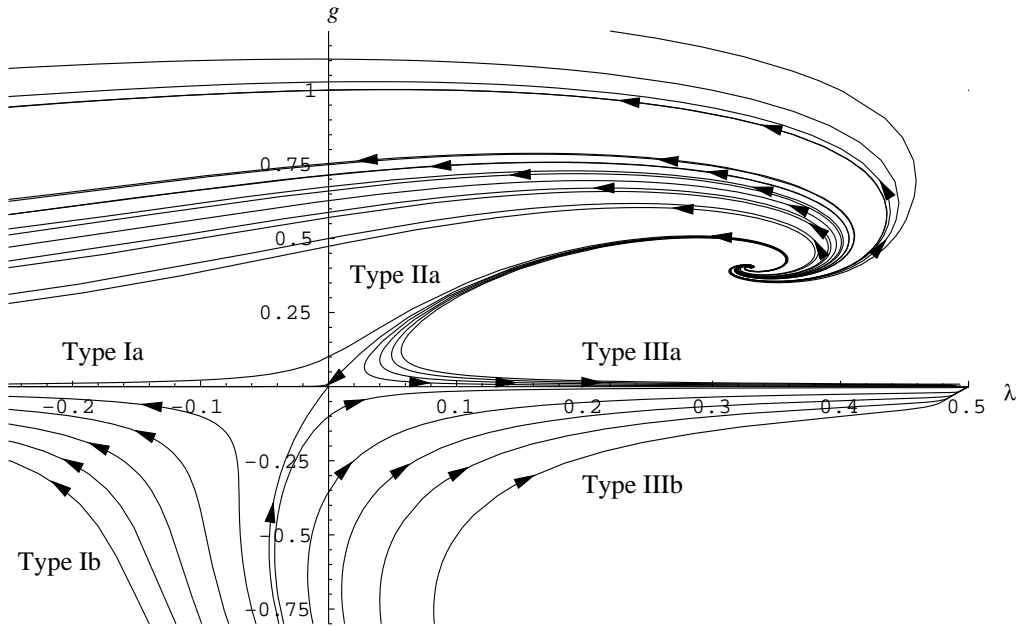}
\caption{\small
RG flow in the $g$-$\lb-$plane. The arrows point in the direction 
of increasing coarse graining, i.e., of decreasing $k$. (From \cite{frank1}.)}
\label{fig0}
\end{figure}

%%%%%%%%%%%%%%%%%%%%%%%%%%%%%%%%%%%%%%%%%%%%%%%%%%%%%%%%%%%%%%%%%
The Einstein-Hilbert truncation \eqref{G3} constitutes the
most prominent truncation studied to date \cite{mr,souma,oliver1,frank1,litimgrav,Codello:2008vh,Benedetti:2010nr}.
In \cite{frank1} the corresponding RG equations (\ref{betaeq}) 
have been analyzed in detail, using both 
analytical and numerical methods. In particular all RG trajectories
 have been classified, and examples
have been computed numerically. The most important classes of 
trajectories in the phase portrait on the $g$-$\lb-$plane 
are shown in fig.~\ref{fig0}. Notably,
all cutoffs tested to date confirm this picture at least qualitatively.
%
%%%%%%%%%%%%%%%%%%%%%%%%%%%%%%%%%%%%%%%%%%%%%%%%%%%%%%%%%%%%%%%
%%%%%%%%%%%%%%%%%%%%%%%%%%%%%%%%%%%%%%%%%%%%%%%%%%%%%%%%%%%%%%%%%

%
The RG flow is found to be dominated by two fixed points $(g^*, \lb^*)$:
the GFP at $g^* = \lb^* =0$, and 
a NGFP with $g^* > 0$ and 
$\lb^* >0$. There are three classes of trajectories emanating from the NGFP:
trajectories of Type Ia and IIIa run towards negative and positive
cosmological constants, respectively, and the single trajectory of
Type IIa (``separatrix'') hits the GFP for $k\to 0$. The
high momentum properties of QEG are governed by the 
NGFP; for $k \to \infty$, in fig.\ \ref{fig0} 
all RG trajectories on the half--plane
$g>0$ run into this point. The fact that at the NGFP the dimensionless coupling constants $g_k, \lambda_k$ approach constant, non-zero values then implies that the dimensionful quantities
run according to
\be
\label{asymrun}
G_k = g^* k^{2-d} \;, \qquad \bar{\lambda}_k = \lambda^*\,k^2 \, .
\ee 
Hence for $k \ra \infty$ and $d > 2$ the dimensionful Newton constant vanishes
while the cosmological constant diverges.

Thus the Einstein-Hilbert truncation does indeed predict the existence of a NGFP
with exactly the properties needed for the Asymptotic Safety construction. Clearly the crucial question now is whether this NGFP is the projection of a fixed point in the exact theory 
 or whether it is merely the artifact of an insufficient approximation.
We now summarize the properties of the NGFP established 
within the Einstein-Hilbert truncation. 
All findings mentioned below
are independent pieces of evidence pointing in the direction 
that QEG is indeed asymptotically safe in 
four dimensions. Except for point (5) all results refer to $d=4$.  

\noindent
{\bf (1) Universal existence:} 
The non-Gaussian fixed point exists for all
cutoff schemes and shape functions implemented to date. It seems impossible to find an admissible cutoff which
destroys the fixed point in $d=4$. This result is highly non-trivial since in
higher dimensions $(d\gtrsim 5)$ the existence of the NGFP depends on the cutoff 
chosen \cite{frank1}.

\noindent
{\bf (2) Positive Newton constant:}
While the position of the fixed point is scheme dependent, all cutoffs yield
{\it positive} values of $g^*$ and $\lambda^*$. A negative $g^*$ might have 
been problematic for stability reasons, but there is no mechanism in the flow
equation which would exclude it on general grounds. 

\noindent
{\bf (3) Stability:}
For any cutoff employed the NGFP is found to be UV
attractive in both directions of the $\lambda$-$g-$plane. Linearizing the
flow equation according to eq.\ (\ref{H2}) we obtain a pair of complex
conjugate critical exponents $\theta_1=\theta_2^*$ with positive real part 
$\theta'$ and imaginary parts $\pm\theta''$. In terms of $t = \ln(k/k_0)$
the general solution to the linearized flow equations reads
\begin{eqnarray}
\label{H4}
\left(\lambda_k,g_k\right)^{\bf T}
&=&\left(\lambda^*,g^*\right)^{\bf T}
+2\Bigg\{\left[{\rm Re}\,C\,\cos\left(\theta''\,t\right)
+{\rm Im}\,C\,\sin\left(\theta''\,t\right)\right]
{\rm Re}\,V\nonumber\\
& &+\left[{\rm Re}\,C\,\sin\left(\theta''\,t\right)-{\rm Im}\,C
\,\cos\left(\theta''\,t\right)\right]{\rm Im}\,V\Bigg\}e^{-\theta' t}\;.
\end{eqnarray}
with $C\equiv C_1=(C_2)^*$ an arbitrary complex number and $V\equiv 
V^1=(V^{2})^*$ the right-eigenvector of ${\bf B}$ with eigenvalue $-\theta_1
=-\theta_2^*$. Eq.~(\ref{H2}) implies that, due to the positivity of 
$\theta'$, all trajectories hit the fixed point as $t$ is sent to infinity. 
The non-vanishing imaginary part $\theta''$ has no impact on the stability. 
However, it influences the shape of the trajectories which spiral into the 
fixed point for $k\rightarrow\infty$. Thus, the fixed 
point has the stability properties needed in the Asymptotic Safety scenario.

Solving the full, non-linear flow equations \cite{frank1} shows that the 
asymptotic scaling region where the linearization (\ref{H4}) is valid 
extends from $k=\infty$ down to about $k\approx m_{\rm Pl}$ with the
Planck mass defined as $m_{\rm Pl}\equiv G_0^{-1/2}$. Here $m_{\rm Pl}$ 
plays a role similar to $\Lambda_{\rm QCD}$ in QCD: it marks the lower 
boundary of the asymptotic scaling region. We set $k_0\equiv
m_{\rm Pl}$ so that the asymptotic scaling regime extends from about 
$t=0$ to $t=\infty$.

\noindent
{\bf (4) Scheme- and gauge dependence:}
Analyzing the cutoff scheme dependence of $\theta'$, $\theta''$, and 
$g^*\lambda^*$ as a measure for the reliability of the truncation,
the critical exponents were found to be reasonably constant within about a 
factor of 2. For $\alpha=1$ and
$\alpha=0$, for instance, they assume values in the ranges $1.4\lesssim
\theta'\lesssim 1.8$, $2.3\lesssim\theta''\lesssim 4$ and $1.7\lesssim
\theta'\lesssim 2.1$, $2.5\lesssim\theta''\lesssim 5$, respectively. The
universality properties of the product $g^*\lambda^*$ are even more
impressive. Despite the rather strong scheme dependence of $g^*$ and 
$\lambda^*$ separately, their product has almost no visible $s$-dependence for
not too small values of $s$. Its value is
\begin{eqnarray}
\label{H5}
g^*\lambda^*\approx\left\{\begin{array}{l}\mbox{$0.12$ for $\alpha=1$}\\
\mbox{$0.14$ for $\alpha=0$\,.}\end{array}\right.
\end{eqnarray}
The difference between the ``physical'' (fixed point) value of the gauge
parameter, $\alpha=0$, and the technically more convenient $\alpha=1$ are at 
the level of about 10 to 20 percent. 

\begin{figure}[t]
\centering
\begin{minipage}{.5\columnwidth}
	\centering
	\includegraphics[width=\columnwidth]{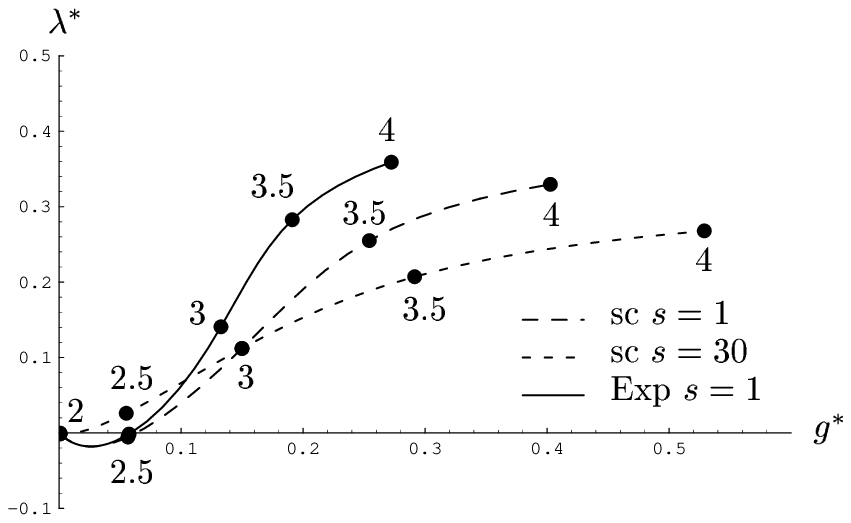}
\end{minipage}
\hfill
\begin{minipage}{.48\columnwidth}
	\centering
	\includegraphics[width=\columnwidth]{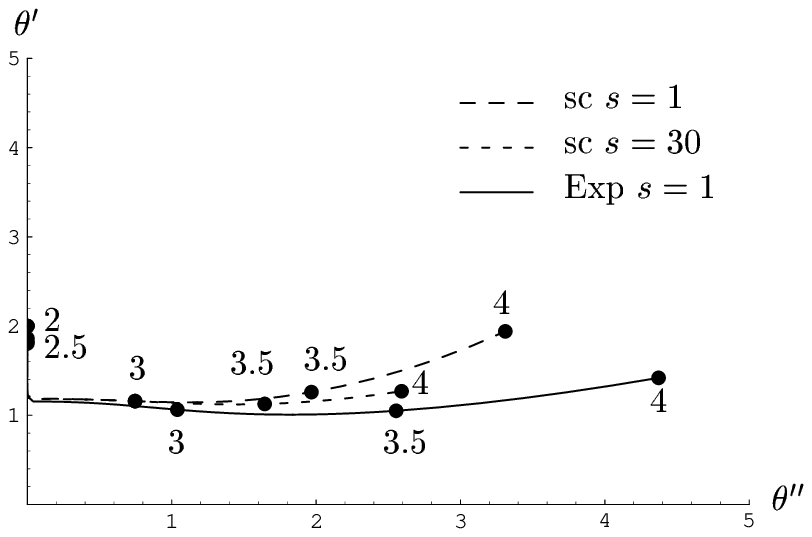}
\end{minipage}
\begin{minipage}{.5\columnwidth}
	\centering
	\includegraphics[width=\columnwidth]{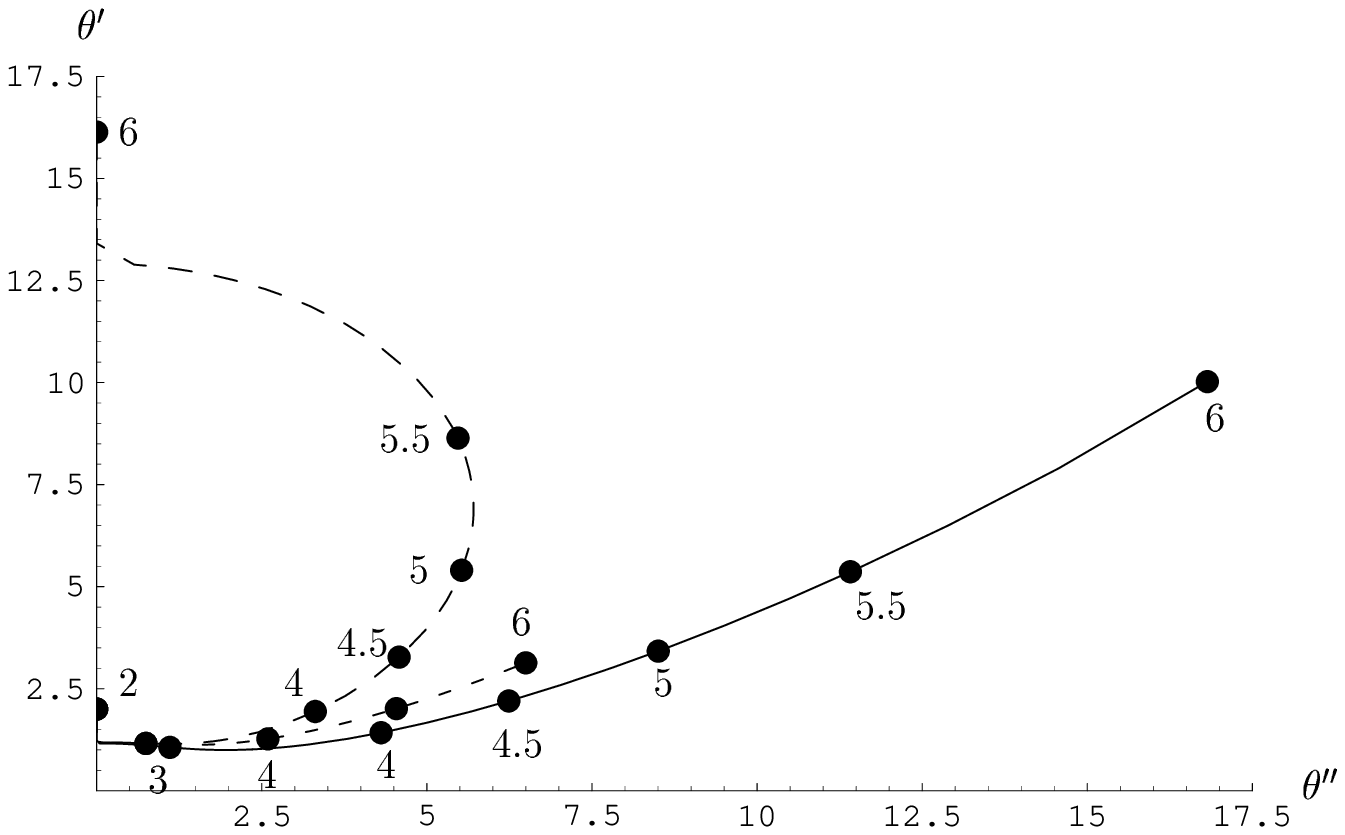}
\end{minipage}
\caption{\label{fixd}{\small Comparison of $\lambda^*,g^*, 
\theta^{\prime}$ and $\theta^{\prime \prime }$ for different cutoff 
functions in dependence of the dimension $d$. Two versions of the sharp 
cutoff (sc) and the exponential cutoff with $s=1$ (Exp) have been employed.
 The upper line shows that for $2+ \varepsilon \le d \le 4$ the cutoff scheme dependence of the
 results is rather small. The lower diagram shows that increasing $d$ beyond 
about 5 leads to a significant difference in the results for $\theta^{\prime}, 
\theta^{\prime \prime}$ obtained with the different cutoff schemes. (From \cite{frank1}.)}}
\end{figure}

\noindent
{\bf (5) Higher and lower dimensions:} The beta functions implied 
by the FRGE are continuous functions of the space-time dimensionality and it is 
instructive to analyze them for $d \neq 4$. 
In ref.\ \cite{mr} it has been shown that for $d = 2 + \varepsilon$, $|\varepsilon| \ll 1 $, the FRGE reproduces Weinberg's \cite{wein} fixed point for Newton's constant, $g^* = \frac{3}{38}\varepsilon$, and also supplies a corresponding fixed point value for the cosmological constant, $\lambda^* = - \frac{3}{38} \Phi^1_1(0) \varepsilon$, with the threshold function given in \eqref{G17}. 
For arbitrary $d$ and a generic cutoff the RG flow is quantitatively 
similar to the 4-dimensional one for all $d$ smaller than a certain 
critical dimension $d_{\rm crit}$, above which the existence or non-existence of the NGFP becomes  cutoff-dependent. The critical dimension is scheme dependent, but for any admissible 
cutoff it lies well above $d=4$. As $d$ approaches $d_{\rm crit}$ from 
below, the scheme dependence of the universal quantities increases 
drastically, indicating that the $R$-truncation becomes insufficient 
near $d_{\rm crit}$. 

 In fig.~\ref{fixd} we show 
the $d$-dependence of $g^*$, $\lb^*$, $\th'$, and $\th''$ for 
two versions of the sharp cutoff (with $s=1$ and $s=30$, respectively)
and for the exponential cutoff with $s=1$. For $2 + \varepsilon \leq d \leq 4$
the scheme dependence of the critical exponents is rather weak;
it becomes appreciable only near $d \approx 6$ \cite{frank1}. 
Fig.~\ref{fixd} suggests that the Einstein-Hilbert 
truncation in $d=4$ performs almost as well as near $d=2$. Its validity can be extended towards larger dimensionalities by optimizing the shape function \cite{litimgrav}.
%

%----------------------------------------------------------------------------
\subsection{$f(R)$-type truncations}
\label{sect:6.4}
%----------------------------------------------------------------------------
The ultimate justification of a given truncation consists in 
checking that if one adds further terms to it, its physical 
predictions remain robust. The first step towards testing the 
robustness of the Einstein-Hilbert truncation near the NGFP
against the inclusion of other invariants has been taken in 
refs.~\cite{oliver2,oliver3} where the beta functions for
the three generalized couplings $g, \lb$ 
and $\beta$ entering into the  $R^2$--truncation of 
eq.~(\ref{G20}) have been derived and analyzed.  

Subsequently, the truncated theory space has been extended to arbitrary functions of the Ricci scalar in \cite{r6,MS1,Codello:2008vh,Rahmede:2011zz}. 
In this truncation ansatz $S_{\rm gf}$ and $S_{\rm gh}$ are 
taken to be classical while, in the language of eq.\ \eqref{G1},
\be
\bar{\Gamma}_k[g] = \int \! d^4x \sqrt{g} \, f_k(R) \, , \quad \widehat{\Gamma}_k[g, \bg] = 0 \, .
\ee
Substituting this ansatz into the flow equation \eqref{E9} results in a rather complicated 
partial differential equation governing the scale-dependence of $f_k(R)$ \cite{MS1}.
Based on this equation, the search for the NGFP on truncation subspaces involving higher powers of the curvature scalar reduces to an algebraic problem. 
Substituting the ansatz\footnote{The Einstein-Hilbert truncation discussed in section \ref{sect:4} corresponds to setting 
$f_k(R) = (16 \pi G_k)^{-1} \left(-R + 2 \Lambda_k \right)$ and using \eqref{EHstruc} rather than setting
$\widehat{\Gamma}[g, \bg] = 0$. Comparing the Einstein-Hilbert ansatz to \eqref{eq:7.20} shows $u_1(k) = - (16 \pi g_k)^{-1}$, 
so that a negative $u_1(k)$ actually corresponds to a positive Newton's constant.}
\be\label{eq:7.20}
f_k(R) = \sum_{n = 0}^N u_n(k) \, k^4 \, (R/k^2)^n \; , \quad N \in \mathbb{N} \, ,
\ee
and expanding the resulting equation in powers of $R$ allows
to extract the non-perturbative beta functions for the dimensionless couplings $u_n(k)$,
\be\label{eq:7.21}
k \p_k u_n(k) = \beta_{u_n}(u_0, \cdots , u_N) \,, \quad n = 0, \cdots,N \, .
\ee 
The fixed point conditions $\beta_{u_n}(u_0^*, \cdots , u_N^*) = 0, n = 0,\ldots,N$ can then be solved numerically.
Notably, the inclusion of higher-derivative terms provided crucial evidence that UV-critical hypersurface of the NGFP known from the Einstein-Hilbert truncation has a finite dimension, indicating
that the Asymptotic Safety scenario is predictive. We shall now summarize the central results obtained within this class of truncations.

\begin{figure}[t]
\begin{minipage}{.49\columnwidth}
	\centerline{\includegraphics[width=\columnwidth]{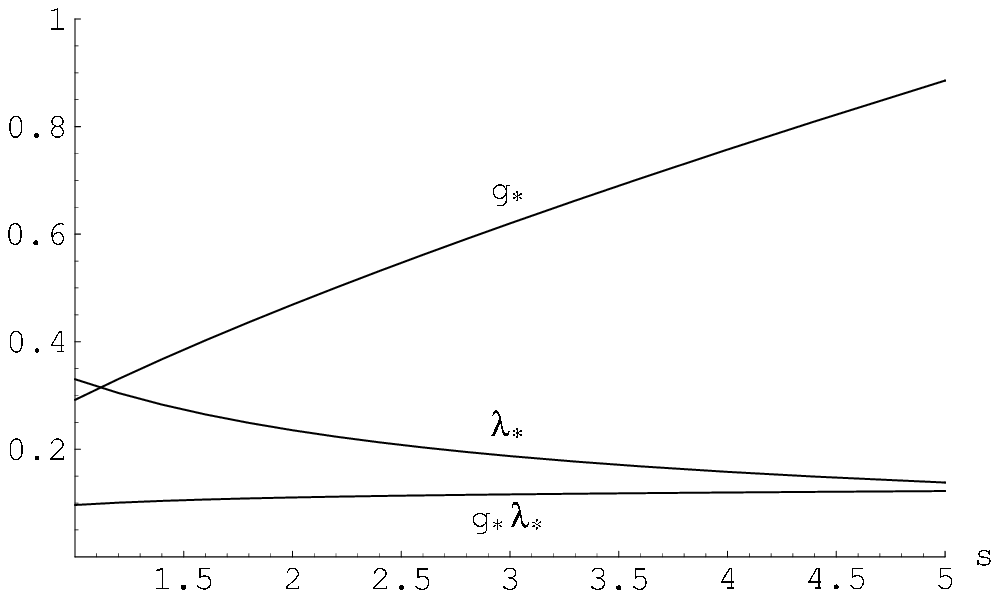}}
	\centerline{(a)}
\end{minipage}
\hfill
\begin{minipage}{.49\columnwidth}
	\centerline{\includegraphics[width=\columnwidth]{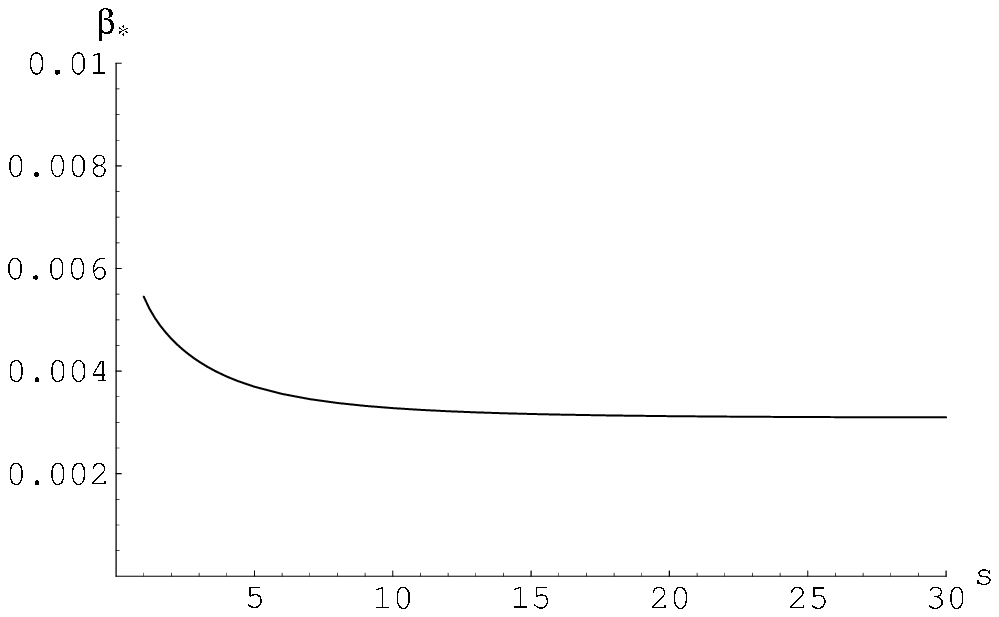}}
	\centerline{(b)}
\end{minipage}
\vspace{0.2cm}
\caption{\small (a) $g^*$, $\lambda^*$, and $g^*\lambda^*$ as functions of $s$ 
for $1\le s\le 5$, and (b) $\beta^*$ as a function of $s$ for $1\le s\le 30$,
using the family of exponential shape functions (\ref{G19}). 
(From ref.~\cite{oliver3}.)}  
\label{plot1}
\end{figure}

\noindent
{\bf (1)  Position of the fixed point $(R^2)$:}
Also with the generalized truncation \eqref{G20} the NGFP is found to exist for all
admissible cutoffs. Fig.\ \ref{plot1} shows its coordinates 
$(\lambda^*,g^*,\beta^*)$ for the family of shape functions (\ref{G19}). 
For every shape parameter $s$, the values of $\lambda^*$ 
and $g^*$ are almost the same as those obtained with the Einstein-Hilbert 
truncation. In particular, the product $g^*\lambda^*$ is constant with a very
high accuracy. For $s=1$, for instance, one obtains 
$(\lambda^*,g^*)=(0.348,0.272)$ from the Einstein-Hilbert truncation and
$(\lambda^*,g^*,\beta^*)=(0.330,0.292,0.005)$ from the generalized truncation.
It is quite remarkable that $\beta^*$ is always significantly
smaller than $\lambda^*$ and $g^*$. Within the limited precision of our
calculation this means that in the 3-dimensional parameter space the fixed
point practically lies on the $\lambda$-$g-$plane with $\beta=0$, 
i.e., on the parameter space of the pure Einstein-Hilbert truncation.

\noindent
{\bf (2) Eigenvalues and -vectors $(R^2)$:}
The NGFP of the
$R^2$-truncation proves to be UV attractive in any of the three directions of
the $(\lambda,g,\beta)-$space for all cutoffs used. The linearized flow in
its vicinity is always governed by a pair of complex conjugate critical
exponents $\theta_1=\theta'+{\rm i}\theta''=\theta_2^*$ with $\theta'>0$ and
 a single real, positive critical exponent $\theta_3>0$. 
For the exponential shape function with $s=1$, for instance, we find
$\theta'=2.15$, $\theta''=3.79$, $\theta_3=28.8$.
The first two are again of the spiral type while the third one is a straight line.

For any cutoff, the numerical results have several quite remarkable
properties. They all indicate that, close to the NGFP, the
RG flow is rather well approximated by the pure Einstein-Hilbert truncation.

\begin{figure}[t]
\begin{minipage}{.49\columnwidth}
	\centerline{\includegraphics[width=\columnwidth]{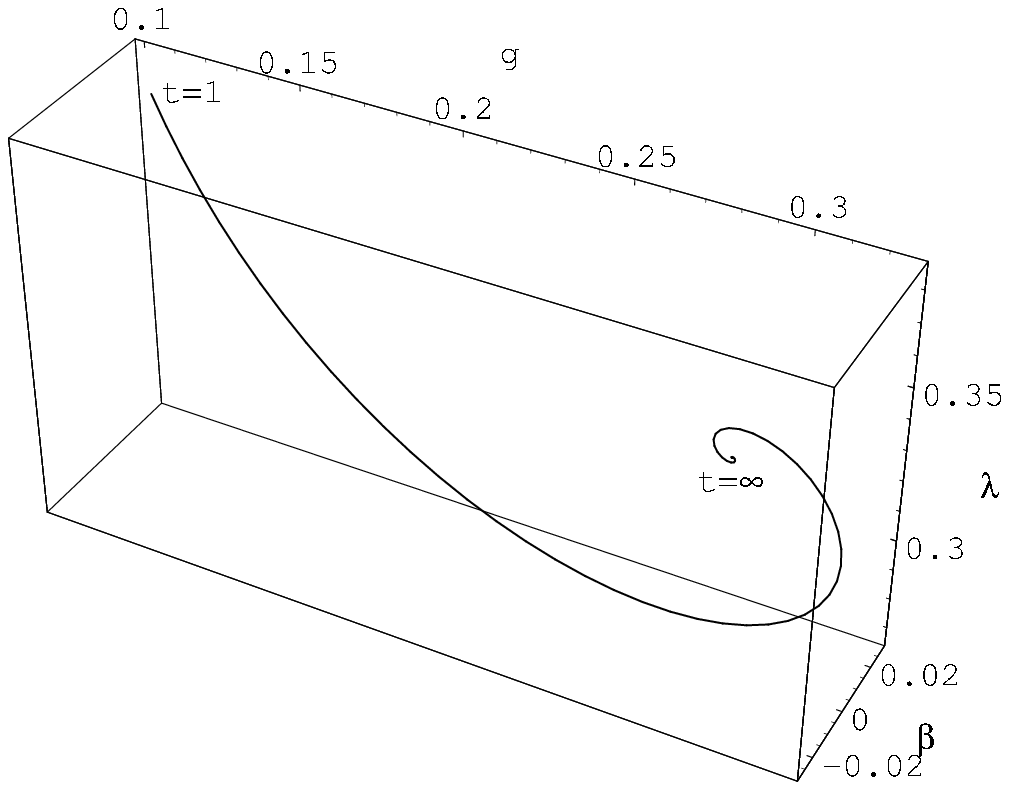}}
	\centerline{(a)}
\end{minipage}
\hfill
\begin{minipage}{.49\columnwidth}
	\centerline{\includegraphics[width=\columnwidth]{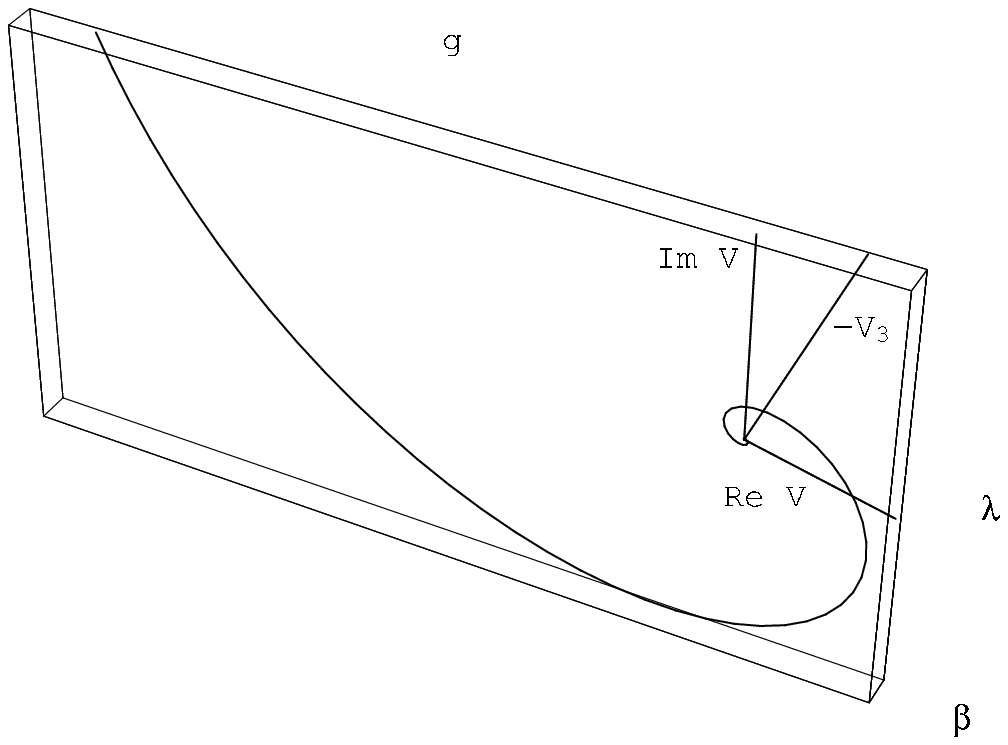}}
	\centerline{(b)}
\end{minipage}
\vspace{0.2cm}
\caption{\small Trajectory of the linearized flow equation obtained from the
$R^2$--truncation for $1\le t=\ln(k/k_0)<\infty$. In (b) we depict
the eigendirections and the ``box'' to which the trajectory is 
confined. (From ref.~\cite{oliver3}.)}  
\label{plot2}
\end{figure}

\noindent
{\bf (a)} The eigenvectors associated with the spiraling directions span
a plane which virtually coincides with the
$g$-$\lambda-$subspace at $\beta=0$, i.e., with the parameter space of the
Einstein-Hilbert truncation. As a consequence, the  corresponding normal modes are essentially the same trajectories as the
``old'' normal modes already found without the $R^2$--term. Also the
corresponding $\theta'$-- and $\theta''$--values coincide within the scheme
dependence.

\noindent
{\bf (b)}
The new eigenvalue $\theta_3$ introduced by the $R^2$--term is
significantly larger than $\theta'$. When a trajectory approaches the fixed
point from below $(t\rightarrow\infty)$, the ``old'' normal modes 
 are proportional to $\exp(-\theta' t)$, but the new
one is proportional to $\exp(-\theta_3 t)$, so that it decays much quicker.
 For every trajectory running into the fixed point we find therefore that once
$t$ is sufficiently large the trajectory lies entirely in the
 $\beta=0$-plane practically.
Due to the large value of $\theta_3$, the new scaling field is very
``relevant''. However, when we start at the fixed point $(t=\infty)$
and lower $t$ it is only at the low energy scale $k\approx m_{\rm Pl}$
$(t\approx 0)$ that $\exp(-\theta_3 t)$ reaches unity, and only then, i.e.,
far away from the fixed point, the new scaling field starts growing
rapidly.

Thus very close to the fixed
point the RG flow seems to be essentially 2-dimensional, and that this
2-dimensional flow is well approximated by the RG equations of the
Einstein-Hilbert truncation. In fig.\ \ref{plot2} we show a typical trajectory
which has all three normal modes excited with equal strength.
All the way down from
$k=\infty$ to about $k=m_{\rm Pl}$ it is confined to a very thin
box surrounding the $\beta=0$--plane.

\begin{figure}[t]
\begin{minipage}{.49\columnwidth}
	\centerline{\includegraphics[width=\columnwidth]{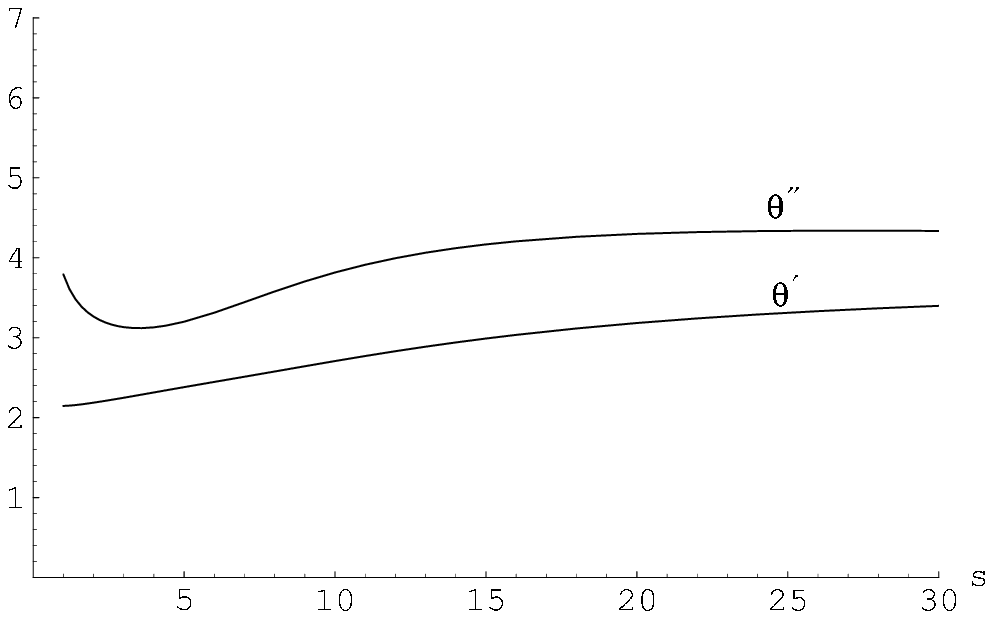}}
	\centerline{(a)}
\end{minipage}
\hfill
\begin{minipage}{.49\columnwidth}
	\centerline{\includegraphics[width=\columnwidth]{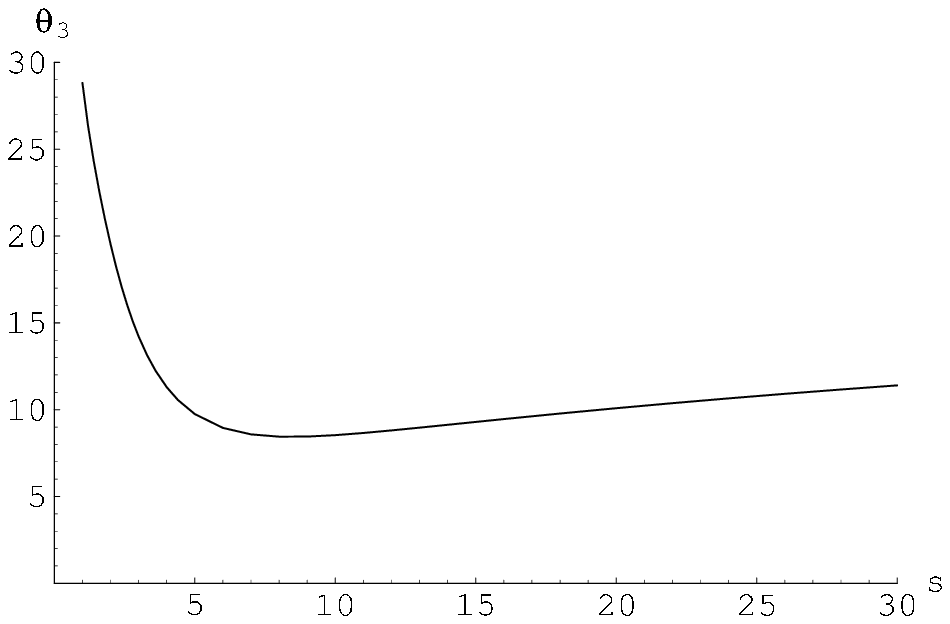}}
	\centerline{(b)}
\end{minipage}
\vspace{0.2cm}
\caption{\small (a) $\theta'={\rm Re}\,\theta_1$ and $\theta''={\rm Im}\,\theta_1$, 
and (b) $\theta_3$ as functions of $s$, using the family of exponential shape 
functions (\ref{G19}). (From \cite{oliver2}.)}  
\label{plot3}
\end{figure}
\noindent
{\bf (3) Scheme dependence $(R^2)$:}
The scheme dependence of the critical exponents and of the product
$g^*\lambda^*$ turns out to be of the same order of magnitude as in the case
of the Einstein-Hilbert truncation. Fig.\ \ref{plot3} shows the cutoff
dependence of the critical exponents, using the family of shape functions
(\ref{G19}). For the cutoffs employed $\theta'$ and $\theta''$ assume values
in the ranges $2.1\lesssim\theta'\lesssim 3.4$ and $3.1\lesssim\theta''
\lesssim 4.3$, respectively. While the scheme dependence of $\theta''$ is
weaker than in the case of the Einstein-Hilbert truncation one finds that it is
slightly larger for $\theta'$. The exponent $\theta_3$ suffers from relatively
strong variations as the cutoff is changed, $8.4\lesssim\theta_3\lesssim
28.8$, but it is always significantly larger than $\theta'$.
The product $g^*\lambda^*$ again exhibits an extremely weak scheme dependence.
Fig. \ref{plot1}(a) displays $g^*\lambda^*$ as a function of $s$. 
It is impressive to see how the cutoff
dependences of $g^*$ and $\lambda^*$ cancel almost perfectly. Fig.
\ref{plot1}(a) suggests the universal value $g^*\lambda^*\approx 0.14$.
Comparing this value to those obtained from the Einstein-Hilbert truncation
we find that it differs slightly from the one based upon
the same gauge $\alpha=1$. The deviation is of the same size as the difference
between the $\alpha=0$-- and the $\alpha=1$--results of the Einstein-Hilbert
truncation.

\noindent
{\bf (4) Dimensionality of ${\cal S}_{\rm UV}$ ($R^2$):}
According to the canonical dimensional analysis, the
(curvature)$^n$-invariants in 4 dimensions are classically 
marginal for
$n=2$ and irrelevant for $n>2$. The results for $\theta_3$ indicate that there
are large non-classical contributions so that there might be relevant operators
perhaps even beyond $n=2$. With the $R^2-$truncation it is clearly not
possible to determine their number $\Delta_{\rm UV}$ in $d=4$. However, as it is
hardly conceivable that the quantum effects change the signs of arbitrarily
large (negative) classical scaling dimensions, $\Delta_{\rm UV}$ should be
finite \cite{wein}. 

The first confirmation of this picture came from the
$R^2$-calculation in $d=2+\varepsilon$ where the dimensional count is shifted
by two units. In this case we find indeed that the third scaling field is
{\it irrelevant} for any cutoff employed, $\theta_3<0$. 
 Using the
$\varepsilon$-expansion the corresponding numerical results
for selected values of the shape parameter $s$ are presented in
table  \ref{Tab.O}. 

\begin{table}[t]
\renewcommand{\arraystretch}{1.5}
\begin{center}
\vspace{0.2cm}
\begin{tabular}
{c c c c c c c}
\hline
$s$ & \,$\lambda_*\,(+{\cal O}(\varepsilon^2)$) & 
\,$g_*\,(+{\cal O}(\varepsilon^2))$ & \,$\beta_*\,(+{\cal O}(\varepsilon))$ & 
$\theta_1\,(+{\cal O}(\varepsilon))$ & $\theta_2\,(+{\cal O}(\varepsilon^2))$
& $\theta_3\,(+{\cal O}(\varepsilon))$\\
\hline\hline 1 & $-0.131\varepsilon$ & $0.087\varepsilon$ & $-0.083$ & $2$ & 
$0.963\varepsilon$ & $-1.968$\\
5 & $-0.055\varepsilon$ & $0.092\varepsilon$ & $-0.312$ & $2$ &
$0.955\varepsilon$ & $-1.955$\\
10 & $-0.035\varepsilon$ & $0.095\varepsilon$ & $-0.592$ & $2$ & 
$0.955\varepsilon$ & $-1.956$\\
\hline\end{tabular}
\end{center}
\renewcommand{\arraystretch}{1}
\caption{\small Fixed point coordinates and critical exponents of the $R^2$-truncation in $2+\varepsilon$ dimensions.
	The negative value $\theta_3<0$ implies that $\cal{S}_{\rm UV}$ is an (only!) 2-dimensional surface in the 3-dimensional
	theory space.}
\label{Tab.O}
\end{table}

For all cutoffs used we
obtain three {\it real} critical exponents, the first two are positive and the
third is negative. This suggests that in $d = 2 + \varepsilon$ the dimensionality of ${\cal S}_{\rm UV}$ could be as
small as $\Delta_{\rm UV}=2$ and characterized by only two free parameters, the
renormalized Newton constant $G_0$ and the renormalized cosmological constant
$\bar{\lambda}_0$, for instance.

\begin{table}[t]
\renewcommand{\arraystretch}{1.5}
\begin{center}
\begin{tabular}{ccccccccc}
\hline
$N$ & $u_0^*$ & $u_1^*$ & $u_2^*$ & $u_3^*$ & $u_4^*$ & $u_5^*$  & $u_6^*$ & $g^* \lambda^*$ \\ \hline\hline
$1$ & $0.00523$ & $-0.0202$ &         &         &         &          &      & $0.127$   \\
$2$ & $0.00333$ & $-0.0125$ & $0.00149$ &         &         &          &    & $0.211$    \\
$3$ & $0.00518$ & $-0.0196$ & $0.00070$ & $-0.0104$ &         &          &   & $0.134$      \\
$4$ & $0.00505$ & $-0.0206$ & $0.00026$ & $-0.0120$ & $-0.0101$ &          &   & $0.118$      \\
$5$ & $0.00506$ & $-0.0206$ & $0.00023$ & $-0.0105$ & $-0.0096$ & $-0.00455$ &  & $0.119$        \\
$6$ & $0.00504$ & $-0.0208$ & $0.00012$ & $-0.0110$ & $-0.0109$ & $-0.00473$ & $0.00238$ & $0.116$ \\\hline
\end{tabular}
\end{center}
\renewcommand{\arraystretch}{1}
\caption{\small Location of the NGFP obtained within the $f(R)$-truncation by expanding the partial differential equation
	in a power series in $R$ up to order
	$R^N$, including $k$-dependent dimensionless coupling constants $u_n(k)$, $n = 0,\cdots,N$. (From \cite{MS1}.)}
\label{t.3}
\end{table}
\noindent
{\bf (5) Position of the non-Gaussian fixed point ($f(R), d = 4$):}
The beta functions \eqref{eq:7.21} also give rise to a NGFP with $g^* > 0, \lambda^* > 0$, 
whose $N$-dependent position is shown in table \ref{t.3}. 
In particular the product $g^* \lambda^* = u_0/(32 \pi (u_1)^2)$ displayed in the last column is remarkably constant. It is in excellent agreement with the Einstein-Hilbert truncation \eqref{H5}, and the $R^2$-truncation, fig.\ \ref{plot1}(a). Only the value obtained in the case $N=2$ shows a mild deviation from the $R^2-$computations, which can, most probably, be attributed to the use of a different gauge-fixing procedure, cutoff shape function, and ansatz for $\widehat{\Gamma}_k$.

\begin{table}[t]
\renewcommand{\arraystretch}{1.5}
\begin{center}
\begin{tabular}{cccccccc}
\hline
$N$ & $\theta^\prime$ & $\theta^{\prime \prime}$ & $\theta_2$ & $\theta_3$ & $\qquad \; \; \theta_4$ & $\theta_5$ & $\theta_6$ \\ \hline\hline
$1$ & $2.38$ & $-2.17$ &         &          &         &         &         \\
$2$ & $1.26$ & $-2.44$ & $27.0$    &          &         &         &         \\
$3$ & $2.67$ & $-2.26$ & $2.07$    & $-4.42$    &         &         &         \\
$4$ & $2.83$ & $-2.42$ & $1.54$    & $-4.28$    &  $-5.09$\phantom{+  4.57 $\I$}   &         &         \\
$5$ & $2.57$ & $-2.67$ & $1.73$    & $-4.40$    &  $-3.97 +  4.57 \I$ & $-3.97 - 4.57 \I$ & \\
\;$6$\; & \;$2.39$\; & \;$-2.38$\; & \;$1.51$\;    & \;$-4.16$\;    &  \;$-4.67 +  6.08 \I$\; & \;$-4.67 - 6.08 \I$\; & \;$-8.67$\; \\\hline
\end{tabular}
\end{center}
\renewcommand{\arraystretch}{1}
\caption{\small Stability coefficients of the NGFP for increasing dimension $N+1$ of the truncation subspace.
	The first two critical exponents are a complex pair $\theta_\pm = \theta^\prime \pm \I \theta^{\prime \prime}$. (From \cite{MS1}.)}
\label{t.4}
\end{table}
\noindent
{\bf (6) Dimensionality of ${\cal S}_{\rm UV}$ ($f(R), d = 4$):}
The critical exponents resulting from the stability analysis of the NGFP emerging from 
the $f(R)$-truncation are summarized in table \ref{t.4}. In particular we find that only three of the eigendirections associated to the NGFP are relevant, i.e., UV attractive. Including higher derivative terms $R^n, n \ge 3$ in the truncation creates irrelevant directions only. Thus, as in the $R^2$- and $R^2 + C^2$-truncation, $\Delta_{\rm UV} = 3$, i.e., the UV critical surface associated to the fixed point is a 3-dimensional submanifold in the truncated theory space. Its dimensionality is stable with respect to increasing $N$. The RG trajectories tracing out this surface are determined by fixing the three relevant couplings. These describe in which direction tangent to $\cS_{\rm UV}$ the trajectory flows away from the fixed point. All remaining couplings, the irrelevant ones, are predictions from Asymptotic Safety. In \cite{Codello:2008vh} these results have been extended up to $N=8$, providing even stronger evidence for the robustness of the RG flow under the inclusion of further invariants.

%-----------------------------------------------------------------------------------
\subsection{The $R^2 + C^2$-truncation in $d=4$}
\label{sect:6.5}
%-----------------------------------------------------------------------------------
The extension of the $R^2$-truncation by including non-scalar curvature terms in
the truncation subspace has been investigated in \cite{HD1,HD2}. This setup 
is based on the ansatz \eqref{R2C2}, and includes the characteristic features of higher-derivative 
gravity as, e.g., a fourth-order propagator for the helicity 2 states. Moreover, it provides an important
spring-board for understanding the role of the counterterms arising in the perturbative 
quantization of gravity in the Asymptotic Safety program.

Including tensor structures like $C_{\mu\nu\rho\sigma} C^{\mu\nu\rho\sigma}$ in the truncation
subspace requires the generalization of the background metrics $\bar{g}$. As discussed at the end
of section \ref{sect:4}, choosing the class of maximally symmetric metrics on $S^d$ as a background
considerably simplifies the evaluation of the truncated flow equation, but comes with the
drawback that the flow is projected on interaction terms built from the Ricci scalar only.
Thus the inclusion of the remaining four-derivative operators requires the use of a more general
class of backgrounds. Ideally, this new class is generic enough to disentangle
 the coefficients multiplying $R^2$ and the tensorial terms, and, most importantly, simple enough to avoid the appearance of non-minimal 
higher-derivative differential operators inside the trace. While the maximally symmetric backgrounds used up to now are insufficient in 
the former respect, a generic compact Einstein background (without Killing
 or conformal Killing vectors and without boundary for 
simplicity), satisfying
$\bar{R}_{\mu\nu}=\tfrac{\bar R}{4} \, \bar{g}_{\mu\nu}$, 
is sufficient to meet both criteria and allows one to determine the non-perturbative 
beta functions of the linear combinations \eqref{lincomb}.

Surprisingly, projecting the flow equation resulting from the ansatz \eqref{R2C2} onto an generic
Einstein background, the differential operators appearing on its RHS organize themselves into
second order differential operators of the Lichnerowicz form
\be
\Delta_{2L} \phi_{\mu\nu}  \equiv -\bar{D}^2 \phi_{\mu\nu} - 2 \bar{R}_{\mu\,\,\,\nu}^{\,\,\,\alpha\,\,\,\beta} \phi_{\alpha\beta}\,,%\\
\; \;
\Delta_{1L} \phi_\mu   \equiv -\bar{D}^2 \phi_\mu - \bar{R}_{\mu\nu} \phi^\nu\,, \; %\\
\; \Delta_{0L} \phi  \equiv -\bar{D}^2 \phi,
\ee
which commute with all the other curvature terms inside the trace. This feature makes the 
traces amenable to standard heat kernel techniques for minimal second order differential
operators. The resulting beta functions for the dimensionless coupling constants
$\lambda_k, g_k, \beta_k$, and $\gamma_k$ are somewhat involved, so that we only highlight 
their main properties 

\noindent
{\bf(1) Existence of the NGFP:} The beta functions of the $R^2+C^2$-truncation also give rise to a NGFP with positive Newtons and cosmological constant
\be\label{R2C2FP}
g^* = 1.960 \, , \quad \lambda^* = 0.218 \, , \quad \beta^* = 0.008 \, , \quad \gamma^* = -0.005 \, , \qquad g^* \lambda^* = 0.427 \, .
\ee
The finite values for $\beta^*$ and $\gamma^*$ also imply a finite value of $\sigma^*$, via eq.\ \eqref{lincomb}. This should be contrasted to the one-loop result $\sigma^*=0$ obtained from the perturbative quantization of fourth-order gravity. Thus the non-perturbative corrections captured by the FRGE shift the fixed point underlying the asymptotic freedom obtained within perturbation theory to the NGFP featuring in the Asymptotic Safety program.

\noindent
{\bf(2) Stability properties:} An important characteristics of the NGFP are its stability properties.
Linearizing the RG flow at the fixed point \eqref{R2C2FP} along the lines of section \ref{sect:2}, the stability coefficients are found as 
\be\label{stab:coeff}
%\begin{split}
\theta_0 = 2.51  \, , \qquad  \theta_1 = 1.69 \, , \qquad
\theta_2 = 8.40  \, , \qquad  \theta_3 =  -2.11 \,  .
%\end{split}
\ee
We observe that the inclusion of the $C^2$-coupling leads to real stability
coefficients. This is in contrast to the complex stability
coefficients and the corresponding spiraling approach of the RG flow
characteristic for $f(R)$-type truncations.

Moreover, eq.\ \eqref{stab:coeff} further indicates that the ``new'' direction added by extending the $R^2$-truncation is UV-repulsive.
Thus, in this 4-dimensional truncation space, $\cS_{\rm UV}$ remains 3-dimensional, as in the case of the
$R^2$-truncation. The condition for a trajectory being inside the UV-critical surface then imposes one constraint
between the coupling constants. In the linear regime at the NGFP, this constraint can be used to express $\gamma_k$, say,
in terms of the other coupling constants contained in the ansatz
\be\label{critsurface}
\gamma_k = -0.116 + 0.030 \, \lambda_k \, g_k^{-1} + 0.049 \, g_k^{-1} + 11.06 \beta_k \, .
\ee

Comparing the results \eqref{R2C2FP} and \eqref{stab:coeff} to their counterparts in the Einstein-Hilbert and $R^2$-truncation 
summarized in the previous subsections,
we conclude that the $C^2$-term leads to a moderate shift on fixed point structure. In particular the universal coupling $g^* \lambda^*$ turns out
to be enhanced by a factor of three, while the stability coefficients of the fixed point turn out to be real.
Thus the $C^2$-term significantly influences the RG flow of the theory. It effects the fixed point structure more drastically then the 
inclusion of the $R^2$-term or working with different cutoff-schemes or gauge-fixing
functions within the Einstein-Hilbert action.

\noindent
{\bf (3) The role of perturbative counterterms:} An interesting twist arises from supplementing the truncation ansatz \eqref{R2C2} with a free scalar field.  
From perturbative viewpoint the two-derivative terms of the resulting action constitute the prototypical example of a gravitational theory which is perturbatively non-renormalizable at one loop, with the four-derivative terms being the corresponding non-renormalizable on-shell counterterms \cite{tHooft:1974bx}. 
At the same time, it has been shown in \cite{perper1} that the RG flow obtained from the two-derivative truncation 
possesses a NGFP whose properties are very close to the one found in the Einstein-Hilbert truncation of pure gravity. 
Thus this setup provides a valuable laboratory, where the influence of perturbative counterterms on the NGFP underlying the Asymptotic Safety scenario can be studied
explicitly.

The beta function for the ``$R^2 + C^2 + {\rm scalar}$''-truncation can be computed completely analogous to the case of pure gravity.
 Notably, they also give rise to an NGFP \cite{HD2}
\be\label{FP:scalar}
g^* = 2.279 \, , \quad \lambda^* = 0.251 \, , \quad \beta^* = 0.010  \, , \quad \gamma^* = - 0.0043 \, , \qquad g^* \lambda^* = 0.571 \, ,
\ee
with stability coefficients
\be\label{stab:scalar}
\theta_0 = 2.67  \, , \qquad  \theta_1 = 1.39 \, , \qquad
\theta_2 = 7.86  \, , \qquad  \theta_3 =  -1.50 \,  .
\ee
Comparing these to the pure gravity results \eqref{R2C2FP}, \eqref{stab:coeff}, we observe that the gravity-matter fixed point has
strikingly similar properties. The universal couplings and stability coefficients change their values by roughly 25{\%}. Thus
the beta functions are dominated by the contribution from the gravitational sector. This result
lends strong support to the assertion that \emph{perturbative counterterms do not play a distinguished role in 
the Asymptotic Safety program}. In particular they are not fatal for the NGFP, so that the gravity-matter theory
remains asymptotically safe despite the inclusion of an perturbative conterterms in the truncation subspace.

\noindent
{\bf Summary:}
The above results strongly suggest that the non-Gaussian 
fixed point occurring in the Einstein-Hilbert truncation is not a 
truncation artifact but rather the projection of a fixed
point in the exact theory space. The fixed point and all its qualitative 
properties are stable against variations of the cutoff
and the inclusion of further invariants in the truncation. It is particularly
remarkable that within the scheme dependence the additional $R^2$--term has
essentially no impact on the fixed point. Moreover, truncations involving higher-order 
polynomials in $R$ or the tensor structure $C^2$ fully confirm the general picture suggested
 by the simple Einstein-Hilbert truncation and provide a strong indication that the corresponding quantum field 
theories are characterized by a finite number of free parameters. 
We interpret the above results and
their mutual consistency as quite non-trivial indications supporting the
conjecture that 4-dimensional QEG indeed possesses a RG
fixed point with precisely the properties needed for Asymptotic Safety.

%----------------------------------------------------------------------------
\section{The multifractal properties of QEG space-times}
\setcounter{equation}{0}
\label{sect:7}
%----------------------------------------------------------------------------
An intriguing consequence arising from the 
scale-dependence of the gravitational effective action is
 that the QEG space-times at short distances 
develop fractal properties \cite{oliver1,oliver2,oliverfrac,frankfrac}.
These manifest themselves at the level of the graviton propagator at high 
energies or in the diffusion processes of test particles. 
One of the striking conclusions reached in refs.\
\cite{oliver1,oliver2} was that the effective dimensionality of space-time 
 equals 4 at macroscopic distances ($\ell\gg\ell_{\rm Pl}$)
but, near $\ell\approx\ell_{\rm Pl}$, it gets dynamically reduced to the value
2. The remainder of this review 
is dedicated to the discussion of the arguments
that underlying this conclusion.

%----------------------------------------------------------------------------
\subsection{The origin of fractality: scale-dependent metrics}
\setcounter{equation}{0}
\label{sect:7.1}
%----------------------------------------------------------------------------
As we have seen, the effective average action $\Gamma_k[g_{\mu\nu}]$ defines a
continuous family of effective field theories, where all quantum fluctuations with momenta larger than
$k$ have been integrated out. Intuitively speaking, the solution 
$\big<g_{\mu\nu}\big>_k$ of the scale dependent field equation 
\begin{eqnarray}
\label{fe}
\frac{\delta\Gamma_k}{\delta g_{\mu\nu}(x)}\Big[\big<g\big>_k\Big]=0
\end{eqnarray}
can be interpreted as the metric averaged over (Euclidean) space-time volumes
of a linear extension $\ell$ which typically is of the order of $1/k$. Knowing
the scale dependence of $\Gamma_k$, i.e., the renormalization group trajectory
$k\mapsto\Gamma_k$, we can in principle follow the solution 
$\big<g_{\mu\nu}\big>_k$ from the ultraviolet $(k\rightarrow\infty)$ to the
infrared $(k\rightarrow 0)$.

It is an important feature of this approach that the infinitely many equations
of (\ref{fe}), one for each scale $k$, are valid {\it simultaneously}. They
all refer {\it to the same} physical system, the ``quantum space-time'', but
describe its effective metric structure on different scales. An observer using
a ``microscope'' with a resolution $\approx k^{-1}$ will perceive the universe
to be a Riemannian manifold with metric $\big<g_{\mu\nu}\big>_k$. At every 
fixed $k$, $\big<g_{\mu\nu}\big>_k$ is a smooth classical metric. But since
the quantum space-time is characterized by the infinity of equations (\ref{fe})
with $k=0,\cdots,\infty$ it can acquire very nonclassical and in particular
fractal features.

Let us describe more precisely what it means to ``average'' over Euclidean
space-time
volumes. The quantity we can freely tune is the IR cutoff scale $k$. The
``resolving power'' of the microscope, henceforth denoted $\ell$, is in
general a complicated function of $k$. (In flat space, $\ell\approx 1/k$.)
In order to understand the relationship between $\ell$ and $k$ we must recall
some steps from the construction of $\Gamma_k[g_{\mu\nu}]$ in section \ref{sect:3}.

The IR cutoff of the average action is implemented by first expressing the
functional integral over all metrics in terms of eigenmodes of $\bar{D}^2$,
the covariant Laplacian formed with the aid of the background metric 
$\bar{g}_{\mu\nu}$. Then a suppression term is introduced which damps the
contribution of all $-\bar{D}^2$-modes with eigenvalues smaller than $k^2$.
Following the steps of section \ref{sect:3} this leads to the scale dependent
functional $\Gamma_k[g_{\mu\nu};\bar{g}_{\mu\nu}]$, and the action with one
argument is again obtained by equating the two metrics:
$\Gamma_k[g_{\mu\nu}]\equiv\Gamma_k[g_{\mu\nu};\bar{g}_{\mu\nu}=g_{\mu\nu}]$.
This is this action which appears in (\ref{fe}). Because of the identification
of the two metrics we see that it is basically the eigenmodes of
$\bar{D}^2=D^2$,
constructed from the argument of $\Gamma_k[g]$, which are cut off at $k^2$.
Since $\big<g_{\mu\nu}\big>_k$ is the corresponding stationary point, we can
say that the metric $\big<g_{\mu\nu}\big>_k$ applies to the situation where
only the quantum fluctuations of $-D^2(\big<g_{\mu\nu}\big>_k)$ with
eigenvalues larger than $k^2$ are integrated out. Therefore there is a 
complicated
interrelation between the metric and the scale at which it provides an
effective description: The covariant Laplacian which ultimately decides about
which modes are integrated out is constructed from the ``on shell'' 
configuration $\big<g_{\mu\nu}\big>_k$, so it is $k$-dependent by itself
already.

From these remarks it is clear now how to obtain the ``resolving power'' 
$\ell$ for a given $k$, at least in principle.
 We start from a fixed RG trajectory
$k\mapsto\Gamma_k$, derive its effective field equations at each $k$, and solve
them. The resulting quantum mechanical counterpart of a classical space-time is 
specified by the infinity of Riemannian metrics $\{\big<g_{\mu\nu}\big>_k\big|
k=0,\cdots,\infty\}$. While the totality of these metrics contains all 
physical information, the parameter $k$ is only a book keeping device 
a priori. In a  second step, it can be given a physical interpretation by 
relating it
to the (proper) length scale of the averaging procedure: One constructs the
Laplacian $-D^2(\big<g_{\mu\nu}\big>_k)$, diagonalizes it, looks how rapidly 
its $k^2$-eigenfunction varies, and ``measures'' the length $\ell$ of typical 
variations with the metric $\big<g_{\mu\nu}\big>_k$ itself. By solving the
resulting $\ell=\ell(k)$ for $k=k(\ell)$ we can in principle reinterpret the
metric $\big<g_{\mu\nu}\big>_k$ as referring to a microscope with a known
position and direction dependent resolving power. The price we have to pay for
the background independence is that we cannot freely choose $\ell$ directly
but rather $k$ only.

We now illustrate this procedure at the level of the
Einstein-Hilbert truncation discussed in section \ref{sect:6.1}.
Without matter, the corresponding field equations
 happen to coincide with the 
ordinary Einstein equation, but with $G_k$ and $\bar{\lambda}_k$ replacing the
classical constants
\be\label{einsteq}
R_{\mu\nu}(\big<g\big>_k)
= \frac{2}{2-d} \, \bar{\lambda}_k \, \big<g_{\mu\nu}\big>_k \, . 
\ee

It is easy to make the $k$-dependence of $\big<g_{\mu\nu}\big>_k$ explicit.
Picking an arbitrary reference scale $k_0$ we may rewrite (\ref{einsteq}) as
$[\bar{\lambda}_{k_0}/\bar{\lambda}_k]\,R^\mu_{\;\;\nu}(\big<g\big>_k)
= \tfrac{2}{2-d} \bar{\lambda}_{k_0}\,\delta^\mu_\nu$. Since $R^\mu_{\;\;\nu}(c\,g)=c^{-1}\,
R^\mu_{\;\;\nu}(g)$ for any constant $c>0$, this relation implies that the
average metric and its inverse scale as
\be\label{metricscaling}
\langle g_{\m\nu}(x) \rangle_k = [\bar{\lambda}_{k_0} / \bar{\lambda}_k] \langle g_{\m\nu}(x) \rangle_{k_0} \, , \qquad 
\langle g^{\m\nu}(x) \rangle_k = [\bar{\lambda}_k/ \bar{\lambda}_{k_0}] \langle g^{\m\nu}(x) \rangle_{k_0} \, .
\ee
These relations are valid provided the family of solutions considered exists
for all scales between $k_0$ and $k$, and $\bar{\lambda}_k$ has the
same sign always.

Denoting the Laplace operators corresponding to the metrics $\langle g_{\m\nu} \rangle_k$ and $\langle g_{\m\nu} \rangle_{k_0}$ by $\Delta(k)$ and $\Delta(k_0)$, respectively, these relations imply
\be\label{Laplacescaling}
\Delta(k) = \left[\bar{\lambda}_k / \bar{\lambda}_{k_0} \right] \Delta(k_0) \, .
\ee

At this stage, the following remark is in order. In the asymptotic scaling regime associated with the NGFP the scale-dependence of the couplings is determined by the fixed point, see eq.\  \eqref{asymrun}.
Choosing $\bar{\lambda}_{k_0}$ in the classical regime, this implies in particular
\be\label{UVscaling}
\langle g_{\m\nu}(x) \rangle_k \propto k^{-2} \qquad (k \rightarrow \infty) \, .
\ee
This asymptotic relation is actually an {\it exact} consequence of Asymptotic Safety, which solely relies on the scale-independence of the theory at the fixed point. This can be seen as follows.
The complete effective average action has the structure \eqref{Gexpansion}. 
If $\bar{u}_\alpha(k)$ has the canonical dimension $d_\alpha$ the
corresponding dimensionless couplings are ${u}_\alpha(k)\equiv k^{-d_\alpha}\,
\bar{u}_\alpha(k)$ and we have
\begin{eqnarray}
\label{fullact}
\Gamma_k[g_{\mu\nu}]&=&\sum_\alpha{u}_\alpha(k)\,k^{d_\alpha}\,P_\alpha[g_{\mu\nu}]\;\,=\;\,
\sum_\alpha{u}_\alpha(k)\,P_\alpha[k^2\,g_{\mu\nu}] \, . 
\end{eqnarray}
In the second equality we used that $P_\alpha[c^2\,g_{\mu\nu}]=c^{d_\alpha}\,
P_\alpha[g_{\mu\nu}]$ for any $c>0$ since $P_\alpha$ has dimension $-d_\alpha$.
If the theory is asymptotically safe at the exact level, all
${u}_\alpha(k)$ approach constant values ${u}_{\alpha}^*$ for 
$k\rightarrow\infty$:
\begin{eqnarray}
\label{fixpact}
\Gamma_{k\rightarrow\infty}[g_{\mu\nu}]&=&\sum_\alpha{u}_{\alpha}^*\,
P_\alpha[k^2\,g_{\mu\nu}] \, .
\end{eqnarray}
Obviously this functional depends on $k^2$ and $g_{\mu\nu}$ only via the
combination $k^2\,g_{\mu\nu}$. Therefore the solutions of the corresponding
field equation, $\big<g_{\mu\nu}\big>_k$, scale proportional to $k^{-2}$, and
this is exactly the scaling behavior (\ref{UVscaling}).

The fact that \eqref{metricscaling} implies a fractal structure
of space-time at short distances can be argued as follows. Since in absence of dimensionful constants of integration
$\bar{\lambda}_k$ is the only quantity in this equation which sets a 
scale, every solution to (\ref{einsteq}) has a typical radius of curvature 
$r_c(k)\propto 1/\sqrt{\bar{\lambda}_k}$. (For instance, the maximally 
symmetric $S^4$-solution has the radius $r_c=r=\sqrt{3/\bar{\lambda}_k}$.) 
If we want to explore the space-time structure at a fixed length scale $\ell$ 
we should use the action $\Gamma_k[g_{\mu\nu}]$ at $k=1/\ell$ because with 
this functional a tree level analysis is sufficient to describe the essential 
physics at this scale, including the relevant quantum effects. Hence, when we 
observe the space-time with a microscope of resolution $\ell$, we will see an 
average radius of curvature given by 
$r_c(\ell)\equiv r_c(k=1/\ell)$. Once $\ell$ is
smaller than the Planck length $\ell_{\rm Pl}\equiv m_{\rm Pl}^{-1}$
we are in the fixed point regime where $\bar{\lambda}_k\propto k^2$ so that 
$r_c(k)\propto 1/k$, or
\begin{eqnarray}
\label{radius}
r_c(\ell)\propto\ell \, . 
\end{eqnarray}
Thus, when we look at the structure of space-time with a microscope of 
resolution $\ell\ll\ell_{\rm Pl}$, the average radius 
of curvature which we measure is proportional to the resolution 
itself. If we want to probe finer details and decrease $\ell$ we automatically
decrease $r_c$ and hence {\it in}crease the average curvature. Space-time seems
to be more strongly curved at small distances than at larger ones. The 
scale-free relation (\ref{radius}) suggests that at distances below the Planck
length the QEG space-time is a special kind of fractal with a self-similar 
structure. It has no intrinsic scale because in the fractal regime, i.e., when 
the RG trajectory is still close to the NGFP, the parameters which usually
set the scales of the gravitational interaction, $G$ and $\bar{\lambda}$, are 
not yet ``frozen out''. This happens only later on, somewhere half way between
the non-Gaussian and the Gaussian fixed point, at a scale of the order of 
$m_{\rm Pl}$. Below this scale, $G_k $ and $\bar{\lambda}_k$ stop running and, as a result,
$r_c(k)$ becomes independent of $k$ so that $r_c(\ell)={\rm const}$ for 
$\ell\gg\ell_{\rm Pl}$. In this regime $\big<g_{\mu\nu}\big>_k$ is 
$k$-independent, indicating that the macroscopic space-time is describable by a
single smooth, classical Riemannian manifold.

%----------------------------------------------------------------------------------
\subsection{Dynamical dimensional reduction of the graviton propagator}
\label{sect:7.3}
%----------------------------------------------------------------------------------
As another consequence of the NGFP, ref.\ \cite{oliver1} observed that the 4-dimensional graviton propagator
undergoes a dynamical dimensional reduction at high energies.
 The corresponding 
argument is based upon the anomalous dimension $\eta_N\equiv- \partial_t
\ln Z_{Nk}$, eq.\ \eqref{G14}, which for the RG
trajectories of the Einstein-Hilbert truncation (within its domain of validity)
have $\eta_N\approx 0$ for $k\rightarrow 0$\footnote{In the case of Type IIIa
trajectories \cite{frank1,h3} the macroscopic $k$-value is still far above
$k_{\rm term}$, i.e., in the ``GR regime'' described in \cite{h3}.}
and $\eta_N\approx -2$ for 
$k\rightarrow\infty$. The smooth change by two units occurs near 
$k\approx m_{\rm Pl}$. 

This information can be used to determine the 
momentum dependence of the dressed graviton propagator for momenta $p^2\gg
m_{\rm Pl}^2$. Expanding 
\begin{eqnarray}
\label{3in2}
\Gamma_k[g]=\left(16\pi G_k\right)^{-1}\int d^4x\,\sqrt{g}\left\{
-R+2\bar{\lambda}_k\right\} + \text{gauge fixing}
\end{eqnarray}
about flat space
and omitting the standard tensor structures we find the inverse propagator
$\widetilde{\cal G}_k(p)^{-1}\propto Z_N(k)\,p^2$. The conventional dressed 
propagator $\widetilde{\cal G}(p)$ contained in $\Gamma\equiv\Gamma_{k=0}$ 
obtains from the {\it exact} $\widetilde{\cal G}_k$ in the limit 
$k\rightarrow 0$.
For $p^2>k^2\gg m_{\rm Pl}^2$ the actual cutoff scale is the physical momentum
$p^2$ itself\footnote{See section 1 of ref. \cite{h1} for a detailed discussion
of ``decoupling'' phenomena of this kind.}
so that the $k$-evolution of $\widetilde{\cal G}_k(p)$ stops at the threshold 
$k=\sqrt{p^2}$. Therefore
\begin{eqnarray}
\label{gp1}
\widetilde{\cal G}(p)^{-1}\propto\;Z_N\left(k=\sqrt{p^2}\right)\,p^2\propto\;
(p^2)^{1-\frac{\eta}{2}}
\end{eqnarray}
because $Z_N(k)\propto k^{-\eta}$ when $\eta$ is 
(approximately) constant. In $d$ dimensions, and for $\eta\neq 2-d$, the
Fourier transform of $\widetilde{\cal G}(p)\propto 1/(p^2)^{1-\eta/2}$ yields 
the following propagator in position space:
\begin{eqnarray}
\label{gp2}
{\cal G}(x;y)\propto\;\frac{1}{\left|x-y\right|^{d-2+\eta}}\;.
\end{eqnarray}
This form of the propagator is well known from the theory of critical 
phenomena, for instance. (In the latter case it applies to large distances.)
Eq. (\ref{gp2}) is not valid directly at the NGFP. For $d=4$ and $\eta=-2$
the dressed propagator is $\widetilde{\cal G}(p)=1/p^4$ which has the following
representation in position space:
\begin{eqnarray}
\label{gp3}
{\cal G}(x;y)=-\frac{1}{8\pi^2}\,\ln\left(\mu\left|x-y\right|\right)\;.
\end{eqnarray}
Here $\mu$ is an arbitrary constant with the dimension of a mass. Obviously
(\ref{gp3}) has the same form as a $1/p^2$-propagator in 2 dimensions.

Slightly away from the NGFP, before other physical scales intervene, the 
propagator is of the familiar type (\ref{gp2}) which shows that the quantity 
$\eta_N$
has the standard interpretation of an anomalous dimension in the sense that
fluctuation effects modify the decay properties of ${\cal G}$ so as to 
correspond to a space-time of effective dimensionality $4+\eta_N$. 
Thus the properties of the RG trajectories imply a remarkable dimensional
reduction: Space-time, probed by a ``graviton'' with $p^2\ll m_{\rm Pl}^2$ is
4-dimensional, but it appears to be 2-dimensional for a graviton with 
$p^2\gg m_{\rm Pl}^2$.

It is interesting to note that in $d$ classical dimensions, where the 
macroscopic space-time is $d$-dimensional, the anomalous dimension at the
fixed point is $\eta=2-d$. Therefore, for any $d$, the dimensionality of the
fractal as implied by $\eta_N$ is $d+\eta=2$ \cite{oliver1,oliver2}.

%-------------------------------------------------------------
\section{Fractal dimensions within QEG}
\setcounter{equation}{0}
\label{sect:7d}
%-------------------------------------------------------------
Having encountered the first evidence for the fractal nature
of the effective QEG space-times, we now quantify their 
properties by computing their spectral, walk, and Hausdorff dimensions
introduced in appendix \ref{sect:7c} in the framework of QEG \cite{oliverfrac,frankfrac}.
 These generalized
dimensions can readily be obtained by studying diffusion
processes of test particles on the effective space-times.

%-------------------------------------------------------------
\subsection{Diffusion processes on QEG space-times}
%-------------------------------------------------------------
Since in QEG one integrates over all metrics, the central idea is to replace the classical return probability of a diffusing test particle $P_g(T)$, eq.\ \eqref{eqA3}, by its expectation value
\be\label{QPgT}
P(T) \equiv \langle P_\gamma(T) \rangle \equiv \int \cD \gamma \cD C \cD \bar{C} \, P_\gamma(T) \, e^{-S_{\rm bare}[\gamma, C, \bar{C}] } \, .
\ee
Here $\gamma_{\m\nu}$ denotes the microscopic metric and $S_{\rm bare}$ is the bare action related to the UV fixed 
point, with the gauge-fixing and the pieces containing the ghosts $C$ and $\bar{C}$ included. For the untraced heat kernel \eqref{heat1}, we define likewise
\be\label{heatexp}
K(x, x^\prime; T) \equiv \langle K_\gamma(x, x^\prime; T) \rangle \, . 
\ee

Following the discussion in subsection \ref{sect:7.1}, these expectation values are most conveniently calculated from the effective average action.
 Since $\Gamma_k$ defines an effective field theory at the scale $k$ we know that $\langle \cO(\gamma_{\m\nu}) \rangle \approx \cO(\langle g_{\m\nu} \rangle_k)$ provided the observable $\cO$ involves only momentum scales of the order of $k$. We apply this rule to the RHS of the diffusion equation, $\cO = - \Delta_\gamma K_\gamma(x, x^\prime; T)$. The subtle issue here is the correct identification of $k$. If the diffusion process involves (approximately) only a small interval of scales near $k$ over which $\bar{\lambda}_k$ does not change much, the corresponding heat equation contains the operator $\Delta(k)$ for this specific, fixed value of $k$: $\p_T K(x, x^\prime; T) = - \Delta(k) K(x, x^\prime; T)$. Denoting the eigenvalues of $\Delta(k_0)$ by $\cE_n$ and the corresponding eigenfunctions by $\phi_n$, this equation is solved by
\be\label{heatk}
K(x, x^\prime; T) = \sum_n \phi_n(x) \phi_n(x^\prime) \exp\Big( - F(k^2) \cE_n T \Big) \, . 
\ee 
Here we introduced the convenient notation $F(k^2) \equiv \bar{\lambda}_k/\bar{\lambda}_{k_0}$. Knowing the propagation kernel, we can time-evolve any initial probability distribution $p(x; 0)$ according to 
\be
p(x; T) = \int d^dx^\prime \sqrt{g_0(x^\prime)} \, K(x, x^\prime; T) \, p(x^\prime; 0)
\ee
 with $g_0$ the determinant of $\langle g_{\m\nu} \rangle_{k_0}$. If the initial distribution has an eigenfunction expansion of the form $p(x; 0) = \sum_n C_n \phi_n(x)$ we obtain
\be\label{prob2}
p(x; T) = \sum_n C_n \phi_n(x) \exp\Big( - F(k^2) \cE_n T \Big) \, .
\ee

If the $C_n$'s are significantly different from zero only for a single eigenvalue $\cE_N$, we are dealing with a single-scale problem and would identify $k^2 = \cE_N$ as the relevant scale at which the running couplings are to be evaluated. In general the $C_n$'s are different from zero over a wide range of eigenvalues. In this case we face a multiscale problem where different modes $\phi_n$ probe the space-time on different length scales. If $\Delta(k_0)$ corresponds to flat space, say, the eigenfunctions $\phi_n = \phi_p$ are plane waves with momentum $p^\m$, and they resolve structures on a length scale $\ell$ of order $1/|p|$. Hence, in terms of the eigenvalue $\cE_n \equiv \cE_p = p^2$ the resolution is $\ell \approx 1/\sqrt{\cE_n}$. This suggests that when the manifold is probed by a mode with eigenvalue $\cE_n$ it ``sees'' the metric $\langle g_{\m\nu} \rangle_k$ for the scale $k = \sqrt{\cE_n}$. Actually, the identification $k = \sqrt{\cE_n}$ is correct also for curved space since, in the construction of $\Gamma_k$, the parameter $k$ is introduced precisely as a cutoff in the spectrum of the covariant Laplacian.

As a consequence, under the spectral sum of \eqref{prob2}, we must use the scale $k^2 = \cE_n$ which depends explicitly on the resolving power of the corresponding mode. Likewise, in eq.\ \eqref{heatk}, $F(k^2)$ is to be interpreted as $F(\cE_n)$:
\be\label{2.19}
\begin{split}
K(x, x^\prime; T) = & \, \sum_n \phi_n(x) \phi_n(x^\prime) \exp\Big(-F(\cE_n) \cE_n T \Big) \\
= & \, \sum_n \phi_n(x)  \exp\Big(-F\big(\Delta(k_0)\big) \Delta(k_0) T \Big) \phi_n(x^\prime) \, .
\end{split} 
\ee
As in \cite{oliverfrac}, we choose $k_0$ as a macroscopic scale in the classical regime, and we assume that at $k_0$ the cosmological constant is small, so that $\langle g_{\m\nu} \rangle_{k_0}$ can be approximated by the flat metric on $\mathbb{R}^d$. The eigenfunctions of $\Delta(k_0)$ are plane waves then and eq.\ \eqref{2.19} becomes
\be\label{Heat:FlatSpace}
K(x, x^\prime; T) = \int \frac{d^dp}{(2 \pi)^d} \, e^{i p \cdot (x-x^\prime)} \, e^{-p^2 F(p^2) T}
\ee
where the scalar products are performed with respect to the flat metric, $\langle g_{\m\nu} \rangle_{k_0} = \delta_{\m\nu}$. The kernel \eqref{Heat:FlatSpace} satisfies $K(x, x^\prime; 0) = \delta^d(x-x^\prime)$ and, provided that $\lim_{p \rightarrow 0} p^2 F(p^2) = 0$, also $\int d^dx K(x, x^\prime; T) = 1$. 

Taking the normalized trace of \eqref{Heat:FlatSpace} within this ``flat space-approximation'' yields \cite{oliverfrac}
\be\label{2.21}
P(T) = \int \frac{d^dp}{(2 \pi)^d} \, e^{-p^2 F(p^2) T} \, .
\ee
Introducing $z = p^2$, the final result for the average return probability reads
\be\label{2.22}
P(T) = \frac{1}{(4\pi)^{d/2}} \, \frac{1}{\Gamma(d/2)} \int_0^\infty dz \, z^{d/2-1} \, \exp\Big(-z F(z) T \Big) \, , 
\ee
where $F(z) \equiv \bar{\lambda}(k^2 = z)/\bar{\lambda}_{k_0}$. In the classical case, $F(z) = 1$, this relation 
reproduces the familiar result $P(T) = 1/(4 \pi T)^{d/2}$. Thus, in this case, the spectral dimension \eqref{DsT} is given by $\cD_s(T) = d$ independently of $T$.
%
%-------------------------------------------------------------
\subsection{The spectral dimension in QEG}
%-------------------------------------------------------------

We shall now discuss the spectral dimension \eqref{DsT} for several other illustrative and important examples.

\noindent
{\bf (A)} To start with, let us evaluate the average return probability \eqref{2.22} for a simplified RG trajectory where the scale dependence of the cosmological constant is given by a power law, with the same exponent $\delta$ for all values of $k$:
\be\label{2.30}
\bar{\lambda}_k \propto k^\delta \quad \Longrightarrow \quad F(z) \propto z^{\delta/2} \, . 
\ee
By rescaling the integration variable in \eqref{2.22} we see that in this case
\be\label{2.31}
P(T) = \frac{\rm const}{T^{d/(2+\delta)}} \, . 
\ee
Hence \eqref{DsT} yields the important result
\be\label{powerspect}
\framebox{\; \; $\cD_s(T) = \frac{2d}{2+\delta}$ \bigg. \; \;} \, .
\ee
It happens to be $T$-independent, so that for $T \rightarrow 0$ trivially
\be\label{2.33}
d_s = \frac{2d}{2+\delta} \, . 
\ee

\noindent
{\bf (B)} Next, let us be slightly more general and assume
that the power law \eqref{2.30} is valid only for squared momenta in a 
certain interval, $p^2 \in [z_1, z_2]$, but $\bar{\lambda}_k$ remains unspecified 
otherwise. In this case we can obtain only partial information about $P(T)$,
namely for $T$ in the interval $[z_2^{-1}, z_1^{-1}]$. The reason is that for $T \in [z_2^{-1}, z_1^{-1}]$
the integral in \eqref{2.22} is dominated by momenta for which 
approximately $1/p^2 \approx T$, i.e., $z \in [z_1, z_2]$. This leads us again to the formula
\eqref{powerspect}, which now, however, is valid only for a restricted range of diffusion times $T$; 
in particular the spectral dimension of interest may not be given by extrapolating \eqref{powerspect} to $T \rightarrow 0$.

\noindent
{\bf (C)} Let us consider an arbitrary asymptotically safe RG trajectory so that its behavior for $k \rightarrow \infty$ is controlled by
the NGFP. For $k \ge m_{\rm Pl}$ the scale-dependence of the cosmological constant is governed by the fixed point, i.e., 
$\bar{\lambda}_k = \lambda_* k^2$, independently of $d$.
This corresponds to a power law with $\delta = 2$, which entails in the {\bf NGFP regime}, i.e., for $T \lesssim 1/m_{\rm Pl}^2$,
\be\label{UVspec}
\cD_s(T) = \frac{d}{2} \qquad \quad \Big( \mbox{NGFP regime} \Big) \, . 
\ee 
This dimension, again, is locally $T$-independent. It coincides with the $T \rightarrow 0$ limit:
\be
d_s = \frac{d}{2} \, .
\ee
This is the result first derived in ref.\ \cite{oliverfrac}. As it was explained there, it is actually an exact consequence of Asymptotic Safety
which relies solely on the existence of the NGFP and does not depend on the Einstein-Hilbert truncation.

\noindent
{\bf (D)} Returning to the Einstein-Hilbert truncation, let us consider the piece of the Type IIIa RG trajectory depicted in Fig.\ \ref{Fig.epl} which lies inside the linear regime of the GFP. Newton's constant is approximately $k$-independent there and the cosmological constant evolves according to
\be\label{2.36}
\bar{\lambda}_k = \bar{\lambda}_0 + \nu G_0 k^d.
\ee
Here $\nu = (4 \pi)^{1-d/2} (d-3) \Phi^1_{d/2}(0)$ is a scheme-dependent constant \cite{mr,frank1}. When
$k$ is not too small, so that $\bar{\lambda}_0$ can be neglected relative to $\nu G_0 k^d$, we are in what we shall call the ``$k^d$ regime''; it is characterized by a pure power law 
$\bar{\lambda}_k \approx k^\delta$ with $\delta = d$. The physics behind this scale dependence is simple and well-known: It
represents exactly the vacuum energy density obtained by
summing up the zero point energies of all field modes integrated out. For $T$ in the range of scales pertaining to the $k^d$ regime we find
\be
\cD_s(T) = \frac{2d}{2+d} \qquad (k^d \; \mbox{regime}) \, . 
\ee
Note that for every $d > 2$ the spectral dimension in the $k^d$ regime is even {\it smaller} than in the NGFP regime.
%-------------------------------------------------------------
\subsection{The walk dimension in QEG}
%-------------------------------------------------------------
In order to determine the walk dimension for the diffusion on the effective QEG space-times
we return to eq.\ \eqref{Heat:FlatSpace} for the untraced heat kernel. We restrict ourselves to a regime with 
a power law running of $\bar{\lambda}_k$, whence $F(p^2) = (Lp)^\delta$ with some constant length-scale $L$.

Introducing $q_\m \equiv p_\m T^{1/(2+\delta)}$ and $\xi_\m \equiv (x_\m - x_\m^\prime) / T^{1/(2+\delta)}$ we can rewrite \eqref{Heat:FlatSpace}
in the form
\be\label{2.45}
K(x, x^\prime; T) = \frac{1}{T^{d/(2+\delta)}} \, \Phi\left( \frac{|x-x^\prime|}{T^{1/(2+\delta)}} \right)
\ee
with the function
\be
\Phi(|\xi|) \equiv \int \frac{d^dq}{(2\pi)^d} \, e^{i q \cdot \xi} \, e^{- L^\delta q^{2+\delta}} \, .
\ee
For $\delta = 0$, this obviously reproduces \eqref{FSHeat}. From the argument of $\Phi$ in \eqref{2.45} we infer that $r = |x-x^\prime|$ scales as $T^{1/(2+\delta)}$ so that the walk
dimension can be read off as
\be\label{walkspect}
\framebox{\; \; $\cD_w(T) = 2+\delta$ \bigg. \; \;} \, . 
\ee
In analogy with the spectral dimension, we use the notation $\cD_w(T)$ rather than $d_w$ to indicate that it might refer to an approximate scaling law which is valid for a finite range of scales only.

For $\delta = 0, 2$, and $d$ we find in particular, for any topological dimension $d$, 
\be\label{2.48}
\cD_w = \left\{ 
\begin{array}{cl}
2   & \mbox{classical regime} \\
4   & \mbox{NGFP regime} \\
2+d \, \,  & k^d\mbox{ regime} \\
\end{array}
\right.
\ee
Regimes with all three walk dimensions of \eqref{2.48} can be realized along a single RG trajectory. Again, the result for the NGFP regime, $\cD_w = 4$, is exact in the sense that it does not rely on the Einstein-Hilbert truncation.

%-------------------------------------------------------------
\subsection{The Hausdorff dimension in QEG}
%-------------------------------------------------------------
The smooth manifold underlying QEG has per se no fractal properties 
whatsoever. In particular, the volume of a $d$-ball $\cB^d$ covering a patch
of the smooth manifold of QEG space-time scales as
\be
V(\cB^d) = \int_{\cB^d} d^dx \sqrt{g_k} \propto (r_k)^d \, . 
\ee 
Thus, by comparing to eq.\ \eqref{Hdd}, we read off that the 
Hausdorff dimension is strictly equal
to the topological one:
\be\label{hdspect}
\framebox{\; \; $d_H = d$ \bigg. \; \;} \, . 
\ee

We emphasize that the effective QEG space-times should {\it not} be visualized as a kind of 
sponge. Their fractal-like properties have no simple geometric interpretation; they are not due
to a ``removing'' of space-time points. Rather they are of an entirely {\it dynamical} nature,
reflecting certain properties of the {\it quantum states} the system ``space-time metric'' can be in.

%---------------------------------------------------------
\subsection{The Alexander-Orbach relation}
%---------------------------------------------------------

For standard fractals the quantities $d_s$, $d_w$, and $d_H$ are not independent but are related by 
\cite{orbach}
\be\label{fracrel}
\frac{d_s}{2} = \frac{d_H}{d_w} \, .
\ee
By combining eqs.\ \eqref{powerspect}, \eqref{walkspect}, and \eqref{hdspect} we see that the same
relation holds true for the effective QEG space-times, at least within the Einstein-Hilbert approximation
and when the underlying RG trajectory is in a regime with power-law scaling of $\bar{\lambda}_k$. For every value
of the exponent $\delta$ we have
\be\label{rel}
\frac{\cD_s(T)}{2} = \frac{d_H}{\cD_w(T)} \, . 
\ee

The results $d_H = d$, $\cD_w = 2 + \delta$ imply that, as soon as
$\delta > d-2$, we have $\cD_w > d_H$ and the random walk is {\it recurrent} then \cite{avra}.
Classically ($\delta = 0$) this condition is met only in low dimensions $d < 2$, but 
in the case of the QEG space-times it is always satisfied in the $k^d$ regime $(\delta = d)$, for example.
So also from this perspective the QEG space-times, due to the specific quantum gravitational dynamics
to which they owe their existence, appear to have a dimensionality smaller than their topological one.

It is particularly intriguing that, in the NGFP regime, $\cD_w = 4$ independently of $d$. Hence the walk
is recurrent $(\cD_w > d_H)$ for $d < 4$, non-recurrent for $d > 4$, and the marginal case $\cD_w = d_H$ is realized
if and only if $d=4$, making $d=4$ a distinguished value. Notably, there is another feature of the QEG space-times which singles out $d=4$: it is the only
dimensionality for which $\cD_s$(NGFP regime)$=d/2$ coincides with the effective dimension
$d_{\rm eff} = d + \eta_* = 2$ derived from the graviton propagator in section \ref{sect:7.3}.

%-------------------------------------------------------------
\section{The scale-dependence of $\cD_s$ and $\cD_w$}
\setcounter{equation}{0}
\label{sect:7e}
%-------------------------------------------------------------
%
We now proceed by discussing the scale-dependence of the spectral and walk dimension, arising within the Einstein-Hilbert truncation.
For this purpose, we consider an arbitrary RG trajectory $k \mapsto (g_k, \lambda_k)$. Along such an RG trajectory there might be isolated intervals of $k$-values where the cosmological constant 
evolves according to a power law, $\bar{\lambda}_k \propto k^\delta$, for some constant exponents $\delta$ which are not necessarily the same on different such
intervals. If the intervals are sufficiently long, it is meaningful to ascribe a spectral and walk dimension to them since $\delta = {\rm const}$  implies $k$-independent values 
$\cD_s = 2d/(2+\delta)$ and $\cD_w = 2 + \delta$.

In between the intervals of approximately constant $\cD_s$ and $\cD_w$, where the $k$-dependence of $\bar{\lambda}_k$ is not a
power law,
the notion of a spectral or walk dimension might not be meaningful. The concept of a {\it scale-dependent} dimension
$\cD_s$ or $\cD_w$ is to some extent arbitrary with respect to the way it connects the ``plateaus'' on which $\delta = {\rm const}$  
for some extended period of RG time. While RG methods allow the computation of the $\cD_s$ and $\cD_w$ values on the various plateaus,
it is a matter of convention how to combine them into continuous functions $k \mapsto \cD_s(k), \cD_w(k)$ which interpolate between
the respective values.

%-------------------------------------------------------------
\subsection{The exponent $\delta$ as a function on theory space}
%-------------------------------------------------------------
In this subsection, we describe a special proposal for a $k$-dependent $\cD_s(k)$ and $\cD_w(k)$ which is motivated by technical simplicity and the general insights
it allows. We retain eqs.\ \eqref{powerspect} and \eqref{walkspect}, but promote $\delta \rightarrow \delta(k)$ to a $k$-dependent quantity 
\be\label{defscale}
\delta(k) \equiv k \p_k \ln(\bar{\lambda}_k) \, .
\ee
When $\bar{\lambda}_k$ satisfies a power law, $\bar{\lambda}_k \propto k^\delta$ this relation reduces to the case of constant
$\delta$.
If not, $\delta$ has its own scale dependence, but no direct physical interpretation should be attributed to it. The particular definition \eqref{defscale} has the special property that it actually can be evaluated without first solving for the RG trajectory. The function
$\delta(k)$ can be seen as arising from a certain scalar function on theory space, $\delta = \delta(g, \lambda)$, whose $k$-dependence results from inserting an RG trajectory: $\delta(k) \equiv \delta(g_k, \lambda_k)$. In fact, \eqref{defscale} implies
$\delta(k) = k \p_k \ln(k^2 \lambda_k) = 2 + \lambda_k^{-1} k \p_k \lambda_k$
so that
$\delta(k) = 2 + \lambda^{-1}_k \beta_\lambda(g_k, \lambda_k)$ upon using the RG equation $k \p_k \lambda_k = \beta_\lambda(g_k, \lambda_k)$. Thus when we consider
$\delta$ as a function on theory space, coordinatized by $g$ and $\lambda$, it reads
\be\label{deltatheoryspace}
\delta(g, \lambda) = 2 + \frac{1}{\lambda} \, \beta_\lambda(g, \lambda) \, . 
\ee
Substituting this relation into
 \eqref{powerspect} and \eqref{walkspect}, the spectral and the walk dimensions become functions
on the $g$-$\lambda$-plane
\be\label{dstheo}
\cD_s(g, \lambda) = \frac{2d}{4 + \lambda^{-1} \beta_\lambda(g, \lambda)} \, , \qquad \cD_w(g, \lambda) = 4 + \lambda^{-1} \beta_\lambda(g, \lambda) \, .
\ee

To evaluate these expressions further, we use the beta functions \eqref{betafcts}.
\begin{figure}[t]
	\centering
	\includegraphics[width=0.77\textwidth]{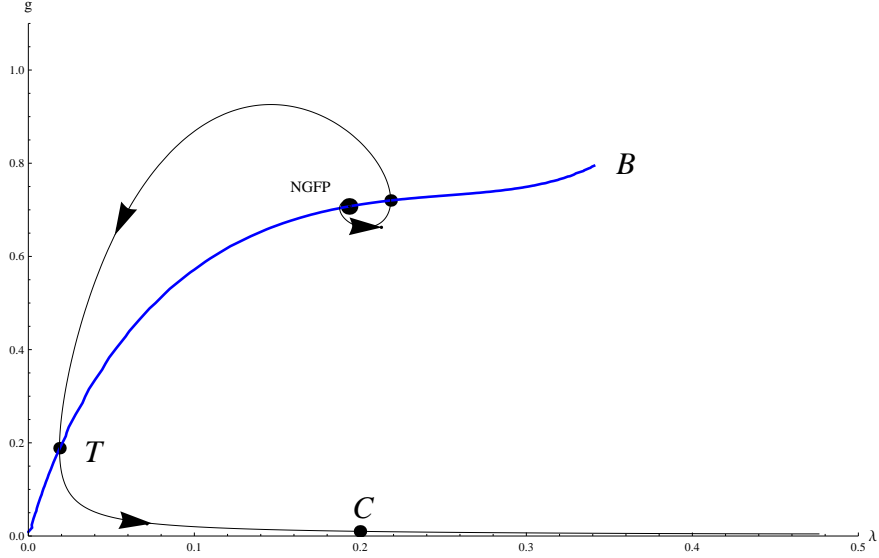}
	\caption{\small The $g$-$\lambda-$theory space with the line of turning points, $\cB$, and a
		typical trajectory of Type IIIa. The arrows point in the direction of decreasing $k$. The big
		black dot indicates the NGFP while the smaller dots represent points at which the RG trajectory switches
		from increasing to decreasing $\lambda$ or vice versa. The point $T$ is the lowest turning point,
		and $C$ is a typical point within the classical regime. For $\lambda \gtrsim 0.35$, the RG flow leaves
		the classical regime and is no longer reliably captured by the Einstein-Hilbert truncation. (From \cite{frankfrac}.)}
  \label{Fig.epl}
\end{figure}
As we already discussed, the scaling regime of a NGFP has the exponent $\delta = 2$. From eq.\ \eqref{deltatheoryspace} we learn that this value is realized at all points $(g, \lambda)$ where $\beta_\lambda =0$. The second condition for the NGFP, $\beta_g = 0$, is not required here, so that we have $\delta = 2$ along the entire line in theory space:
\be\label{2.75}
\cB = \Big\{ \, (g, \lambda) \, \Big| \, \beta_\lambda(g, \lambda) = 0 \, \Big\} \, . 
\ee
For $d=4$ the curve $\cB$ is shown as the bold blue line in fig.\ \ref{Fig.epl}. Both the GFP $(g, \lambda) = (0, 0)$ and the NGFP, $(g, \lambda) = (g_*, \lambda_*)$, are located on this curve. Furthermore, the turning points $T$ of all Type IIIa trajectories are also situated on $\cB$, and the same holds for all the higher order turning points which occur when the trajectory spirals around the NGFP. This observation leads us to an important conclusion: The values $\delta = 2 \Longleftrightarrow  \cD_s = d/2, \cD_w = 4$ which (without involving any truncation) are found in the NGFP regime, actually also apply to all points $(g, \lambda) \in \cB$, provided the Einstein-Hilbert truncation is reliable and no matter is included. 

%-------------------------------------------------------------
\subsection{The spectral and walk dimensions along a RG trajectory}
%-------------------------------------------------------------
We proceed by investigating how the spectral and walk dimension of the effective QEG space-times change along a given RG trajectory.
Our interest is in scaling regimes where $\cD_s$ and $\cD_w$ remain (approximately) constant for a long interval of $k$-values.
For the remainder of this subsection, we will restrict ourselves to the case $d=4$ for concreteness. 

We start by numerically solving the coupled differential equations \eqref{betaeq} 
with the beta functions \eqref{betafcts} evaluated with the optimized cutoff for a series of initial conditions keeping $\lambda_{\rm init} = \lambda(k_0) = 0.2$ fixed and successively lowering
$g_{\rm init} = g(k_0)$. The result is a family of RG trajectories where the classical regime becomes more and more pronounced. These solutions are substituted into \eqref{dstheo}, which give $\cD_s(t; g_{\rm init}, \lambda_{\rm init})$ and $\cD_w(t; g_{\rm init}, \lambda_{\rm init})$ in dependence of the RG-time $t \equiv \ln(k)$ and the RG trajectory. One can verify explicitly, that substituting the RG trajectory into the return probability \eqref{2.22} and computing the spectral dimension from \eqref{DsTlim} by carrying out the resulting integrals numerically gives rise to the same picture.

\begin{figure}[t]
	\centering
	\includegraphics[width=0.45\textwidth]{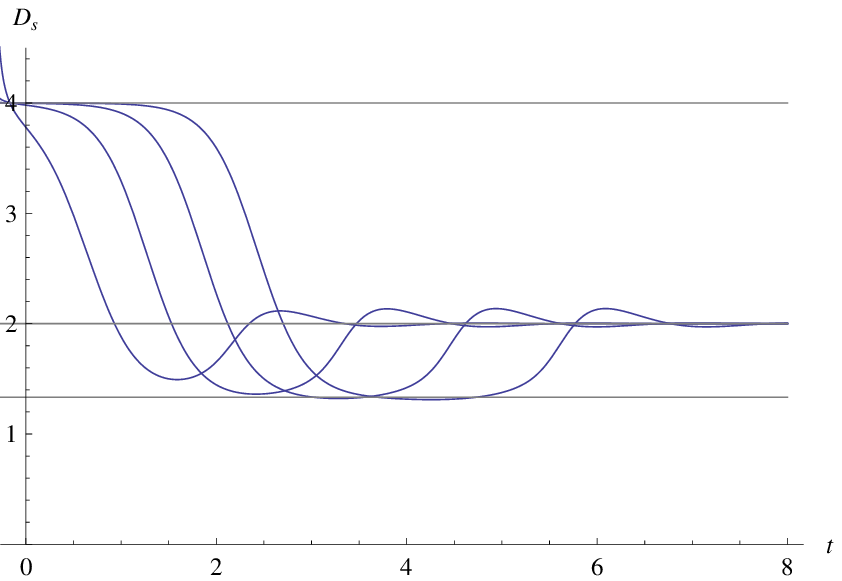} \, \, \, 
	\includegraphics[width=0.45\textwidth]{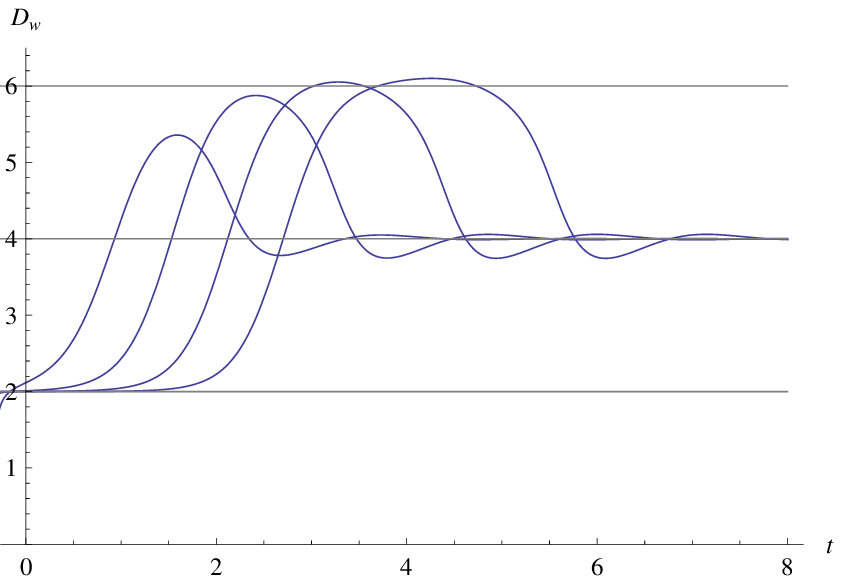} \\[1.2ex]
	\includegraphics[width=0.45\textwidth]{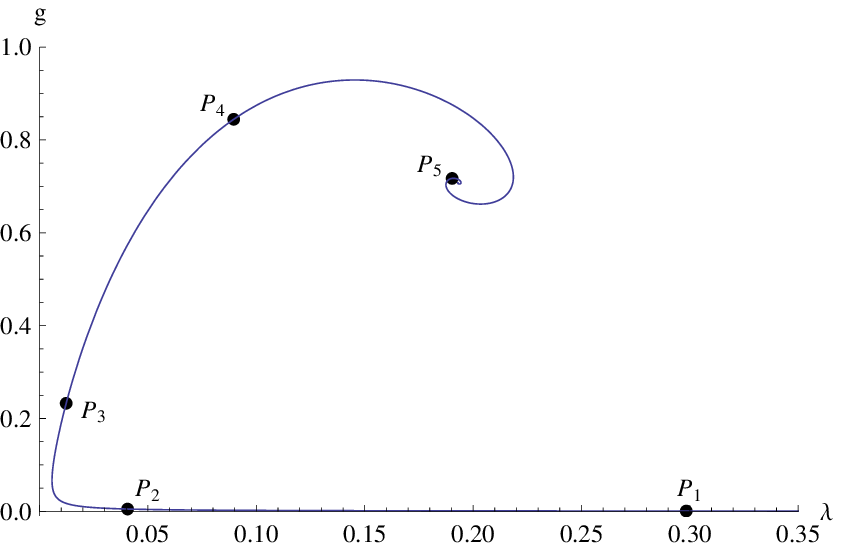}
\caption{\small The $t \equiv \ln(k)$-dependent spectral dimension (upper left) and walk dimension (upper right) along
	illustrative solutions of the RG equations \eqref{betaeq} in $d=4$. The trajectories develop three plateaus: the
	classical plateau with $\cD_s =4, \cD_w = 2$, the semi-classical plateau where $\cD_s = 4/3, \cD_w = 6$ and the NGFP
	plateau with $\cD_s = 2, \cD_w = 4$. These plateau values are indicated by the gray horizontal lines and connected
	by crossover parts. The lower figure shows the location of these plateaus on the RG trajectory: the classical, $k^4$,
	and NGFP regime appear between the points $P_1$ and $P_2$, $P_3$ and $P_4$, and above $P_5$, respectively. (From \cite{frankfrac}.)}
\label{Fig.spec}
\end{figure}

Fig.\ \ref{Fig.spec} then shows the resulting spectral dimension, the walk dimension, and the localization of the plateau-regimes on the RG trajectory in the top-left, top-right and lower diagram, respectively. In the top diagrams, $g_{\rm init}$ decreases by one order of magnitude for each shown trajectory, starting with the highest value to the very left.
As a central result, fig.\ \ref{Fig.spec} establishes that the RG flow gives rise to {\it three} plateaus where $\cD_s(t)$ and $\cD_w(t)$ are approximately constant: \\
{\bf (i)} For small values $k$, below $t \simeq 1.8$, say, one finds a {\it classical plateau} where $\cD_s = 4, \cD_w = 2$ for a long range of $k$-values. Here $\delta = 0$, indicating that the cosmological constant is indeed constant. \\
{\bf (ii)} Following the RG flow towards the UV (larger values of $t$) one next encounters the {\it semi-classical plateau} where $\cD_s = 4/3, \cD_w = 6$. In this case $\delta(k) = 4$ so that $\bar{\lambda}_k \propto k^4$ on the corresponding part of the RG trajectory. \\
{\bf (iii)} Finally, the {\it NGFP plateau} is characterized by $\cD_s = 2, \cD_w = 4$, which results from the scale-dependence of the cosmological constant at the NGFP $\bar{\lambda}_k \propto k^2 \Longleftrightarrow  \delta = 2$. \\

At this stage, it is worthwhile to see which parts of a typical RG trajectory realize the scaling relations underlying the plateau-values of $\cD_s$ and $\cD_w$. This is depicted in the lower diagram of fig.\ \ref{Fig.spec} where we singled out the third solution with $g_{\rm init} = 10^{-3}$ for illustrative purposes. In this case the classical plateau is bounded by the points $P_1$ and $P_2$ and appears well below the turning point $T$, while the semi-classical plateau is situated between the points $P_3$ and $P_4$ well above the turning point. The NGFP plateau is realized by the piece of the RG trajectory between $P_5$ and the NGFP. The turning point $T$ is not situated in any scaling region but appears along the crossover from the classical to the semi-classical regime of the QEG space-times. For $t < 0$, the spectral dimension (walk dimension) increases (decreases) rapidly. In this region, the underlying RG trajectory is evaluated outside the classical regime at points $\lambda \gtrsim 0.35$. In this region of the theory space, the Einstein-Hilbert truncation is no longer trustworthy, so that this rapid increase of $\cD_s$ is most likely an artifact, arising from the use of an insufficient truncation.

The plateaus observed above become more and more extended the closer the trajectories turning point $T$ gets to the GFP, i.e., the smaller the IR value of the cosmological constant. The first RG trajectory with the largest value $g_{\rm init} = 0.1$ does not even develop a classical and semi-classical plateau, so that a certain level of fine-tuning of the initial conditions is required in order to make these structures visible. Interestingly enough, when one matches the observed data against the RG trajectories of the Einstein-Hilbert truncation \cite{h3,entropy} one finds that the 
``RG trajectory realized by Nature'' displays a very extreme fine-tuning of this sort. The coordinates of the turning point are approximately $g_{T} \approx \lambda_{T} \approx 10^{-60}$ and it is passed at the scale $k_{T} \approx 10^{-30} m_{\rm Pl} \approx 10^{-2} {\rm eV} \approx (10^{-2} {\rm mm})^{-1}$, so that there will be very pronounced plateau structures in this case.

%-------------------------------------------------------------
\subsection{Matching the spectral dimensions of QEG and CDT}
\label{sect:7f}
%-------------------------------------------------------------
The key advantage of the spectral dimension $\cD_s(T)$ is that it may 
be defined and computed within various a priori unrelated approaches to
quantum gravity. In particular, it is easily accessible in Monte Carlo simulations
of the Causal Dynamical Triangulations (CDT) approach in $d=4$ \cite{ajl34} and $d=3$ \cite{Benedetti:2009ge} as well as
in Euclidean Dynamical Triangulations (EDT) \cite{laiho-coumbe}. This feature allows a direct comparison between
 $\cD_s^{\rm CDT}(T)$ and $\cD_s^{\rm EDT}(T)$ obtained within the discrete approaches and $\cD_s^{\rm QEG}(T)$ capturing the fractal properties of the QEG effective space-times.
We conclude this section by reviewing this comparison for $d=3$, following ref.\ \cite{frankfrac}.\footnote{For a similar fit in
Ho\v{r}ava gravity see ref.\ \cite{Horavafit}.} In particular we shall determine the
specific RG trajectory of QEG which, we believe, underlies the numerical data obtained in \cite{Benedetti:2009ge}.

\begin{figure}[t]
	\centering
	\includegraphics[width=0.5\textwidth]{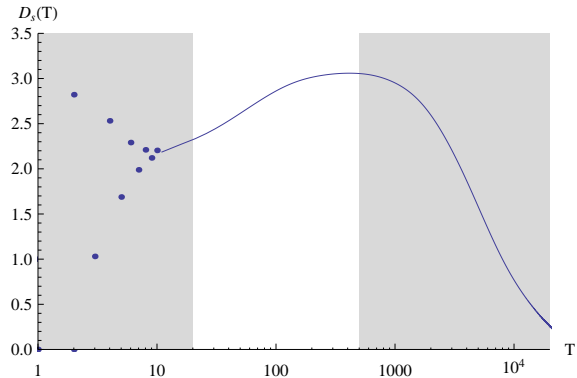}
\caption{\small Spectral dimension $\cD_s^{\rm CDT}(T)$ determined from random walks on a CDT space-time
	built from $N = 200$k simplices \cite{Benedetti:2009ge}.}
\label{Fig.CDT}
\end{figure}

Let us start by looking into the typical features of the spectral dimension $\cD_s^{\rm CDT}(T)$ obtained from the simulations. A prototypical data set showing 
$\cD_s^{\rm CDT}(T)$ as function of the length of the random walk $T$ is given in fig.\ \ref{Fig.CDT}.
The resulting curve is conveniently split into three regimes: \\
{\bf (i)} For $T \le 20$, corresponding to the left gray region in fig.\ \ref{Fig.CDT}, $\cD_s^{\rm CDT}(T)$ undergoes rapid oscillations. These originate from the discrete structure of the triangulation to which the short random walks are particular sensitive. \\
{\bf (ii)} For long random walks with $T \ge 500$, the data shows an exponential fall-off. This feature is due to the compact nature of the triangulation, which implies that for long random walks $\cD_s^{\rm CDT}(T)$ is governed by the lowest eigenvalue of the Laplacian on the compact space. This regime is marked by the right gray region in fig.\ \ref{Fig.CDT}. \\
{\bf (iii)} Between these two regimes, $\cD_s^{\rm CDT}(T)$ is affected neither by the discreteness nor the compactness of the triangulation. Since for $\cD_s^{\rm QEG}(T)$, determined by the flat-space approximation \eqref{2.21}, we do not expect any of these effects to appear, we use this middle region to compare the $T$-dependent spectral dimensions arising from the two, a priori different, approaches. 

The comparison between the CDT-data and the QEG effective space-times is carried out as follows: \\
{\bf (i)} First, we numerically construct a RG trajectory $g_k(g_0, \lambda_0), \lambda_k(g_0, \lambda_0)$ depending on the initial conditions $g_0, \lambda_0$, by solving the flow equations \eqref{betaeq}. \\
{\bf (ii)} Subsequently, we evaluate the resulting spectral dimension $\cD_s^{\rm QEG}(T; g_0, \lambda_0)$ of the corresponding effective QEG space-time.
This is done by first finding the return probability $P(T; g_0, \lambda_0)$, eq.\ \eqref{2.22}, for the RG trajectory under consideration and then substituting the resulting expression into \eqref{DsT}.  Besides on the length of the random walk, the spectral dimension constructed in this way also depends on the initial conditions of the RG trajectory. \\ 
{\bf (iii)} Finally, we determine the RG trajectory underlying the CDT-simulations by 
fitting the parameters $g_0, \lambda_0$ to the Monte Carlo data. The corresponding best-fit values are obtained via an ordinary least-square fit, minimizing the squared Euclidean distance 
\be\label{lsf}
(\Delta \cD_s )^2 \equiv \sum_{T = 20}^{500} \, \left( \cD_s^{\rm QEG}(T; g_0^{\rm fit}, \lambda_0^{\rm fit}) - \cD_s^{\rm CDT}(T) \right)^2 \, , 
\ee
 between the (continuous) function $\cD_s^{\rm QEG}(T; g_0, \lambda_0)$ and the points $\cD_s^{\rm CDT}(T)$. We thereby restrict ourselves to 
the random walks with discrete, integer length $20 \le T \le 500$, which constitute the white part of fig.\ \ref{Fig.CDT} and correspond to the regime {\bf (iii)} discussed above. 

\begin{table}
\renewcommand{\arraystretch}{1.5}
\begin{center}
\begin{tabular}{c c c c }
\hline
        & \quad \qquad $g_0^{\rm fit}$ \qquad \quad & \quad \qquad $\lambda_0^{\rm fit}$ \qquad \qquad & \quad \quad  \quad $(\Delta \cD_s )^2$ \quad  \quad \quad \\ \hline\hline
 \quad $70$k\quad &  $0.7 \times 10^{-5}$ & $ 7.5 \times 10^{-5}$  & $0.680$ \\
 \quad$100$k\quad	&  $8.8 \times 10^{-5}$ & $39.5 \times 10^{-5}$  & $0.318$ \\
% $140$k &  $13.4 \times 10^{-5}$& $51 \times 10^{-5}$   & $0.076$ & $0.606$ \\
 \quad$200$k\quad &  $13 \times 10^{-5}$  & $61 \times 10^{-5}$    & $0.257$ \\ \hline
\end{tabular}
\end{center}
\renewcommand{\arraystretch}{1}
\caption{\small Initial conditions $g_0^{\rm fit}, \lambda_0^{\rm fit}$ for the RG trajectory providing
	the best fit to the Monte Carlo data \cite{Benedetti:2009ge}. The fit-quality $(\Delta \cD_s )^2$,
	given by the sum of the squared residues, improves systematically when increasing the number of simplices
	in the triangulation.}
\label{Table.1}
\end{table}
The resulting best-fit values $g_0^{\rm fit}, \lambda_0^{\rm fit}$ for the triangulations with $N = 70.000$, $N=100.000$, and $N=200.000$ simplices are collected in table \ref{Table.1}.
Notably, the sum over the squared residuals in the third column of the table  
 improves systematically with an increasing number of simplices. By integrating
 the flow equation for $g(k), \lambda(k)$ for the best-fit initial conditions one furthermore observes that the points $g_0^{\rm fit}, \lambda_0^{\rm fit}$ are actually located 
on {\it different} RG trajectories. Increasing the size of the simulation $N$ leads to a mild, but systematic increase of the distance between the turning point $T$ and the GFP of the corresponding best-fit trajectories.

\begin{figure}[t]
	\centering
	\includegraphics[width=0.45\textwidth]{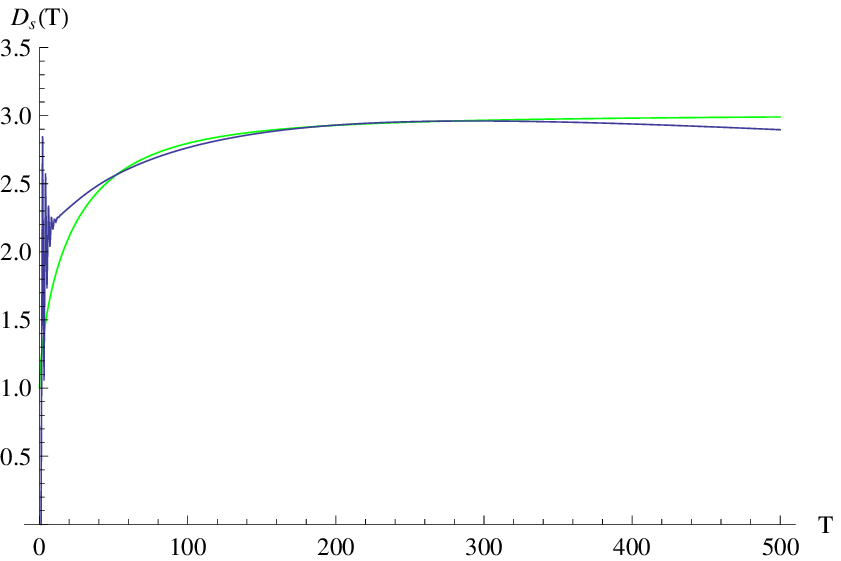} \;\;
	\includegraphics[width=0.45\textwidth]{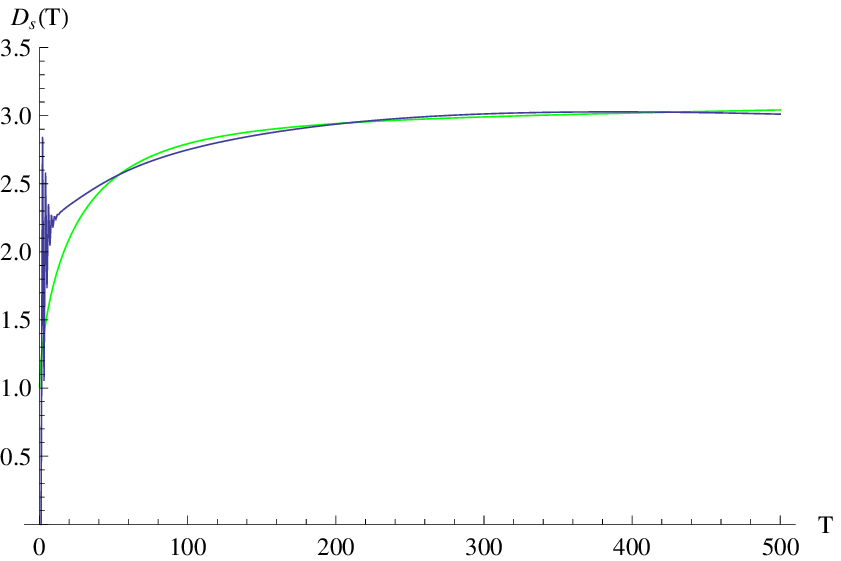} \\
	\includegraphics[width=0.45\textwidth]{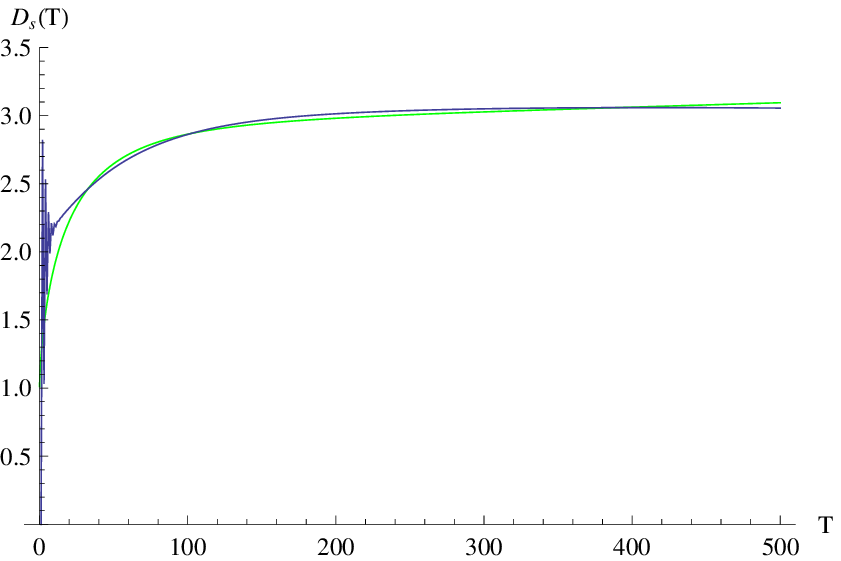} \;\;
	\includegraphics[width=0.45\textwidth]{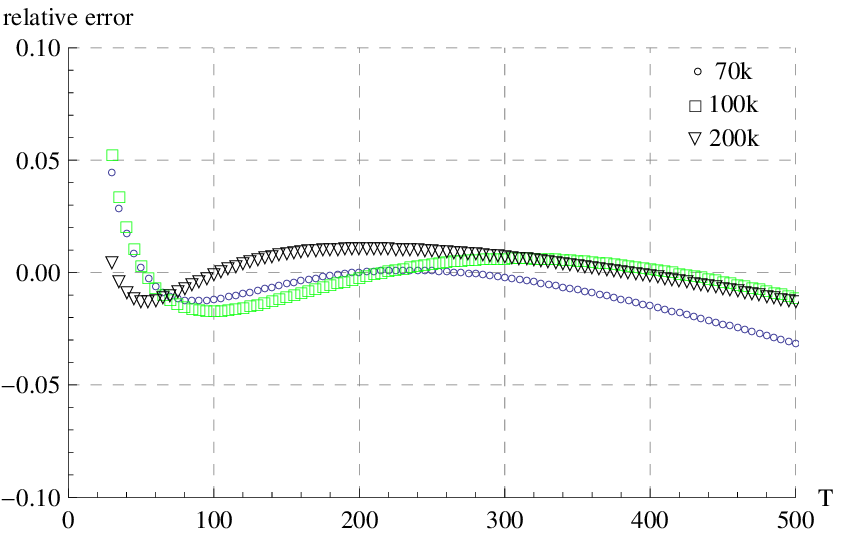}
\caption{\small Comparison between the 3-dimensional CDT data-sets $70$k (upper left), $100$k (upper right),
	and $200$k (lower left) obtained in \cite{Benedetti:2009ge} (blue curves) and the best fit values for
	$\cD_s^{\rm QEG}(T; g_0^{\rm fit}, \lambda_0^{\rm fit})$ (green curves). The relative errors for the fits
	to the CDT-datasets with $N = 70.000$ (circles), $N=100.000$ (squares) and $N = 200.000$ (triangles) simplices
	are shown in the lower right. The residuals grow for very small and very large durations $T$ of the random walk,
	consistent with discreteness effects at small distances and the compactness of the simulation for large values
	of $T$, respectively. The quality of the fit improves systematically for triangulations containing more
	simplices. For the $N=200$k data the relative error is $\approx 1\%$. (From \cite{frankfrac}.)}
\label{p.fit1}
\end{figure}

Fig.\ \ref{p.fit1} then shows the direct comparison between the spectral dimensions obtained by the simulations (blue curves) and the best-fit QEG trajectories (green curves)
for $70$k, $100$k and $200$k simplices in the upper left, upper right and lower left panel, respectively. 
 This data is complemented by the relative error 
\be
\epsilon \equiv - \frac{\cD_s^{\rm QEG}(T; g_0^{\rm fit}, \lambda_0^{\rm fit}) - \cD_s^{\rm CDT}(T)}{\cD_s^{\rm QEG}(T; g_0^{\rm fit}, \lambda_0^{\rm fit})}
\ee
for the three fits in the lower right panel. The $70$k data still shows a systematic deviation from the classical value $\cD_s(T) = 3$ for long random walks, which is not present in the QEG results. This mismatch decreases systematically for larger triangulations where the classical regime becomes more and more pronounced. Nevertheless and most remarkably we find that for the $200$k-triangulation that $\epsilon \lesssim 1\%$, throughout. All three sets of residues thereby show a systematic oscillatory structure. These originate from tiny oscillations in the CDT data which are not reproduced by $\cD_s^{\rm QEG}(T)$. Such oscillations commonly appear in systems with discrete symmetries \cite{calcagni-reviews} and are thus likely to be absent in the continuum computation. As a curiosity, we observe that the QEG result matching the most extensive simulation with $N = 200$k ``overshoots'' the classical value $\cD_s(T) = 3$, yielding $\cD_s^{\rm QEG}(T) > 3$ for $T \gtrsim 450$. At this stage, the RG trajectory is evaluated outside the classical regime in a region of theory space where the Einstein-Hilbert approximation starts to become unreliable. It is tempting to speculate that larger triangulations may also be sensitive to quantum gravity effects at distances beyond the classical regime.

\begin{figure}
	\centering
	\includegraphics[width=0.75\textwidth]{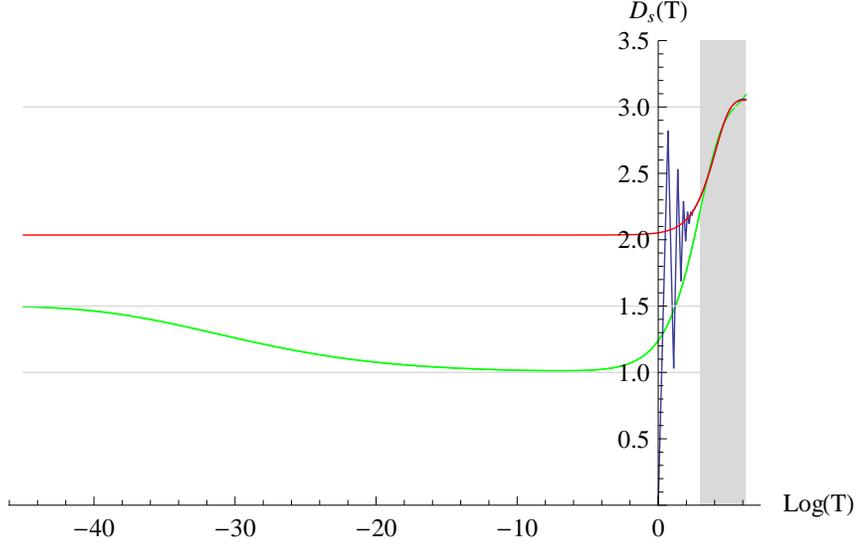}
\caption{\small 
Extrapolation of the CDT-data for the spectral dimension measured from Monte Carlo simulations with 200.000 simplices \cite{Benedetti:2009ge} (blue curve) to infinitesimal random walks.
The green and the red curve show the QEG-prediction $\cD_s^{\rm QEG}(T; g_0^{\rm fit}, \lambda_0^{\rm fit})$ and the best-fit function \eqref{CDT-fit} employed in \cite{Benedetti:2009ge}, respectively. Notably, the scaling regime corresponding to the NGFP is reached
for $\log(T) < -40$, which is well below the distance scales probed by the Monte Carlo simulation.
}
\label{p.fit2}
\end{figure}

We conclude this section by extending $\cD_s^{\rm QEG}(T; g_0^{\rm fit}, \lambda_0^{\rm fit})$ obtained from the $200$k data to the region of very short random walks $T < 20$. The result is depicted in fig.\ \ref{p.fit2} which displays the CDT-data $\cD_s^{\rm CDT}(T)$ (blue curve), the best-fit QEG result $\cD_s^{\rm QEG}(T; g_0^{\rm fit}, \lambda_0^{\rm fit})$ (green curve) and the best-fit function
\be\label{CDT-fit}
\cD_s^{\rm CDT-fit}(T) = a + b \e^{- c T} \, , \qquad a = 3.05 \, , \; \; b = -1.02 \, , \; \; c = 0.017 \, , 
\ee
employed in \cite{Benedetti:2009ge} (red curve) as a function of $\log(T)$.
Similarly to the 4-dimensional case discussed in fig.\ \ref{Fig.spec}, the function $\cD_s^{\rm QEG}(T; g_0^{\rm fit}, \lambda_0^{\rm fit})$ obtained for $d=3$ develops three plateaus where the spectral dimension is approximately constant over a long $T$-interval. For successively decreasing duration of the random walks, these plateaus correspond to the classical regime $\cD_s^{\rm QEG}(T) = 3$, the semi-classical regime where $\cD_s^{\rm QEG}(T) \approx 1$ and the NGFP regime where $\cD_s^{\rm QEG}(T) = 3/2$. The figure illustrates that $\cD_s^{\rm CDT}(T)$ probes the classical regime and part of the first crossover towards the semi-classical regime only. This is in perfect agreement with the assertion \cite{Benedetti:2009ge} that \emph{the present simulations do not yet probe structures below the Planck scale.} The different predictions for the spectral dimensions obtained for infinitesimal random walks solely arise from different extrapolation schemes.

%-------------------------------------------------------------
\section{Concluding remarks}
\setcounter{equation}{0}
\label{sect:7g}
%-------------------------------------------------------------
In this article we reviewed the basic concepts underlying the gravitational Asymptotic Safety program and summarized evidence supporting the existence of the non-Gaussian renormalization group fixed point (NGFP) at the heart of this scenario. The NGFP constitutes a highly non-trivial feature of the gravitational renormalization group flow and defines a consistent and predictive quantum theory of gravity, Quantum Einstein Gravity or QEG for short. Based on the continuum average action approach \cite{mr}, we then outlined our present understanding of the multifractal features characteristic for the effective QEG space-times. These structures are at least to some extend captured by the spectral, walk, and Hausdorff dimension seen by a fictitious diffusion process set up on the effective space-times. Their explicit computation allows a direct comparison between the results obtained from the continuum functional renormalization group and other approaches to quantum gravity like the discrete Causal Dynamical Triangulations \cite{ajl1,ajl2,ajl34,Ambjorn:2009ts,Kommu:2011wd} and Euclidean Dynamical Triangulation \cite{laiho-coumbe} programs. We close our review with the following comments:

\noindent
{\bf (1)} We stress that the construction of an effective average action for gravity reviewed in
section \ref{sect:3} represents a {\it ``background independent''} approach to quantum gravity.
Somewhat paradoxically, this background independence is achieved by means of the background field
formalism: One fixes an arbitrary background, quantizes the fluctuation field in this background,
and afterwards adjusts $\bar{g}_{\mu \nu}$ in such a way that the expectation value of the
fluctuation vanishes: $\bar{h}_{\mu \nu} = 0$. In this way the background gets fixed dynamically.
The results of refs.\ \cite{creh1,creh2} suggest that a proper implementation of ``background
independence'' might be crucial for Asymptotic Safety.

\noindent
{\bf (2)} The combination of the effective average action with the background field method has been
successfully tested within conventional field theory. In QED and Yang-Mills type gauge theories it
reproduces the known results and extends them into the non-perturbative domain \cite{ym,ymrev}.

\noindent
{\bf (3)} The coexistence of Asymptotic Safety and perturbative non-renormalizability is well understood.
In particular upon fixing $\bg_{\mu \nu} = \eta_{\mu \nu}$ and expanding the trace on its RHS in powers
of $G$ the FRGE reproduces the divergences of perturbation theory; see ref.\ \cite{livrev} for a detailed
discussion of this point.

\noindent
{\bf (4)} It is to be emphasized that in the average action framework the RG flow, i.e., the vector
field $\vec{\beta}$, is completely determined once a theory space is fixed. As a consequence, the choice
of theory space determines the set of fixed points $\Gamma^*$ at which asymptotically safe theories can
be defined. Therefore, in the Asymptotic Safety scenario the bare action $S$ related to $\Gamma^*$ is a
{\it prediction} of the theory rather than an ad hoc postulate as usually in quantum field theory.
Ambiguities could arise only if there is more than one suitable NGFP.

\noindent
{\bf (5)} According to the results available to date, the Einstein-Hilbert action of classical General Relativity seems not to play any distinguished role in the Asymptotic Safety context, at least not at the conceptual level. The only known NGFP on the theory space of QEG has the structure $\Gamma^* = \mbox{Einstein-Hilbert action} + \mbox{``more''}$ where ``more'' stands for both local and non-local corrections. So it seems that the Einstein-Hilbert action is only an approximation to the true fixed point action, albeit an approximation which was found to be rather reliable for many purposes.

\noindent
{\bf (6)} A solution to the FRGE, as such, does not define a regularized functional integral;
a priori it is not even clear whether or not a path integral representation exists for a given
RG trajectory. In \cite{elisa1} it was shown how, upon specification of a UV regularized measure,
the information contained in $\Gamma_{k\rightarrow\infty}$ may be used to obtain the cutoff dependence
of the bare action. A thorough understanding of this relationship is essential in comparing
the average action to other approaches, such as Monte Carlo simulations of discrete models of gravity,
for instance.

\noindent
{\bf (7)} Any quantum theory of gravity must reproduce the successes of classical General Relativity. As for QEG, it cannot be expected that this will happen for all RG trajectories in $\cS_{\rm UV}$, but it should happen for some or at least one of them. Within the Einstein-Hilbert truncation it has been shown \cite{h3} that there actually do exist trajectories (of Type IIIa) which have an extended classical regime and are consistent with all observations.

\noindent
{\bf (8)} In the classical regime mentioned above the dynamics of space-time geometry is unaffected by quantum effects to a very good approximation. In this regime the familiar methods of quantum field theory in curved classical space-times apply, and it is clear therefore that effects such as Hawking radiation or cosmological particle production are reproduced by the general framework of QEG with matter.

\noindent
{\bf (9)} Coupling free massless matter fields to gravity, 
it turned out \cite{perper1} that the fixed point continues to exist under very 
weak conditions concerning the number of various types 
of matter fields (scalars, fermions, etc.). No fine tuning 
with respect to the matter multiplets is necessary. In particular 
 Asymptotic Safety does not seem to require 
any special constraints or symmetries among the matter fields such as 
supersymmetry, for instance. 

\noindent
{\bf (10)} Since the NGFP seems to exist already in pure gravity it is 
likely that a widespread prejudice about 
gravity may be incorrect: its quantization seems not to  
require any kind of unification with the other fundamental 
interactions. 

\noindent
{\bf (11)}
To the best of our knowledge, the effective space-times arising within QEG carry a multifractal structure \cite{frankfrac}. Some of these features are captured by the spectral, walk, and Hausdorff dimension seen by a fictitious diffusion process set up on the effective space-time. The resulting Hausdorff dimension is constant and equal to the topological dimension of the (background) space-time. Thus the fractal properties do not originate from the QEG space-times ``loosing points'' at short distances but rather represent a genuine dynamical effect of quantum field theory. In contrast to the Hausdorff dimension the spectral dimension and the walk dimension seen by the diffusion process depend on the diffusion time. Fig.\ \ref{Fig.spec} identifies \emph{three} regimes in which these generalized dimensions are constant for a wide range of scales, the classical, semi-classical and NGFP regime, which are connected by short crossovers.  

\noindent
{\bf (12)} In section \ref{sect:7f} we performed a direct comparison between the spectral dimension of the 3-dimensional effective QEG space-times with the one measured in Causal Dynamical Triangulations (CDT) \cite{Benedetti:2009ge}. Notably, the best-fit RG trajectory reproduces the CDT data with approximately 1\% accuracy for the range of diffusion times where the simulation data is reliable. Notably current Monte Carlo data neither probes the semi-classical plateau nor the scaling regime of the NGFP. Determining the spectral dimension for infinitesimal random walks by extrapolating the leading quantum corrections to the classical regime may therefore miss the imprints of the NGFP on the structure of space-time.

\noindent
\textbf{Acknowledgements}\\
We thank D.~Benedetti and J.~Henson for sharing the Monte Carlo data underlying their work \cite{Benedetti:2009ge} with us.
We are also grateful to A.~Nink for a careful reading of the manuscript. The research of F.S.\ is supported by the Deutsche Forschungsgemeinschaft (DFG)
within the Emmy-Noether program (Grant SA/1975 1-1).

%-------------------------------------------------------------
\begin{appendix}
%\label{App:1}
%-------------------------------------------------------------

%-------------------------------------------------------------
\section{Generalized dimensions on classical manifolds}
\setcounter{equation}{0}
\label{sect:7c}
%-------------------------------------------------------------
Investigating random walks and diffusion processes on fractals, one is led to introduce 
various notions of fractal dimensions, such as the spectral or walk dimension \cite{avra}.
These notions also prove useful when characterizing properties of space-time in quantum gravity, and 
we will review these concepts in the remainder of this section.
%-------------------------------------------------------------
\subsection{The spectral dimension}
%-------------------------------------------------------------
To start with, consider the diffusion process where a spin-less test particle performs a Brownian random walk on
an ordinary Riemannian manifold with a fixed classical metric $g_{\mu\nu}(x)$. It is described by the heat kernel
$K_g(x, x^\prime; T)$ which gives the probability density for a transition of the particle from $x$ to $x^\prime$ during 
the fictitious time $T$. It satisfies the heat equation
\be %\label{heateq}
\p_T K_g(x, x^\prime; T) = - \Delta_g K_g(x, x^\prime; T) \, ,
\ee
where $\Delta_g = - D^2$ denotes the Laplace operator. In flat space, this equation is easily solved by
\be\label{heat1}
K_g(x, x^\prime; T) = \int \frac{d^dp}{(2\pi)^d} \, \e^{i p \cdot (x-x^\prime)} \, \e^{-p^2 T} 
\ee
In general, the heat kernel is a matrix element of the operator $\exp(- T \Delta_g)$. In the random walk picture
its trace per unit volume,
\be\label{eqA3}
P_g(T) = V^{-1} \int d^dx \sqrt{g(x)} \, K_g(x, x; T) \equiv V^{-1} \, \Tr \, \exp(- T \Delta_g) \, , 
\ee
has the interpretation of an average return probability. Here $V \equiv \int d^dx \sqrt{g(x)}$ denotes the total volume. It is well known that $P_g$ possesses an asymptotic early time expansion (for $T \rightarrow 0$) of the form $P_g(T) = (4 \pi T)^{-d/2} \sum_{n=0}^\infty A_n T^n$, with $A_n$ denoting the Seeley-DeWitt coefficients. From this expansion one can motivate the definition of the spectral dimension $d_s$ as the $T$-independent logarithmic derivative
\be \label{DsTlim} %\label{dflatspace}
d_s \equiv \left. - 2 \frac{d \ln P_g(T)}{d \ln T} \right|_{T = 0} \, . 
\ee
On smooth manifolds, where the early time expansion of $P_g(T)$ is valid, the spectral dimension agrees with the topological dimension $d$ of the manifold.

Given $P_g(T)$, it is natural to define a, in general $T$-dependent, generalization of the spectral dimension by
\be\label{DsT}
\cD_s(T) \equiv - 2 \frac{d \ln P_g(T)}{d \ln T}\, .
\ee
 According to \eqref{DsTlim}, we recover the true spectral dimension of the space-time by 
considering the shortest possible random walks, i.e., by taking the limit $d_s =  \lim_{T \rightarrow 0} \cD_s(T)$.
Note that in view of a possible comparison with other (discrete) approaches to quantum gravity
 the generalized, scale-dependent version \eqref{DsT} plays a central role.

%-------------------------------------------------------------
\subsection{The walk dimension}
%-------------------------------------------------------------
Regular Brownian motion in flat space has the celebrated property
that the random walker's average square displacement increases linearly
with time: $\langle r^2 \rangle \propto T$. Indeed, performing the integral
\eqref{heat1} we obtain the familiar probability density
\be\label{FSHeat}
K(x, x^\prime; T) = (4 \pi T)^{-d/2} \exp\left(- \frac{\sigma(x, x^\prime)}{2 T} \right)
\ee
with $\sigma(x, x^\prime) = \half |x - x^\prime |^2$ half the squared geodesic distance between the points $x, x^\prime$. 
Using \eqref{FSHeat} yields the expectation value $ \langle r^2 \rangle \equiv \langle x^2 \rangle = \int d^dx \, x^2 \, K(x, 0; T) \propto T$.

Many diffusion processes of physical interest (such as diffusion on fractals) are anomalous in the sense that this linear relationship is generalized to a power law
$\langle r^2 \rangle \propto T^{2/d_w}$ with $d_w \not = 2$. The interpretation of the so-called walk dimension $d_w$ is as follows. The trail left by the random walker is a random object, which is interesting in its own right. It has the properties of a fractal, even in the ``classical'' case when the walk takes place on a regular manifold. The quantity $d_w$ is precisely the fractal dimension of this trail. Diffusion processes are called regular if $d_w = 2$, and anomalous when $d_w \not = 2$.  

%-------------------------------------------------------------
\subsection{The Hausdorff dimension}
%-------------------------------------------------------------
Finally, we introduce the Hausdorff dimension $d_H$. Instead of working with its mathematically rigorous definition in terms of the Hausdorff measure and all possible covers of the metric space under consideration, the present, simplified definition may suffice for our present purposes. On a smooth set, the scaling law for the volume
$V(r)$ of a $d$-dimensional ball of radius $r$ takes the form 
\be\label{Hdd}
V(r) \propto r^{d_H} \, .
\ee
The Hausdorff dimension is then obtained in the limit of infinitely small radius,
\be
d_H \equiv \lim_{r \rightarrow 0} \frac{\ln V(r)}{\ln r} \, .
\ee
 Contrary to the spectral or walk dimension whose definitions are linked to dynamical diffusion 
 processes on space-time, there is no such dynamics associated with $d_H$. 

%-------------------------------------------------------------
\end{appendix}
%-------------------------------------------------------------

%------------------ Bibtex Bibliography -------------------

%----------------------------------------------------------
%
\end{spacing}
\end{document}